\documentclass[11pt]{article}
\usepackage{jheppub}
\pdfoutput=1
\usepackage{psfrag}
\usepackage{array}
\usepackage{amssymb}
\usepackage{amsmath}
\usepackage{amsfonts}
\usepackage{amsthm}
\usepackage{mathtools}
\usepackage{graphicx}
\usepackage{xcolor}
\usepackage{placeins}
\usepackage[labelsep=quad]{subcaption}
\usepackage[utf8]{inputenc}

\usepackage{tikz}
\newcommand{\beq}{\begin{equation}}
\newcommand{\eeq}{\end{equation}}
\newcommand{\beqn}{\begin{eqnarray}}
\newcommand{\eeqn}{\end{eqnarray}}
\newcommand{\pa}{\partial}

\newcommand{\bw}{\boldsymbol{\omega}}
\usepackage{comment}

\def\gf{{\cal G}}
\def\tgf{\tilde{{\cal G}}}

\def\zpi{{\cal Z}}

\def\wf{{\cal W}}

\def\dm2{{D-2 \over 2}}
\def \half{\frac{1}{2}}
\def\dflat{\mathcal{D}^{\flat}}

\def\zbar{\Bar{Z}}
\graphicspath{ {./graphics/}}

\title{The S-matrix and boundary correlators in flat space}
\author[a,1]{Diksha Jain,\note{diksha.2012jain@gmail.com}}
\author[b,2]{Suman Kundu,\note{kundusuman1994@gmail.com
}}
\author[a,3]{Shiraz Minwalla,\note{minwalla@theory.tifr.res.in}}
\author[a,4]{Onkar Parrikar,\note{parrikar@theory.tifr.res.in}}
\author[a,5]{Siddharth G. Prabhu,\note{siddharth.g.prabhu@gmail.com}}
\author[a,c,6]{Pushkal Shrivastava.\note{pushkalshrivastava@g.harvard.edu}}
\affiliation[a]{Department of Theoretical Physics, Tata Institute
	of Fundamental Research, Homi Bhabha Rd, \\Mumbai 400005, India}
\affiliation[b]{Department of Particle Physics and Astrophysics, Weizmann Institute of Science, \\Rehovot 76100, Israel}
\affiliation[c]{Department of Physics, Harvard University, 17 Oxford Street, Cambridge, MA 02138, USA}

\date{}
\begin{document}

\abstract{We consider the path integral of a quantum field theory in Minkowski spacetime with fixed boundary values (for the elementary fields) on asymptotic boundaries. We define and study the corresponding boundary correlation functions obtained by taking derivatives of this path integral with respect to the boundary values. The S-matrix of the QFT can be extracted directly from these boundary correlation functions after smearing. We interpret this relation in terms of coherent state quantization and derive the constraints on the path-integral as a function of boundary values that follow from the unitarity of the S-matrix. We then study the locality structure of boundary correlation functions. In the massive case, we find that the boundary correlation functions for generic locations of boundary points are dominated by a saddle point which has the interpretation of particles scattering in a small elevator in the bulk, where the location of the elevator is determined dynamically, and the S-matrix can be recovered after stripping off some dynamically determined but non-local ``renormalization'' factors. In the massless case, we find that while the boundary correlation functions are generically analytic as a function on the whole manifold of locations of boundary points, they have special singularities on a sub-manifold, points on which correspond to light-like scattering in the bulk. We completely characterize this singular scattering sub-manifold, and find that the corresponding residues of the boundary correlations at these singularities are precisely given by S-matrices. This analysis parallels the analysis of bulk-point singularities in AdS/CFT and generalizes it to the case of multi-bulk point singularities.}

\maketitle

\section{Introduction}

In AdS space, the on-shell bulk action (quantum mechanically the bulk path integral), can be computed as a functional of boundary values of the bulk fields. This quantity is,  perhaps, the best studied observable in the AdS/CFT correspondence, and is identified with the generating functional of correlation functions of local operators in  the dual CFT. The path integral as a functional of boundary values is also well defined in flat space,  both for theories of gravity as well as for quantum field theories.  It is natural to wonder about the interpretation and structural properties of this flat space observable.

There is at least one context in which the flat spacetime path integral as a functional of boundary values is closely related to a familiar, interesting, and very well studied flat space observable. Consider Lorentzian space cut off at two time slices - one in the early past at time $-T$, and the second in the late future at time $T$ (in the rest of this paper we refer to such a cut off spacetime as a `slab'). The path integral as a functional of fields on the past and future boundaries, computes a transition amplitude between arbitrary initial and final states, and so can be used to reconstruct the $S$ matrix (provided it exists) of the theory under study. 

Given this situation, it is natural to wonder 
\begin{itemize}
\item[$\mathcal{Q}$1.] What is the precise nature of the boundary conditions for the flat space version of the path integral as a functional of boundary values? How precisely can the S matrix be extracted from this object?
\item[$\mathcal{Q}$2.] Does this path integral contain additional information, apart from the S matrix?
\item[$\mathcal{Q}$3.] What are the properties of the analogues of CFT correlation functions - i.e. position dependent Taylor coefficients of the path integral?
\end{itemize}
In this paper, we investigate each of these questions. 

Let us start with the question $\mathcal{Q}$1. There are two natural boundary conditions one might consider imposing on our slab path integral. First, we might wish to specify the fields $\phi$ on the boundary of the slab (we call this the Dirichlet problem). Second, we might choose to specify the positive energy data of $\phi$ in the past, but its negative energy data in the future (we call this the `in - out' problem). The Dirichlet problem computes transition amplitudes between wave functionals in the Schrodinger basis (see Appendix  \ref{dirap} for details), while the in-out problem computes the same amplitudes in the coherent space basis (see subsection \ref{cbhm} and Appendix \ref{Harmonic} for a review). Even though these two amplitudes are distinct functionals of their respective boundary data\footnote{Of course they can be related to each other by performing the `Fourier Transform' that affects the change of basis.}, we find (see Appendix \ref{relpi}) that the S matrix can be extacted  from either these quantities in a very similar manner. 

The relationship of the in-out problem to the S matrix was studied in classic work by Arefeva, Faddeev, and Slavnov (AFS)\cite{Arefeva:1974jv,Faddeev:1980be,Balian:1976vq,PhysRevD.18.373, PhysRevD.37.1485}. 
Working in momentum space and at leading order in the large $T$ limit, AFS demonstrated that the S matrix can be identified with the Taylor coefficients of the  path integral as a functional of past positive and future negative energy data. In this paper, we establish a similar result for the Dirichlet problem, as we now explain. 

It is convenient to first work in Euclidean space. Consider the path integral $Z[\beta, \bar\beta]$ computed on the Euclidean slab spacetime, as a functional of the field values on the past/future boundaries, $\beta$ and ${\bar \beta}$.  Define the boundary correlation functions to be the Taylor series coefficients of this expansion:
\begin{equation} \label{gbdry} 
G_{\text{bdry}}(\{x_i\}, \{y_i\}):= \prod_{i=1}^{n}\frac{\delta }{\delta \beta(x_i)}\prod_{j=1}^{m}\frac{\delta }{\delta \bar{\beta}(y_j)}  Z[\phi]\Big|_{\beta,\bar{\beta}= 0}.
\end{equation} 
The functions $G_{\text{bdry}}(\{x_i\}, \{y_i\})$ (see Appendix \ref{eapt} for sample computations) are the flat space analogues of boundary CFT correlators of the AdS/CFT correspondence,
and are objects of independent potential interest in attempts to discover an (as yet unknown) independent holographic dual of bulk physics in flat spacetimes. In this paper, we study the boundary Euclidean correlator, $G_{\text{bdry}}(\{x_i\}, \{y_i\})$ ,  in the large $T$ limit and analytically continue this object to Lorentzian space. In \S\ref{smbo} we use methods similar to those employed in \cite{Harlow:2011ke} to demonstrate that this analytically continued object is related to the S matrix in a very clean way. In particular, we derive the formula \eqref{smatrixl}, which  asserts that the S matrix is a sort of `Fourier Transform' of the Lorentzian boundary correlation functions,  $G_{\rm bdry}$. The `Fourier modes' in this transform are the (restriction to the boundary of) the free scattering states whose S matrix we wish to compute 
\footnote{\eqref{smatrixl} - like all the formulas in this paper -  has been derived only for the special case of scalar fields. However, we see no obstruction to generalizing 
\eqref{smatrixl} - and most of the other formulae presented in this paper -to the study of more general scattering processes.}. 

Our formula \eqref{smatrixl} may be thought of as a rewriting of the LSZ formula in position space. It is also a very close analogue of the AFS formula reviewed above. It may initially seem surprising that the same operation 
(smearing with onshell wave functions) performed on two different objects (the in-out and Dirichlet path integrals as a functional of their boundary data) can yield the same final answer (the S matrix). In \S \ref{coherent} and Appendix \ref{relpi}, we explain how this works. 

Next, we turn to question $\mathcal{Q}$2. Clearly $Z[\beta, \bar \beta]$, carries more information than just the S matrix. In particular, it carries the information about the vacuum wave functional (in the Schrodinger basis in the case of the Dirichlet problem, and in the coherent state basis in the case of the in-out problem). In \S\ref{coherent}, we conjecture a relationship between certain singularities in a particular analytic continuation of the vacuum wave functional and the S matrix (see \S\ref{crvwf} for details). The conjecture described in \S\ref{crvwf} is similar, in several respects, to the conjectured relation presented in \cite{Maldacena:2011nz,Raju:2012zr,Arkani-Hamed:2017fdk,Arkani-Hamed:2018kmz,Benincasa:2018ssx,Goodhew:2020hob,Baumann:2022jpr,Pajer:2020wxk} between the singularities in an appropriately analytically continued Hartle-Hawking wave functional in de-Sitter space and the flat space S matrices. 

We turn next to question $\mathcal{Q}$3, i.e. to the study of the properties of the generating functional of `boundary correlators', $Z[\beta, \bar \beta]$ (we will return to aspects of question $\mathcal{Q}$2 later in this introduction).  Let us start with the in-out problem. We have explained above that, in this case, 
$Z[\beta, \bar \beta] = \langle \psi_{out}| U |\psi_{in}\rangle$ where $|\psi_{in}\rangle$ and $\langle \psi_{out}|$ are states in the coherent space basis and $U$ is the time evolution operator. Inserting a complete set of states into the unitarity equation $U^\dagger U=1$,  in \S\ref{ucpi} we obtain the  nonlinear equation \eqref{unitsm} that $Z[\beta, \bar \beta]$ is constrained to obey. We then rederive the same equation starting with Cutkowski's rules and using the AFS relationship to rewrite these rules gives a constraint on $Z[\beta, {\bar \beta}]$. This second derivation applies equally well to the Dirichlet problem, so we conclude that the Dirichlet path integral as a functional of boundary values also obeys \eqref{unitsm}. We view this equation as a potentially useful constraint on attempts to formulate a flat space version of holography. Any proposal for an independent definition of the bondary correlators  $G_{\text{bdry}}(\{x_i\}, \{y_i\})$ must obey \eqref{unitsm}. 

Continuing on question $\mathcal{Q}$3, we now turn to the study of the local properties of the `boundary correlators' $G_{\text{bdry}}(\{x_i\}, \{y_i\})$ (see \eqref{gbdry}) for both the Dirchlet as well as the in-out path integrals. 
In \S \ref{pimp} we first study a theory containing only massive particles of minimum mass $m$. We study the correlators with the mass held fixed, in the limit that $T$ (a measure of the size of our cut off spacetime) is taken to infinity. As $m T \rightarrow \infty$, we demonstrate that evaluation of $G_{\text{bdry}}(\{x_i\}, \{y_i\})$ is dominated by saddle point configurations, given by particle trajectories that start out from the initial time slice, meet and interact at a bulk point in a manner that conserves momentum, and then carry on to the final time slice \footnote{The scattering processes take place in an `elevator' of finite size. This is certainly the case for tree diagrams with contact interactions. Once we incorporate all relevant aspects of dynamics, however, this also turns out to be the case for exchange and loop diagrams. We illustrate how this works in an example. In the case of $2 \rightarrow n$ scattering,  exchange diagrams, involving intermediate particles of a mass greater than the sum of the masses of the scattering particles, appear to receive significant contributions from configurations in which the two interaction points are separated on macroscopic scales (see \S \ref{es}), i.e. length scales of order $T$. Whenever this happens, however, the intermediate particle is always 
unstable  Correcting its propagator to account for this fact (by summing the appropriate self-energy graphs) removes these contributions (see \S\ref{unst}). In the case of flat space (where $T$ is the largest parameter in the problem), these considerations remove the naively divergent contributions of the `spurious saddles' discussed in the context of AdS/CFT in \cite{Komatsu:2020sag}.}. The construction described above applies for all points in an open set in the space of boundary insertions. At arbitrary boundary locations (within the relevant open set), $G_{\rm bdry}$  is given by the S matrix of the associated scattering process, times a kinematically determined nonlocal `renormalization' factor (see \S \ref{hr}). Although the renormalization in question is a product of factors, one for each scattering particle, the values of these factors depend on the location of the bulk scattering point, which, in turn, depends on the locations of all boundary insertions, and so are nonlocal. The discussion of this paragraph has many similarities with the saddle point study of holographic correlators of very heavy operators in $AdS/CFT$ presented in \cite{Komatsu:2020sag}, with $R_{AdS}$ in \cite{Komatsu:2020sag} playing the role of $T$ in our analysis, and the limit $m R_{AdS} \rightarrow \infty$ in \cite{Komatsu:2020sag} playing the role of the limit $mT \rightarrow \infty$ in our analysis.\footnote{The authors of \cite{Komatsu:2020sag} mainly studied Euclidean correlators. The `exchange' saddle points studied in \S 3.3 of that paper (and their flat space analogues) exist only when the mass of the exchanged particle is smaller than twice the masses of the external particles (note this is the converse of the Lorentzian condition). In Appendix \ref{esap} we check that these saddles are not analytic continuations 
of the Lorentzian exchange saddles referred to above. A given set of masses 
and insertion locations admit either Lorentzian saddles or the Euclidean saddles or neither: Euclidean and Lorentzian saddles do not simultaneously 
exist, and never compete with each other for dominance. We would like to thank 
S. Raju and B. Van Rees for discussions on this point.}

In \S\ref{sec:massless} we next turn to an analysis of the local structure of $G_{\rm bdry}$ in the case of massless bulk fields (again our analysis applies to both the Dirichlet and the in-out path integrals). In this case the contribution of individual (for simplicity tree level) diagrams to $G_{\rm bdry}$ is not accurately determined by a saddle point analysis, even in the limit $T \to \infty$. Indeed, the contribution to $G_{\rm bdry}$ of any individual tree diagram scales with $T$ in a simple power law manner. This power law dependence multiplies a function of boundary value locations, which (by a Landau-type analysis, see \S \ref{singular}) can be shown to be analytic upto pole-type singularities. These singularities occur at boundary points that are so located that they are able to scatter (in a momentum-conserving manner) in the bulk. We see that  (the massless limit of) the saddle point equations we encountered for massive fields reappear in the massless case - but this time as equations that determine the locations of singularities of $G_{\rm bdry}$, rather than as equations that determine the value of $G_{\rm bdry}$ at generic locations\footnote{The reason that these saddle point equations do not admit solutions at generic locations of boundary points - in the case of massless particles - is explained in \S \ref{cod}.}.
We show in \S\ref{egsing} that the residues of these pole-type singularities are proportional to the S matrices of the corresponding scattering processes. The analytic structure of $G_{\rm bdry}$ is, therefore, very similar to that of boundary correlators in AdS/CFT: recall that these correlators are generically analytic, but have bulk point singularities on the `scattering sheet' at locations that are, once again, determined by a Landau-type analysis, and whose residues are, once again, proportional to the corresponding flat space S matrices \cite{maldacena2017looking,Gary:2009ae,Chandorkar:2021viw}.

Continuing with the analysis of massless boundary correlators, we name the subspace of boundary insertions at which $G_{\rm bdry}$ blows up the `scattering sub manifold'. 
It turns out (see \S \ref{cod}) that the codimension, $c$,  of this singular manifold (within the space of possible boundary insertions) is unity when the number of boundary insertions $m$ obeys $m \leq D+1$. When $m$ is greater than $D+1$, on the other hand,  the co-dimension of the scattering manifold equals $m-D$. In \S\ref{Nmatrix}, \S\ref{nprop} and \S\ref{geom} we explain the intricate geometry of scattering sub-manifolds which is elegantly controlled entirely by the eigenvalues and eigenvectors of an $m\times m$ matrix $N$, whose entries are the squared distances between the insertion points. We show that the co-dimension $m-D$ scattering submanifold can be thought of
as the intersection of various codimension one scattering subsheets.

 In every case, we find a precise relationship between the coefficients of the singularities and the flat space 
S matrix (see e.g. \eqref{bsgf} for the case of codimension one singularities). This relationship is particularly explicit 
when the singularity has unit co-dimension. As a check, in the particular case of the tree level four point scattering in $D=4$, generated by a non-derivative contact term,  we present an exact computation of Euclidean bulk correlators in terms of $D$ functions in \S \ref{exactcor} \footnote{In Appendix \ref{dflat} we generalize to the computation of correlators for $n$ point scattering generated by contact interactions in $D$ dimensions.}. We then analytically continue this exact answer to the `scattering sheet', check that our answer does indeed develop a singularity on this sheet, and verify that the coefficient of the singularity is given precisely by the bulk S matrix, in agreement with our general analysis (and in a manner that is very similar to the 
analogous discussion, \cite{Gary:2009ae}, in AdS space). 

As far as we are aware, the bulk point singularities of the AdS/CFT correspondence have only been studied in the case $m \leq D+1$. When  $m>D+1$ we expect the discussion of bulk point singularities in AdS/CFT to parallel the discussion presented in \S\ref{cgr1} of this paper. In particular the singularities, in this case, will occur on a codimension $m-D$ slice in the space of boundary insertions, and will capture the information 
of an $m-D-1$ parameter set of inequivalent bulk S matrices.  We 
leave the detailed study of these  AdS multi-bulk point singularities to future work. 

In the final section of our paper, \S\ref{ccft}, we present some brief preliminary remarks about the relationship between our results and those of the programme of celestial holography. In particular, we highlight two qualitative 
differences. The less important difference is that the celestial holography programme scales their boundary to infinity while focusing attention on scattering processes that take place in a finite neighborhood of a particular spacetime point (`the origin'), while our correlators, in general, receive contributions from scattering processes that happen anywhere in spacetime (e.g. three fourths the way to one of the boundaries). The more interesting difference (in the case of massless particles) is that the correlators $G_{\rm bdry}$ presented in this paper are generically analytic functions with singularities whose 
residues are determined by bulk S matrices, while the 
correlators of the celestial holography programme appear to be delta function supported precisely at the location 
of the poles in our correlators. We do not completely understand the reason for this interesting difference. Indeed this difference takes us back to question $\mathcal{Q}$2. In the case of massless particles, our correlator $G_{\rm bdry}$ appears to carry considerably more information than just bulk S matrices. Bulk S matrices determine the residues of singularities of $G_{\rm bdry}$; the nonsingular parts of this correlator appear to be new information\footnote{Note, however, that analytic functions on compact spaces are completely determined by their singularities. For this reason, it is possible that there is no genuinely new information in the regular or `off shell' part of 
$G_{\rm bdry}$. It would be interesting to further investigate this question.}.

Of course, our paper is far from the first to study the flat space path integral and its relationship to the S matrix.  We have already mentioned the foundational work of  Arefeva, Faddeev, and Slavnov \cite{Arefeva:1974jv, Faddeev:1980be,Balian:1976vq,PhysRevD.18.373, PhysRevD.37.1485} (see also \cite{Kim:2023qbl,Gonzo:2022tjm} for a recent discussion). In \cite{Fabbrichesi:1993kz}, the  S-matrix for scalar fields (in eikonal approximation) in the presence of a gravitational field was determined using the gravitational boundary action at null infinity. More recently, the reformulation of S matrices in terms of boundary correlators lies at the heart of the extremely well-studied Celestial Holography program (see e.g.  \cite{Pasterski:2021rjz, Raclariu:2021zjz} for a review). The action as a functional of boundary values is also the object of study in the program of holography of information for massive fields in asymptotically flat spacetimes, initiated in \cite{Laddha:2022nmj}.

\textbf{Note}: While this manuscript was under preparation, \cite{Kim:2023qbl} appeared on the arXiv, which has substantial overlap with sections \S \ref{smbo} and \S \ref{coherent} of this paper. In particular, the authors of \cite{Kim:2023qbl} also relate the S-matrix to the path integral with fixed boundary values, and then use this relation to give a simple picture for soft theorems satisfied by the S-matrix in the context of gauge theories.

\section{The S-matrix and boundary correlator} \label{smbo}

In this section, we start with some formal arguments which suggest a simple and direct relation between the S-matrix and the Dirichlet path-integral. 

All through this paper, we study the S matrices of scalar operators $\phi$. 
We work with fields $\phi$ whose action on the vacuum creates stable particles of mass $m$ (we will study both the generic case $m\neq 0$ as well as the case $m=0$ that will display some special features). All through this paper we work with fields that are normalized so that all $Z$ factors (wave function renormalization factors) are unity; in other words, the fields we study are  normalized such that their two-point functions take the form 
\begin{equation}\label{twopointfuncts} 
\langle T( \phi(x) \phi(y) ) \rangle = \int \frac{d^D p}{(2 \pi)^D} \frac{e^{i p(x-y)}}{p^2+m^2} 
 + \ldots 
 \end{equation} 
 where the $\ldots$ denotes terms in the two-point functions that are either regular or have 
 cut (but no pole) type singularities when viewed in momentum space.

\subsection{S-matrix as a boundary observable}
In Lorentzian signature, our starting point will be the LSZ formula for the S-matrix
\beq
S(\{\vec{p}_i, \vec{q}_j\})= \prod_{i}\int_{M} d^{d+1}x_i e^{ip_i.x_i}\left(\pa_i^2-m^2\right) \prod_{j}\int_{M} d^{d+1}y_j e^{-iq_j.y_j}\left(\pa_j^2-m^2\right)G_{TO}(\{x_i,y_j\}),
\eeq
where $M$ is all of Minkowski spacetime, and $G_{TO}$ is the time-ordered correlation function. The $\{p_i= (\sqrt{\vec{p}_i^2+m^2},\vec{p}_i)\}$ label in-coming momenta while the $q_j$ label out-going momenta 
(Through this paper we use the mostly positive metric convention on flat space). This formula for the S-matrix suggests a natural boundary interpretation. We consider the region of Minkowski spacetime bounded by some large cutoff surface. In general, we can consider a pill-box type cutoff, which has space-like caps at large positive and negative values of time and a time-like radial boundary at some large value of the radius. The details of the shape of this cutoff surface will not be important for our purposes; the only thing that is important is that we want the cutoff to go off to infinity at the end of the day. We will call this boundary $B$. 

Our main goal is to argue that the S-matrix can be written in terms of correlation functions of some boundary observables, i.e., $G_{\text{bdry}}$ defined previously. We will present the argument for a general spacetime, although our interest is primarily in Minkowski spacetime. Consider a set of positive energy modes $f_I(x)$ in a spacetime $M$ which solve the free, massive Klein-Gordon equation:
\beq
(\nabla^2 - m^2) f_I(x) = 0.
\eeq
In equations, the statement that $f_I(x)$ are all positive energy solutions means that, for every constant timelike momentum vector $k$ with a positive time component (i.e., positive energy), 
\begin{equation}\label{whatposenerg} 
\int d^Dx  e^{i k.x} f_I(x) =0, 
\end{equation} 
where $I$ is a label for a complete set of positive energy modes.\footnote{Our conventions are as follows: 
$k_\mu$ is the eigenvalue of the operator $- i \partial_\mu$. 
Therefore $k^0=\omega$ is the eigenvalue of $i \partial_0$.  A mode that 
behaves like $e^{-i \omega t}$, with positive $\omega$, has positive
$k^0$, and so has positive energy.} For instance, in Minkowski spacetime, we could take $f_I$ to be positive energy plane-wave solutions, and then $I$ would label the incoming spatial momenta. Similarly, we can define a set of negative energy modes $\bar{f}_J$, where the index $J$ labels outgoing momenta.\footnote{In the Lorentzian case, the in-coming and out-going modes are just complex conjugates of each other i.e. $\bar{f}_I(x,t) = f_I^*(x,t)$.}  Another possibility is to consider solutions that approach a specific point $Y \in B$ on the boundary; in this case, $I$ would label the location of this boundary point. A third possibility would be to use the `conformal primary basis' commonly employed in studies of Celestial Holography \footnote{It would be interesting to specialize the analysis of this paper in the conformal primary basis, and compare in detail with the results of the Celestial holography programme. We leave this to future work. We thank P. Mitra for this suggestion.}. Note that the fact that $f_I$ are positive energy solutions tells us that the modes $f_I$ are 
necessarily complex: this is true even when the scalar field $\phi$ is real. In the section on coherent state quantization below we will find a physical explanation for the fact that scattering involves the 
consideration of complex fields. 

Now, consider the function
\beq
S(\{I_i\},\{J_j\}) = \prod_{i=1}^n \int d^{d+1}x_i\sqrt{g(x_i)}\, f_{I_i}(x_i)(\nabla^2_i- m^2) \prod_{j=1}^m \int d^dy_j\sqrt{g(y_j)}\, \bar{f}_{J_j}(y_j)(\nabla^2_j- m^2)G_{TO}(\{x_i, y_i\}),
\eeq
where one amputates the external propagators in the correlation function and then puts these legs on-shell by appending the wave-functions $f_{I_i}(x_i)$ or $\bar{f}_{J_j}(y_j)$ respectively. Clearly, the LSZ formula for the S-matrix means that it is an observable of this type, but this discussion is more generally true for any choice of positive energy modes $f_I$ (i.e., not necessarily plane waves). 

Corresponding to the modes $f_I(x)$, we define a basis of boundary ``extrapolate operators'' $O_I$:
\beq \label{ext1}
O_I(B) = \int_{B} \sqrt{h}\,n^{\mu}\bw_{\mu}(f_{I},\phi),\;\;\bar{O}_I(B) = \int_{B} \sqrt{h}\,n^{\mu}\bw_{\mu}(\bar{f}_{I},\phi)
\eeq
where $h$ is the induced metric on $B$, $n^{\mu}$ is the outward pointing normal vector, and the current $\bw_{\mu}$ is given by:
\beq
\bw_{\mu}(\alpha,\beta) = \Big(\alpha^*\pa_{\mu}\beta - (\pa_{\mu}\alpha^*)\beta\Big).
\eeq 
$\bw_{\mu}$ is the free, symplectic flux current. Note that the $\phi$ field in equation \eqref{ext1} is the fully interacting field. Had $\phi$ satisfied the free equation of motion, then $O_I$ would have been a topological operator, independent of the shape of $B$. However, when the surface $B$ goes off to infinity, then one expects the effects of interactions to die off (exponentially in $m$ times the size of the boundary), and so the extrapolate operators defined above become approximately independent of the precise details of the shape of the cutoff surface. Indeed, the difference between the extrapolate operators defined with respect to two different surfaces $B$ and $B'$ (where for simplicity, we take $B'$ to lie inside $B$)  is given by
\beqn
O_I(B)  - O_I(B') &=& \int_{\text{int}(B\cup -B')} \pa_{\mu}(\sqrt{g}\;\bw^{\mu}(f_I,\phi)) \nonumber\\
&=&\int_{\text{int}(B\cup -B')} \sqrt{g}\,\left(f^*_I \nabla^2 \phi - \nabla^2 f_I^* \phi\right)\nonumber\\
&=&\int_{\text{int}(B\cup -B')} \sqrt{g}\,f^*_I\left( \nabla^2 \phi - m^2 \phi\right).
\eeqn
If we take $B'$ to be the empty surface, we get
\beq\label{ext2}
O_I(B) = \int_{\text{int}(B)} \sqrt{g}\,f^*_I\left( \nabla^2 - m^2 \right)\phi.
\eeq
Therefore, inserting the boundary operators $\{O_{I_i}\}$ in the path-integral computes a cutoff version of $S(I_1,\cdots, I_n)$, which approaches $S$ in the limit the boundary is taken to infinity. Note that here the path integral is being done over the fields on the entire spacetime, without imposing any specific boundary conditions at $B$. Note that from equation \eqref{ext2} that if we have two sets of free modes $\{f_I\}$ and $\{g_M\}$ related to each other via the transformation:
\beq
f_I = \sum_M c_{I,M}g_M,
\eeq
then the corresponding extrapolate operators are related by
\beq
O_I = \sum_{M} c^*_{I,M} O_M.
\eeq

It is worth comparing our definition of extrapolate operators with the case of Anti-de Sitter spacetime,
\beq
ds^2 = \frac{dr^2}{r^2} + r^2 \eta_{\mu\nu}dx^{\mu}dx^{\nu},
\eeq
with the asymptotic boundary placed at large $r$. If we take $f_I$ to be:
\beq
(\nabla^2 - m^2 )f_I=0,\;\; \lim_{r\to \infty} f_{y}(r,x^{\mu}) = r^{\Delta -d}\delta^{d}(x^{\mu}-y^{\mu}),
\eeq 
then the definition of boundary extrapolate operators given here agrees precisely with the standard extrapolate dictionary in AdS/CFT, and the corresponding bulk observable $S(I_1,\cdots, I_n)$ agrees with CFT correlation functions. To see this, note that the standard extrapolate dictionary states that
\beq
\phi(r,x) \sim \frac{1}{(2\Delta -d)} r^{-\Delta}\mathcal{O}(x),
\eeq
where $\mathcal{O}$ are CFT operators (normalized such that their correlation functions agree with the GKPW dictionary). Plugging this into our expression for boundary extrapolate operators gives
\beq
O_{y} = \lim_{r_0 \to \infty}\int d^dx\; r^d (f_y r\pa_r \phi - r\pa_r f_y \phi) = \mathcal{O}(y).
\eeq
Thus, the above definition for boundary operators agrees with the standard definition in AdS/CFT and gives a way of generalizing to other spacetimes. For instance, in the case of Minkowski spacetime which is of interest presently, if we take $f_I$ and $\bar{f}_I$ to be plane waves:
\beq\label{mode}
f_{\vec{p}} = e^{- i \omega t + i\vec{p}\cdot \vec{x}},\;\;\; \bar{f}_{\vec{p}}=e^{i\omega t -i\vec{p}\cdot \vec{x}},\;\;\;\cdots\;\;\;(\omega = \sqrt{\vec{p}^2+m^2}),
\eeq
then $S$ becomes the S-matrix. The boundary extrapolate operators in this case are the creation and annihilation operators at infinity which create the incoming and outgoing particles. 

Let us summarize the conclusions of this subsection. The S matrix for a collection of particles, 
in initial states $(I_1, \ldots,I_n)$ and final states and $(J_1, \ldots ,J_m)$, is given by
the `holographic' formula 
\begin{eqnarray}\label{Smatrixcorre}
S(\{I_i\},\{J_j\}) &=&\left\langle \prod_{i =1}^{n} O_{I_i}(B)\prod_{j =1}^{m} \bar{O}_{J_j}(B)\right\rangle\nonumber\\
&=&\left\langle \prod_{i =1}^{n} \int_{B} \sqrt{h}\,n^{\mu_i}\bw_{\mu_i}(f_{I_i},\phi) \prod_{j =1}^{m}  \int_{B} \sqrt{h}\,n^{\mu_j}\bw_{\mu_j}(\bar{f}_{J_j},\phi)\right\rangle\nonumber\\
&=& \prod_{i =1}^{n} \int_{B} \sqrt{h}\,n^{\mu_i}(f_{I_i}\partial_{\mu_i}  -  \partial_{\mu_i} f_{I_i}) \prod_{j =1}^{m}  \int_{B} \sqrt{h}\,n^{\mu_j}(\bar{f}_{J_j}\partial_{\mu_j}  -  \partial_{\mu_j} \bar{f}_{J_j})G_{TO}(\{x_i,y_j\})\nonumber\\
\end{eqnarray}
(in the last line of \eqref{Smatrixcorre}, all free derivatives act on the Time ordered Greens function $G_{TO}(\{x_i,y_j\})$). This formula is holographic in the sense that the formula for the S matrix for scattering in the bulk of flat space is given in terms of correlators evaluated at the boundary of (an IR cut-off version of) 
flat space. 

The formula \eqref{Smatrixcorre} has been derived in Minkowski space. However, since this formula is 
given in terms of time ordered correlators - and since time-ordered correlators are the analytic 
continuation of Euclidean correlators - the RHS of \eqref{Smatrixcorre} can be continued to Euclidean 
space in a straightforward manner
\beq\label{Smatrixeuclid}
S_E(\{I_i\},\{J_j\}) = \prod_{i =1}^{n} \int_{B} \sqrt{h}\,n^{\mu_i}(f_{I_i}\partial_{\mu_i}  -  \partial_{\mu_i} f_{I_i}) \prod_{j =1}^{m}  \int_{B} \sqrt{h}\,n^{\mu_j}(\bar{f}_{J_j}\partial_{\mu_j}  -  \partial_{\mu_j} \bar{f}_{J_j})G_E(\{x_i,y_j\})
\eeq
where $G_E$ is the Euclidean Greens function for the bulk insertion of $n+m$ $\phi$ fields and $S_E$ is the Euclidean continuation of the flat space S-matrix. Once again, all free derivatives in \eqref{Smatrixeuclid} act on $G_E$.

Note that, in Euclidean space, the modes $f_I$ and $\bar{f}_I$ are analytic continuations of the corresponding Lorentzian modes and go as $e^{-\omega \tau}$ and $e^{+\omega \tau}$ respectively. In Lorentzian signature, $f_I$ and $\bar{f}_I$ were simply complex conjugates of each other, but in Euclidean signature, they are instead related to each other by `Euclidean complex conjugation', 
\begin{equation} 
f_I(\tau) = {\bar f}_I^*(-\tau),
\end{equation} 
which involves complex conjugation plus Euclidean time reflection.

\subsection{Path integral with fixed boundary values}

 The goal of this section is to relate the S-matrix to the bulk Euclidean path-integral with Dirichlet boundary conditions on a boundary surface $B$
 \beq \label{PI}
Z[\beta_0] = \int_{\phi|_{B} = \beta_0}[D\phi]\,e^{-S[\phi]},
\eeq
in the limit where the boundary $B$ is taken off to infinity. We will use the Euclidean formula \eqref{Smatrixeuclid} as the starting point for the analysis of this section. Note that, following the standard GKPW dictionary in AdS/CFT, it is natural to expect $Z[\beta_0]$ to be a generating functional for ``boundary correlation functions'' in the putative QFT holographically dual to flat space. From this point of view, it is natural to define: \footnote{We emphasize that, at this stage, the subscript in the quantity  $G_{\text{bdry}}$ is simply notation. In this paper $G_{\text{bdry}}$ is a well defined bulk quantity, 
that may or may not have an independent definition in a (yet to be discovered) boundary dual 
to gravity (and QFT) in flat space.} 
\beq\label{gb}
G_{\text{bdry}}(\{x_i\}, \{y_i\}):= \prod_{i=1}^{n}\frac{\delta }{\delta \beta_0(x_i)}\prod_{j=1}^{m}\frac{\delta }{\delta \beta_0(y_j)}  Z[\beta_0]\Big|_{\beta_0= 0}
\eeq

One can compute this observable in terms of diagrams involving bulk-to-boundary and bulk-to-bulk propagators, much like Witten diagrams in AdS/CFT (see Appendix \ref{eaaafobv}). More importantly, one can relate this observable to the S-matrix. Firstly, if we take derivatives of the expression in equation \eqref{PI} with respect to $\beta_0$, we get
\beq \label{rhsder} 
 \prod_{i=1}^{n}\frac{\delta }{\delta \beta_0(x_i)}\prod_{j=1}^{m}\frac{\delta }{\delta \beta_0(y_j)} Z[\beta_0] =  \int_{\phi|_{B} = 0}[D\phi]\, \prod_{i= 1}^{n}n^{\mu_i}\pa_{\mu_i}\phi(x_i)\prod_{j= 1}^{m}n^{\mu_j}\pa_{\mu_j}\phi(y_j)\,e^{-S[\phi]}.
\eeq
In Appendix \ref{eapt} we present an explicit perturbative verification of \eqref{rhsder}, to low orders, in the particular example of 
$\phi^4$ theory.

We will now demonstrate that the RHS of \eqref{rhsder} actually equals the RHS of \eqref{Smatrixeuclid} 
In order to see this we multiply both sides of \eqref{rhsder} by 
$$ \prod_{i=1}^{n} f_{I_i}(x_i) \prod_{j=1}^m {\bar f}_{J_j}(y_j)$$
and integrate both sides w.r.t. all arguments to obtain
\beqn
 &&\int_{\phi|_{B} = 0}[D\phi]\,\prod_{i=1}^n\left(\int_{B} d^dx_i\sqrt{h(x_i)} f_{I_i}n^{\mu_i}\pa_{\mu_i}\phi(x_i)\right)\, \prod_{j=1}^m\left(\int_{B} d^dy_j\sqrt{h(y_j)} \bar{f}_{J_j}n^{\mu_j}\pa_{\mu_j}\phi(y_j)\right)\,e^{-S[\phi]} \nonumber\\
&=& \int_{\phi|_{B} = 0}[D\phi]\,\prod_{i =1}^{n} \int_{B} \sqrt{h}\,n^{\mu_i}\bw_{\mu_i}(\bar{f}_{I_i},\phi) \prod_{j =1}^{m}  \int_{B} \sqrt{h}\,n^{\mu_j}\bw_{\mu_j}(f_{J_j},\phi)\,e^{-S[\phi]}
\eeqn
where we have used the fact that the path integral is done with Dirichlet boundary conditions $\phi=0$ on the boundary, which allows us to write the above expression in terms of the symplectic flux through the boundary. The right-hand side above now has insertions of the boundary extrapolate operators for the mode functions $f_I$; but we are not yet done, because the path integral is still done with Dirichlet boundary conditions, whereas the observables $S(\{I_i\},\{J_j\})$ are defined via a path integral which does not have such Dirichlet boundary conditions. However, since the symplectic flux is approximately conserved at infinity, instead of performing the integral on $B$, we can perform it on a different surface $B'$ slightly inside $B$:
\beqn
&&\int_B \prod_{i=1}^{n} d^dx_i f_{I_i}(x_i) \prod_{j=1}^m d^dy_j{\bar f}_{J_j}(y_j)\,G_{\text{bdry}}(y_1,\cdots,y_n)\nonumber\\
&=& \int_{\phi|_{B} = 0}[D\phi]\,\prod_{i =1}^{n} \int_{B'} \sqrt{h}\,n^{\mu_i}\bw_{\mu_i}(\bar{f}_{I_i},\phi) \prod_{j =1}^{m}  \int_{B'} \sqrt{h}\,n^{\mu_j}\bw_{\mu_j}(f_{J_j},\phi)\,e^{-S[\phi]}
\eeqn
Now we can send the Dirichlet wall $B$ off to infinity first, and then send $B'$ to infinity. The right-hand side is now precisely a correlation function of boundary extrapolate operators, namely the observable $S_E(I_1,\cdots, I_n)$. We have therefore argued that 
\beq\label{smatrix}
\begin{split}
 S_E(\{I_i\},\{J_j\}) &=  \int \prod_{i=1}^{n} d^dx_i f_{I_i}(x_i) \frac{\delta }{\delta \beta_0(x_i)}\prod_{j=1}^{m}
d^dy_j{\bar f}_{J_j}(y_j)
\frac{\delta }{\delta \beta_0(y_j)} Z[\beta_0]\Big|_{\beta_0= 0}\\
&= \int \prod_{i=1}^{n} d^dx_i f_{I_i}(x_i)\prod_{j=1}^{m}
d^dy_j{\bar f}_{J_j}(y_j) \,  G_{\rm bdry} (\{x_i\}, \{y_j\}).   
\end{split}
\eeq
In the case where we take our mode functions to be plane waves, then the left-hand side above is the Euclidean continuation of the S-matrix, and thus we have established a relation between the S-matrix and the bulk path integral with fixed boundary values. 
\footnote{The formula \eqref{Smatrixeuclid}, the starting point for the analysis of this subsection, 
only applies when the length scale $R$ associated with the boundary surface that appears in this formula is much larger than the inverse of the momentum scale associated with scattering. In this subsection, we have shown that \eqref{Smatrixeuclid} equals \eqref{smatrix}, provided that the 
`path integral as a function of boundary values' that appears on the RHS of \eqref{smatrix} is evaluated
on a surface with length scale $R' \gg R$. It follows that \eqref{smatrix} only applies provided that 
the momentum scale associated with the scattering process in question is much larger than $1/R'$. 
For the scattering of massive particles, this condition is met provided $mR' \gg 1$. The scattering 
of massless particles is more interesting. In this case, the condition is met provided (suitable Lorentz invariant combinations of momenta) obey the schematic equation $p R' \gg 1$. In this case, a scale 
transformation can be used to set $R'=1$. Once we do this the condition $p R' \gg 1$ tells us that 
the S matrix is sensitive only to the very short distance properties of the path integral 
as a function of boundary values of flat space theory with IR cut off of order unity. We will
return to this point in a subsequent section.} 

Note that the positive energy solutions $f_{I_i}(x)$, which are inserted at early times in 
\eqref{smatrix}, behaved like $e^{- i \omega_i t}$ in Lorentzian space, and so scale like 
$e^{-\omega_i \tau}$ in Euclidean space. Similarly, the negative energy solutions ${\bar f}_{I_i}(x)$, which are inserted at late times in \eqref{smatrix}, behaved like $e^{i \omega_i t}$ in Lorentzian space, and so scale like $e^{\omega_i \tau}$ in Eucldiean space. It follows that both the insertions proportional to $f_{I_i}(x)$ and the insertions proportional to ${\bar f}_{I_i}(x)$, blow up in the limit that the 
`final state' Euclidean surface is taken to late Euclidean time, and the initial state Euclidean 
surface is taken to early Euclidean time. More precisely, if $T$ is the time separation between the 
insertion of the initial and final states, then the product of the wave function factors in 
\eqref{smatrix} scale like $e^{TE}$ where $E$ is the net sum of energies of the initial particles 
(which, by energy conservation, also equals the net sum of energies for final particles). 
This apparently divergent $T$ dependence is exactly canceled by the fact that the smeared Euclidean correlator that appears on the RHS of \eqref{smatrix}, decays with $T$ like $e^{-ET}$. These two 
effects cancel each other out, as is clear from the fact that the S matrix (RHS of \eqref{smatrix})
is a cut off (and so $T$) independent quantity. 

In summary, the formula \eqref{smatrix} gives us a simple expression for the Euclidean continuation of the
S matrix in terms of the bulk Euclidean path integral (classically the bulk action) as a function of 
boundary values, smeared against the Euclidean continuation of the free scattering solutions. 
In the next subsection, we will examine the analytic continuation of the RHS of \eqref{smatrix} to Lorentzian space.

\subsection{Analytic Continuation of the Euclidean Path Integral} \label{bc}

In the previous subsection, we derived equation \eqref{smatrix} that relates the Euclidean continuation of the S matrix to the 
Dirichlet path integral in Euclidean space. Analytically continuing both 
sides of \eqref{smatrix} yields 
\beq\label{smatrixl}
 S(\{I_i\},\{J_j\}) =  \int \prod_{i=1}^{n} d^dx_i f_{I_i}(x_i)\prod_{j=1}^{m}
d^dy_j{\bar f}_{J_j}(y_j) \,  G_{\rm bdry} (\{x_i\}, \{y_j\}).   
\eeq
where the mode functions $f_{I_i}(x_i)$ and ${\bar f}_{J_j}(y_j)$ are the analytic continuation of the corresponding Euclidean modes. For the special case of plane wave smearing functions, they are given by $e^{- i \omega t_{in} + i \vec{p}. \vec{x}}$ and $e^{ i \omega t_{f} - i \vec{q}. \vec{y}}$ respectively. 

The boundary correlators $G_{\rm bdry} (\{x_i\}, \{y_j\})$ that appear in 
\eqref{smatrixl} are the Taylor coefficients in the expansion of the 
Lorentzian Dirichlet path integral with one subtlety. Recall that the formula 
\eqref{smatrix} applied only in the limit $T \rightarrow \infty$. \eqref{smatrixl} holds provided we perform the analytic continuation to Lorentzian space after $T$ is taken to infinity. Equivalently, we are instructed to compute the finite time Dirichlet path integral in the `almost  Lorentzian time' $t(1-i\epsilon)$\footnote{Our notation is such that had $\epsilon$ been zero, we would have been working in actual Lorentzian time.}, and then take the limit $\epsilon \to 0$, 
$T \to \infty$ with $\epsilon T \to \infty$ (the last condition effectively ensures that $T$ is taken to infinity before the $\epsilon \to 0$, i.e. that 
we first take $T$ to infinity and then continue to Lorentzian space). 
\footnote{We thank Per Kraus and Richard Myers for very useful comments - on a preliminary version of this manuscript - that forced us to think through the 
material of this subsection more carefully, and also to clearly delineate 
the difference between the Dirichlet and in-out path integrals.}

The physical implications of the limit described in the previous paragraph can be understood as follows. Perturbative contributions to $G_{\rm bdry} (\{x_i\}, \{y_j\})$ are given by products of bulk to boundary propagators and bulk to bulk propagators, sewn together at interaction vertices whose locations are integrated over.  Consider, for instance, one such expression which involves two interaction vertices. Let us first consider the part of the integral in which the two interaction vertices (let's call them $1$ and $2$)  are both located `near to the future boundary' (more precisely,  that the temporal locations, $t_1$ and $t_2$,  of these two interaction vertices obey  $\omega \epsilon (T-t_i) \ll 1$). In this case, future bulk to boundary propagators, as well as the $1 \rightarrow 2$ bulk to bulk propagators are both effectively those of the genuine (i.e $\epsilon =0$) Lorentzian Dirichlet problem. On the other hand, past bulk to boundary propagators (which propagate over times of order $T$) differ significantly from those of the $\epsilon=0$ Lorentzian problem (this is because $\epsilon \omega T \gg 1)$. In Appendix \ref{relpi} we demonstrate that these bulk to boundary propagators become that of the in-out problem \footnote{This is a consequence of the fact that once we take the effect of $i\epsilon$ into account,  solutions behave like
\beq
\phi  \sim  \phi_+ e^{-i\omega t-\epsilon\omega t} + \phi_- e^{+i\omega t + \epsilon \omega t}
\eeq
at early/late times. At late times $m t \gg 1$, the $\phi_+$ mode decays exponentially because of the $e^{-\epsilon \omega t}$ term, while the $\phi_-$ mode is growing. It follows that these propagators reduce to those that we would have used in the in-out problem (see the next section).}. 
Consequently, the part of the integral in which $1$ and $2$ both lie near the future boundary is a sort of hybrid object, with some Lorentzian Dirichlet propagators and other Lorentzian in-out propagators \footnote{In the special case that all boundary insertions are in the future (this is relevant for the computation of the ground state wave function) the region with both interaction points near the future boundary becomes a pure Lorentzian Dirichlet problem.}. Similar remarks apply to all configurations in which atleast one of the two interactions lies near one of the boundaries. In the case that all interaction vertices lie far away (in units of $1/(\epsilon\omega)$ from both boundaries, however, all propagators are effectively in-out. As we explain in the next section, this is precisely the range of integration variables that makes the dominant contribution to the computation of the S matrix (roughly this is the case because the S matrix captures energy conserving processes: the temporal location of the `interaction elevator' is a zero mode for such processes, which thus receive their dominant 
contribution from elevator locations far from both boundaries). It follows, 
in other words, that the diagrams that give the dominant (in the large $T$ limit) contribution to the S matrix are effectively those of the in-out problem. See Appendix \ref{relpi} for more details. 

Motivated by this discussion, we turn, in the next section, to a direct study of the in-out problem. 

\section{The S matrix from the coherent space path integral} \label{coherent}

In this section we review aspects of the AFS proposal \cite{Arefeva:1974jv,Faddeev:1980be,Balian:1976vq,PhysRevD.18.373, PhysRevD.37.1485} that relates S matrices to the in-out path integral. 

The basic rationale behind the AFS proposal is extremely simple. By definition, the S matrix is the amplitude for a given multi free particle initial state to evolve into another given multi free particle final state. As free particle states are very naturally captured in coherent state quantization, the amplitude described above is very simply given in terms of a coherent state path integral. 

In \S \ref{cbhm} below we recall the in-out path integral (for scalar fields) 
that computes transition amplitudes in a coherent state basis. The path integral is given as a functional of positive energy data in the past, and 
negative energy data in the future. The action that appears in this path 
integral comes with a very particular boundary term that ensures well definedness of the variational principle with these boundary conditions 
(see Appendix \ref{Harmonic12}). Though we set this path integral up on a slab type spacetime, we suggest a generalization to arbitrary spacetimes. 
In \S\ref{freef} and \S\ref{scalars} we perform a sample computation within this framework. Specifically we compute the path integral as a functional of positive past and negative future energy boundary data (in momentum space) - up to quartic order in boundary interactions-  in the simple example of a scalar field theory with a  $\phi^4$ interaction.  In \S\ref{ccheck} we then insert the result of \S \ref{scalars} into the RHS \eqref{smatrixl}, and check that we indeed find the  S matrix. In other words, we verify that the path integral as a functional of boundary data does, indeed, capture the 
S matrix is the manner that AFS envisaged. However the path integral in question 
carries additional information beyond that of the `onshell' S matrix: it also captures information about offshell processes. As an example, in \S\ref{crvwf}, we demonstrate that the path integral captures the vacuum wave functional of our theory in a coherent space basis. The fact that the same formal object captures both the S matrix and the vacuum wave function suggests that these structures are intimately related. We conjecture that more generally (i.e. beyond our particular simple example), the onshell S matrix may be obtained as the coefficient of the singularities of the analytic continuation of the `offshell' vacuum state wave functional, in a manner very similar to the dS discussion of 
\cite{Maldacena:2011nz,Raju:2012zr,Arkani-Hamed:2017fdk,Arkani-Hamed:2018kmz,Benincasa:2018ssx,Goodhew:2020hob,Baumann:2022jpr,Pajer:2020wxk}.

\subsection{Coherent State Quantization in quantum field theory}\label{cbhm}

In Appendix \ref{Harmonic12}, we recall how coherent state quantization works in the context of the simple example of a single harmonic oscillator. In this subsection, we generalize the discussion 
of Appendix \ref{Harmonic12} to the study of QFT. This generalization is 
straightforward as (within perturbation theory) a quantum field theory 
may simply be regarded as a collection of harmonic oscillators, one for every momentum mode. 

 In order to perform the coherent space quantization of a scalar field we expand  
\beqn \label{posandneg} 
\phi(t,x) &=& \int \frac{d^d k}{(2 \pi)^d}\frac{1}{\sqrt{2 \omega_k}} \left( z_k(t) e^{ i \vec{k}.\vec{x}} + {\bar z}_k ( t) e^{- i \vec{k}.\vec{x}}\right)\nonumber
\eeqn
where $\omega_k = \sqrt{\vec{k}^2 + m^2 }$ and $d = D-1$ where $D$ is the full spacetime dimension. The equation \eqref{posandneg} can be recast as 
\beqn \label{posandnegrc} 
\phi(t,x) &=&  \phi^+(t,x) +  \phi^-(t,x)\\
\phi^{+}(t,x) &=&\int \frac{d^d k}{(2 \pi)^d}\frac{1}{\sqrt{2 \omega_k}}  z_k( t) e^{ i \vec{k}.\vec{x}}\nonumber \\
\phi^{-}(t,x) &=&\int \frac{d^d k}{(2 \pi)^d}\frac{1}{\sqrt{2 \omega_k}} 
 {\bar z}_k (t) e^{-  i \vec{k}.\vec{x}}\nonumber
\eeqn
Here $\phi^\pm$ are the positive/ negative energy parts of the field.\footnote{ In the special case of the free theory, quantization gives the operator solution
\beqn
\phi(t,x) &=& \int \frac{d^d k}{(2 \pi)^d}\frac{1}{\sqrt{2 \omega_k}} \left( a(k) e^{-i \omega_k t + i \vec{k}.\vec{x}} + a^{\dagger}(k) e^{i \omega_k t - i \vec{k}.\vec{x}}\right)\nonumber\\
&=& \phi^+(t,x) + \phi^-(t,x)\nonumber
\eeqn
where  $a(k)$ and $ a^{\dagger}(k)$ are the usual annihilation and creation operators respectively. Hence $$z_k(t) = a(k) e^{-i \omega_k t}$$} Generalizing \eqref{Onshellaction}, we find that the action for our path integral is given by 
\beq \label{Onshellaction1}
{\cal S} = -i \int \frac{d^{d} k}{(2 \pi)^{d}}  
\left[ \frac{\bar{z}_{k}^{f} z_{k}^{f}+\bar{z}_{k}^{i} z_{k}^{i}}{2} \right]  - i \int \frac{d^{d} k}{(2 \pi)^{d}}  dt \left[ \frac{\Dot{\bar{z}}_k z_k-\Dot{z}_k\bar{z}_k}{2} 
-i \omega_k z_k {\bar z}_k \right]  + S_{int} 
\end{equation}
where the second term above is the integral of the free Lagrangian, the term $S_{int}$ captures 
the effect of (non-derivative) interactions while the first term captures the `counter-terms' which depend on the boundary conditions. The boundary conditions are $\bar{z}_k (t_f = T) = \bar z_{k}^{f}$ and ${z}_k (t_i= -T) = z_{k}^{i}$. Notice that these boundary conditions amount to the following boundary conditions on the positive and negative energy parts of the scalar field $\phi(t,x)$.
\beqn
\beta(x) &=& \phi^{+}(-T,x) =\int \frac{d^d k}{(2 \pi)^d}\frac{1}{\sqrt{2 \omega_k}} \left( z_{k}^{i} e^{ i \vec{k}.\vec{x}} \right)\\
\bar\beta(x)&=& \phi^{-}(T,x) = \int \frac{d^d k}{(2 \pi)^d}\frac{1}{\sqrt{2 \omega_k}} \left(
 {\bar z}_{k}^{f}  e^{- i \vec{k}.\vec{x}}\right)
\eeqn
From the above relations, we can get a relation between $z_k$ and Fourier modes of $\beta(x)$, which we denote by $\beta_{k}$\footnote{$\beta_{\vec{k}} = \int d^dx \, e^{- i \vec{k}.\vec{x}} \beta(\vec{x})$}.
\beq \label{betabarz} 
\beta_k = \frac{1}{(2\pi)^d\sqrt{2 \omega_k}} z_k^i, \qquad \qquad \bar{\beta}_k = \frac{1}{(2\pi)^d\sqrt{2 \omega_k}} \bar{z}_k^f
\eeq

\subsection{Transition amplitude for free propagation} \label{freef}

We can use \eqref{Onshellaction1} to compute the path integral as a function of initial positive energy data and final negative energy data. At quadratic order, interactions can be ignored. The second
term in \eqref{Onshellaction1} vanishes once we use the equations of motion. The full result comes from the first term and is given by
\beq\label{ac2pt}
i \mathcal{S} =\int \frac{d^d k}{(2 \pi)^d}\, z_{k}^{i} {\bar z}_{-k}^{f}  e^{-2i\omega_k  T } = \int d^d k \,(2 \pi)^d 2\omega_k \beta_k \bar\beta_{-k} e^{-i\omega_k  (2T) } 
\eeq
To compute the two-point function (S-matrix between one initial and one final state), we use the momentum space version of equation \eqref{smatrix} and obtain:
\beq
\frac{\delta^2 (i\mathcal{S})}{\delta\beta_{k_1} \delta\bar\beta_{k_2}} =  2 \omega_{k_1}  (2 \pi)^d \delta^d (\vec{k_1} + \vec{k_2}) e^{-i\omega_{k_1} (2 T) } 
\eeq
where the exponent is the free evolution and $\omega_{k_1} \delta^d (\vec{k_1} + \vec{k_2})$ is the usual momentum conserving delta function. Now recall that the inner product of two  one-particle states, 
normalized so that 
\beq\label{normalization}
 |\mathbf{k}\rangle = \sqrt{2 \omega_k} a^\dagger_{\mathbf{k}}|0\rangle
 \eeq
and with 
\begin{equation}\label{comrel} 
 [a({\vec k}) , a^\dagger({\vec k}')] = (2 \pi)^{d} \delta^{^{d}} ({\vec k}-{\vec k}')
 \end{equation} 
is given by the formula
\beq \label{freein} 
\langle \mathbf{k_1}|\mathbf{k_2}\rangle = 2 \omega_{k_1}  (2 \pi)^d \delta^d (\vec{k_1} + \vec{k_2})
\eeq
Consequently, we find that 
\begin{equation} 
\frac{\delta^2 \mathcal{S}}{\delta\beta_{k_1} \delta\bar\beta_{k_2}} 
= \langle \mathbf{k_1}|\mathbf{k_2}\rangle  e^{-i\omega_k  (2T)}.
\end{equation} 
This is precisely the expected answer for the S matrix (transition amplitude between initial and final one particle states) in the free theory.


\subsection{S matrices including interactions: One example} \label{scalars} 

It is not difficult to generalize the computation of the previous subsubsection
to first order in the coupling, for the case of a scalar theory interacting 
via a $\phi^4$ interaction. The Lagrangian is given by:
\beq\label{lagsc}
\mathcal L = \pa_\mu \phi \pa^\mu \phi + \frac{m^2}{2}\phi^2 + \frac{\lambda}{4!} \phi^4
\eeq
In Appendix \ref{Harmonic} we computed the 
action as a function of boundary values for quantum mechanics with quartic potential at first order in the coupling. The formulas of Appendix \ref{4ho} can be generalized to obtain the on-shell action for the above QFT Lagrangian. The expansion of 
our result to fourth order in $\beta_k$ and ${\bar \beta}_{k}$ has terms of homogeneity $\beta_k^4$, $\beta_k^3 {\bar \beta}_{k}$, $\beta_k^2 {\bar \beta}_{k}^2$, $\beta_k {\bar \beta}^3_{k}$ and ${\bar \beta}^4_{k}$. The term 
of homogeneity $\beta_k^2 {\bar \beta}_{k}^2$ turns out to be proportional to 
\beq \label{sonri}
i\mathcal{S}_\lambda \propto -\int \prod_i d^d k_i ~\frac{i \lambda \left(e^{- 2i T \left(\omega _1+\omega _2\right)}-e^{-2i T \left(\omega _3+\omega_4\right)}\right)}{(\omega _1+\omega _2-\omega _3-\omega _4)} (2 \pi)^d \delta^d  \left(\vec k_1 + \vec k_2-\vec k_3-\vec k_4\right)\bar\beta_{k_3} \bar\beta_{k_4} \beta_{k_1} \beta _{k_2}  
\eeq
\eqref{sonri} can be rewritten as 
\beq \label{sonriw1}
i\mathcal{S}_\lambda \propto -\int \prod_i d^d k_i ~\frac{i \lambda \left(e^{- 2i T E_{in} }-e^{-2 i T E_{out}}\right)}{(E_{in} -  E_{out})} (2 \pi)^d \delta^d  \left( \sum_{i} \vec k_i^{\rm in} -
\sum_{j} \vec k_j^{\rm out} \right) \prod_j \bar\beta_{k_j} \prod_i \beta_{k_i} 
\eeq
where $E_{in}$ ($E_{out}$) is the total energy of in-going (out-going) particles. When written in this form, the result for $\mathcal{S}$ applies not just to terms of 
homogeneity $\beta^2 {\bar \beta}^2$, but all other terms as well (e.g. the terms of homogeneity $\beta^4$). 

\subsection{Consistency with \eqref{smatrixl}}\label{ccheck}   
 
We will now examine the consistency of the results of this subsection with 
\eqref{smatrixl}. \eqref{sonriw1} tells us that, in the context of the example of this section, 
\begin{equation} \label{actposit} 
\begin{split}
&\prod_{i=1}^{n}  \frac{\delta }{\delta \beta_0(x_i)}\prod_{j=1}^{m} 
\frac{\delta }{\delta {\bar \beta}_0(y_j)} Z[\beta_0]\Big|_{\beta_0= 0}\\
 &=  \int \prod_{i=1}^{n}  \frac{d^dk_i}{(2\pi)^d} e^{-i {\vec k_i}.{\vec x_i} }
\int \prod_{j=1}^{m}  \frac{d^dk_j }{(2\pi)^d} e^{i {\vec k_j}.{\vec y_j} }
\left( -\frac{i \lambda \left(e^{-2 i T E_{in} }-e^{-2 i T E_{out}}\right)}{(E_{in} -  E_{out})} (2 \pi)^d \delta^d  \left( \sum_{i} \vec k_i^{\rm in} -
\sum_{j} \vec k_j^{\rm out} \right) \right)\\
\end{split} 
\eeq
where we have used the fact that $\beta_k (\bar \beta_k)$ are Fourier modes of the boundary values $\beta_0(x_i) (\bar \beta_0(x_i))$.

Consider the free particle states 
\begin{equation} \label{fps} 
f(x) = \int \frac{ d^d k}{(2 \pi)^d}  g(k) 
e^{- i \omega_k t + i {\vec k}. x }
\end{equation} 
where ${\vec k}$ are the spatial components of $k$ and $\omega$ is the energy (recall that the `fixed time' boundary conditions of this subsection give us a natural split into space and time). Let us now evaluate  \eqref{smatrix} for the scattering of initial states labelled by the 
wave functions $g_i(k)$ ($i=1 \ldots n$)  to the final states labelled by the wave functions $g_j(k)$ ($j=1 \ldots m$). We find 
\begin{equation} \label{s123}
\begin{split} 
&\int \prod_{i=1}^{n} \frac{d^dk_i g_i(k_i) }{(2\pi)^d} \prod_{j=1}^{m}  \frac{d^dk_j g^*(k_j)  }{(2\pi)^d} e^{i(E_{in} + E_{out}) T) } \left(- \frac{i \lambda \left(e^{-2 i T E_{in} }-e^{-2 i T E_{out}}\right)}{(E_{in} -  E_{out})}  (2 \pi)^d \delta^d  \left( \sum_{i} \vec k_i^{\rm in} -
\sum_{j} \vec k_j^{\rm out} \right) \right)\\
&= \lambda \int \prod_{i=1}^{n} \frac{d^dk_i g_i(k_i) }{(2\pi)^d} \prod_{j=1}^{m}  \frac{d^dk_j g^*(k_j)  }{(2\pi)^d} ~(2 \pi)^{d+1}~\frac{
 \sin \left(T( E_{out} - E_{in}) \right) } {\pi (E_{out} - E_{in})}
 \delta^d  \left( \sum_{i} \vec k_i^{\rm in} -
\sum_{j} \vec k_j^{\rm out} \right) 
\end{split} 
\end{equation}
Now in the large $A$ limit, the function $\sin (Ax)/ \pi x$ tends to a $\delta$ function (the integral of this function over the variable $x$ is unity for every value of $A$, so in the limit $T \to \infty$ we find that the S matrix of the scattering wave packets above is given by 
\begin{equation} \begin{split} 
&\int \prod_{i=1}^{n} \frac{d^dk_i g_i(k_i) }{(2\pi)^d} \prod_{j=1}^{m}  \frac{d^dk_j g^*(k_j)  }{(2\pi)^d} ~[i \mathcal{T}(k_i, k_j)]
\end{split} 
\end{equation} 
where the $\mathcal{T}$ matrix is
\beq \label{singscat} 
\mathcal{T}(k_i, k_j) = -i \lambda (2 \pi)^{d+1}  \delta^{d+1}  \left(\sum_i k_i^{\rm in}- \sum_j k_j^{\rm out}\right)
\eeq
in agreement with our general expectations.

\subsection{Conjectured Relationship with the Vacuum Wave Function} \label{crvwf} 

In the previous subsubsection, we saw that the process of convoluting 
the on-shell wave functions ($f_I$ and $f^*_J$) with the coefficient function that appears in \eqref{sonriw1}, yields an answer that is non-vanishing in the large $T$ limit only when energy is conserved (i.e. when the sum of free initial energies equals the sum of free final energies). While the explicit workout of the previous subsection was in the context of a 
particular example, the requirement that the result of such an exercise should always 
agree with the exact formula \eqref{smatrix} suggests that this general feature is exact (i.e. works in every theory and at all orders in perturbation theory). 

For physical questions, however, one can be interested in the quantity \eqref{sonriw1} itself rather than its convolutions with free wave functions. For this purpose, we need to take the large $T$ limit of the coherent space path integral itself (without convoluting with 
incident wave functions). One question of physical interest that one can answer by studying 
\eqref{sonriw1} is the following: What is the wave function of the vacuum of our theory in the coherent space representation? It is, of course, well known that the unnormalized vacuum wave function is the coefficient of the leading large $T$ dependence of the past to future path integral with the past source $\beta$ set equal to (for instance) zero (strictly speaking this statement is true when 
$T$ has a small Euclidean part that damps out the contribution from all 
excited states). Restated, the vacuum wave functional is the path integral, evaluated as a function of future insertions (but with no insertions in the past). In the current context this path integral, when evaluated as a function of ${\bar z}_k$, yields the Schrodinger wave functional of the vacuum in the 
coherent space basis. In equations 
\begin{equation}\label{wavefunc} 
\psi(\{{\bar z^f}_k \}) \equiv \langle \{ z_k \} | \psi \rangle = Z[\{  {\bar z}_{f} \}, \{ z_i=0 \}  ] 
\end{equation} 
From \eqref{betabarz},  we see that the part of the path integral that is a function only 
of ${\bar \beta}$, but is independent of $\beta$ is directly the Schrodinger wave functional of the vacuum in the coherent space representation. 
\footnote{Note that we have defined the wave function \eqref{wavefunc} as the inner product with 
respect to the coherent state $|z>$ whose normalization is given by \eqref{codef} (these coherent states are not unit normalized. This is also the convention used, for instance, in the Quantum Field Theory textbook by Itzykson and Zuber \cite{Itzykson:1980rh}).} 

It is interesting to check how this works out to order $\lambda$ in the example of the 
theory defined by the Lagrangian \eqref{lagsc}, i.e. the example studied in detail 
in the previous subsection. 

Let us first see how \eqref{wavefunc} works in the free theory, i.e. at $\lambda=0$. The free theory is a non-interacting collection of harmonic oscillators, so it is sufficient to check that \eqref{wavefunc} works for any one of those oscillators. Note that for the free harmonic oscillator,  $\langle z|0 \rangle=1$ (this is a special case of 
\eqref{normalization}). It follows that, with our conventions, the wave function of the vacuum for a harmonic oscillator (and hence also for the product of wave functions of many harmonic oscillators) is simply unity. This is also what we find from \eqref{ac2pt}, since the action in \eqref{ac2pt} has no term proportional to ${\bar \beta}^2$ and hence the partition function ($e^{i \mathcal S}$) gives unity. 

Let us now see how, in this example, \eqref{wavefunc} continues to work at  $\lambda$. 
From \eqref{sonriw1} (and using \eqref{betabarz} we find that the wave function of the 
vacuum in this theory, to leading nontrivial order in $\lambda$, is given by the large $T$
limit of 
\beq \label{wavefunctscth} 
\psi( \{ {\bar \beta} _k \} )=  \exp \left( - \int \prod_i d^d k_i ~\frac{i \lambda \left(1-e^{-2 i T \left(\omega _1+\omega _2+\omega _3+\omega _4\right)}\right)}{(\omega _1+\omega _2+\omega _3+\omega _4)}\delta^d  \left(\vec k_1 + \vec k_2 + \vec k_3 +\vec k_4\right)\bar\beta_{k_3} \bar\beta_{k_4} \bar\beta_{k_1} \bar\beta _{k_2} \right) 
 \eeq
  
 In order to find the large $T$ limit of $\left(e^{-2 i T E_{in} }-e^{-2 i T E_{out}}\right)$, it is useful to rotate Minkowski time infinitesimally to Euclidean space, i.e. to make the replacement $T \rightarrow T-i \epsilon T$. We find that the term with the larger of $E_{in}$ and $E_{out}$ is subdominant compared to the other term (the one with the smaller value of energy). In other words, at large time 
$$ \left(e^{-2 i T E_{in} }-e^{-2 i T E_{out}}\right)
\rightarrow e^{-2 i T E_{min} }.$$
In the current context $E_{min}=0$ so that \eqref{wavefunctscth} simplifies to 
\beq \label{wavefunctscthnew} 
\psi( \{ {\bar \beta} _k \} )=  \exp \left( - \int \prod_i d^3 k_i ~\frac{i \lambda}{(\omega _1+\omega _2+\omega _3+\omega _4)}\delta^d  \left(\vec k_1 + \vec k_2 + \vec k_3 +\vec k_4\right)\bar\beta_{k_3} \bar\beta_{k_4} \bar\beta_{k_1} \bar\beta _{k_2} \right) 
 \eeq
In Appendix \ref{expc} we check the single harmonic oscillator analogue of  \eqref{wavefunctscthnew}, by independently evaluating the vacuum wave function using standard first-order Hamiltonian perturbation theory. In the same appendix, we also compute the Dirichlet path integral as a functional of 
future boundary data, and check that this computation correctly evaluates 
the Schrodinger representation of the same wave function.

Notice that the path integral that computes the vacuum wave functional is clearly off-shell even in the large $T$ limit, in the sense that unperturbed energy is clearly not conserved (clearly the energy of four particles cannot match the energy of nothing). Indeed, the fact that this 
path integral is off-shell explains the structure of the answer \eqref{wavefunctscthnew}. 
An off-shell spontaneous particle production can only happen over a time period of order $1/ \sum {\omega_i}$ before the final time ($t= T$). This fact explains why the amplitude for this process is proportional to $1/ \sum {\omega_i}$, and is, in fact, essentially 
independent of $T$. 

Note that the height of the $\delta$ function in energy
that appears in the formula for the on-shell S matrix \eqref{singscat} is actually equal to $\frac{T}{\pi}$. This factor of $T$ is a reflection of the fact that 
the on-shell scattering processes are not localized  
within a finite band around initial or final times, but can happen at any time. The quantity $1/ \sum {\omega_i}$
is the wave function analogue of $\pi \delta(E_{in}-E_{out})$ in the S matrix. 

In more generality than the particular example investigated above, the physical discussion of this paragraph suggests that the coherent space representation of the $\log$ of the vacuum wave function will always be singular in the limit $\sum_i \omega_i=0$, and that this singularity is always that of a pole of unit order. The discussion of the previous paragraph motivates 
the following conjecture. Consider the vacuum wave function and make the following replacement in the singular part of this wave function 
\begin{equation}\label{replacement}
\frac{1}{\sum_i \omega_i} \rightarrow \frac{1}{\sum_i \omega_i - i \epsilon}
\end{equation} 
After making this replacement, let the wave function take the form 
\begin{equation}\label{wafi}
\psi({\bar \beta})= \exp \left( \sum_n \int  \prod_{i=1}^n  d^d k_i  \frac{T_n(k_i) }{\sum_{i} \omega_i-i \epsilon} 
\prod_{i=1}^n \bar\beta(k_i) \right)
\end{equation} 

We conjecture that the analytic continuation of the coefficient functions $T_n(k_i)$ to negative values of energies for some particles, is the T matrix of the corresponding particles, with the particles whose energies have been analytically continued being treated as initial states\footnote{The analytic continuation we have in mind is the following. We work in $d+1$ dimensions. 
and parameterize the momentum of each particle by a unit spatial vector 
${\hat n}_i$ together with  $\omega_i= \sqrt{|{\vec k}_i|^2+ m_i^2}$ ($\omega_i$ is a measure of the magnitude of the vector ${\vec k}_i$). Our analytic continuation 
leaves ${\hat n}_i$ unchanged, but takes $\omega_i$ to $-\omega_i$ for a subset of the participating particles.
}, and the particles whose energies have not been analytically continued being regarded as final states. Of course, the analytic continuation described above is heavily reminiscent of crossing symmetry; our conjecture effectively is that crossing symmetry does more than relate various on-shell S matrices to each other; it also
relates these S matrices to the off-shell (coherent space) wave functional of the vacuum. 
Of course, this conjecture works to order $\lambda$ in the explicit example studied in this 
section (\eqref{wavefunctscthnew} and \eqref{sonriw1}). It would be very interesting to perform further (loop-level) checks of this conjecture. However, we leave this exercise to future work. 

Note that the conjecture above relates the vacuum wave function of the theory to its S-matrix. If the conjecture is correct, then this would also imply that the spatial entanglement structure of the vacuum state in the QFT is directly related to its S-matrix. Making this relationship manifest would be interesting.

\section{Unitarity constraints on the  path integral} \label{ucpi} 

In the previous sections we have discussed the relationship of the S-matrix to the Dirichlet and in-out path integrals (see \eqref{smatrixl}). In  \S\ref{duc}, we use the equation that captures the unitarity of the S-matrix to derive the same non-trivial identity \eqref{unitsm} for each of these path integrals as a function of boundary values. In \S \ref{rio}, we then rederive the same identity for in-out path integral using coherent space completeness relation. Next, we explain that \eqref{unitsm} is exact for the in-out path integral even at finite values values of $T$, but is valid only at large $T$ in the case of the Dirichlet problem. 

\subsection{Derivation of the Unitarity Constraint} \label{duc}

Unitarity constrains the S matrix to obey the equation 
\begin{equation} \label{unitbasic}
	S^{\dagger} S = \mathbb{I}.
\end{equation}
Let us work in the basis of plane waves, and take the matrix element of this equation between a
ket (which we choose to be an  $n$ particle state, where the particle momenta are given by  ${\mathbf q}_j$,  $j= 1 \ldots n$) and 
a bra, (which we choose to be an $m$ particle state, where the particle momenta are given by 
${\mathbf p}_i$, $i = 1 \ldots m$.  We find
\beq\label{sunit}
\begin{split}
\langle \{\mathbf{p}_i\} | S^{\dagger} S |\{\mathbf{q}_j\} \rangle &= 2 \omega_{\{\vec p\}} (2\pi)^d \delta (\{\vec p_i\}, \{\vec q_j\})\\
\end{split}
\eeq
The quantity $2 \omega_{\{\vec p\}}(2\pi)^d \delta (\{\vec p_i\}, \{\vec q_j\})$ that appears on the RHS of \eqref{sunit} is zero if $m \neq n$ (because the operator $I$ in the RHS of 
\eqref{sunit} represents free propagation and particle number is conserved during free propagation. When $m=n$, on the other hand, this quantity equals the product of 
$n$ factors of the form \eqref{freein}, which are then summed over all possible permutations (this is a consequence of Bose symmetry). In equations 
\beq
2 \omega_{\{\vec p\}}(2\pi)^d\delta (\{\vec p_i\},\{\vec q_j\}) = \delta_{m n} \sum_{\rm perm.} \prod_{i = 1}^{n} 2\omega_{p_i}(2\pi)^d\delta^d(\vec{p}_i -\vec q_j)
\eeq

 Notice that, despite appearances,  the RHS is Lorentz invariant. This can be seen from the identity 
\beq
\left(2\omega_{\vec p}(2\pi)^d\delta^d(\vec{p} -\vec q)\right) (2\pi)\delta(p^2 + m^2)= (2\pi)^{d+1}\delta^{d+1}(p-q) 
\eeq

We now insert a resolution of identity in between $S^\dagger$ and $S$. The resolution of identity for a single Harmonic Oscillator takes the form 
\begin{equation}\label{rish} 
\mathbb{I}= \sum_n |n\rangle \langle n|=  : |0\rangle e^{a^\dagger a } \langle 0| : \equiv \sum_{n} \frac{a^{\dagger n}|0\rangle \langle 0| a^n}{n!}
\end{equation} 
where the symbol $: |0\rangle e^{a^\dagger a } \langle 0| :$ is defined by the final expression in \eqref{rish}.
Particle Fock Space is a tensor product of harmonic oscillator Hibert Spaces, one for each value of the 
momentum. As a consequence the resolution of identity appropriate to Fock Space is\footnote{Expanding the exponential, and ignoring states with equal momenta (as they are measure zero) 
turns the resolution of identity into the likely more familiar equation 
\beq\label{iden}
\sum_\lambda \int \frac{d^d k}{(2\pi)^d}\frac{1}{2 \omega_{k,\lambda}}|\mathbf{k}_\lambda \rangle \langle \mathbf{k}_\lambda | = \mathbb{I}
\eeq
(here $\lambda$ labels the number of particles in the intermediate state). However \eqref{rifs} is the conceptually more correct equation. It turns out that, for our current purposes, it is also technically more convenient. 
For this reason we will use \eqref{rifs} rather than \eqref{iden}.}   
\begin{equation}\label{rifs} 
\mathbb{I}= : |0\rangle e^{\int \frac{ d^d k}{(2 \pi)^d } a^\dagger_{\vec k} a_{\vec k}  } \langle 0|: 
\end{equation} 
Inserting this equation between $S^\dagger$ and $S$ in \eqref{sunit} we find 
\begin{equation}\label{sunitnew}
 \langle \{\mathbf{p}_i\} | S^{\dagger}
: |0\rangle e^{\int \frac{ d^d k}{(2 \pi)^d }  a^\dagger_{\vec k} a_{\vec k}  } \langle 0|: 
 S |\{\mathbf{q}_j\} \rangle = 2 \omega_{\{\vec p\}}(2\pi)^d\delta (\{\vec p_i\}, \{\vec q_j\})
\end{equation}
In \eqref{smatrix} if we replace the positive energy modes $f_I(x)$ by
\beq
f_{\vec k_i}(x_i) = e^{- i \omega_{\vec k_i} (-T) - i \vec k_i. \vec x_i}
\eeq
and work with the final surface $t=T$ and the initial surface $t=-T$ we obtain 
\beq\label{smatrixp}
\begin{split}
  \langle \{\mathbf{p}_i\}| S |\{\mathbf{q}_j\} \rangle &= e^{i T  \left( \sum\limits_i \omega_{\vec p_i} + \sum\limits_j \omega_{\vec q_j} \right) }  \int \prod_{i=1}^{n} d^dx_i \prod_{j=1}^{m}
d^dy_j e^{- i \vec p_i. \vec x_i}  e^{i \vec q_j. \vec y_j}
\prod_{i} {\delta \over \delta \bar\beta(\vec{x}_i)} \, \prod_{j} {\delta \over \delta \beta(\vec y_j)} 
Z[\beta]\Big|_{\beta, \bar\beta= 0}\\
\end{split}
\eeq
In terms of the Fourier transform of the boundary values of the scalar field, 
\begin{equation}\label{bek}
\beta_{\vec{k}} = \int d^dx \, e^{- i \vec{k}.\vec{x}} \beta(\vec{x})
\end{equation} 
\eqref{smatrixp} becomes 
\beq\label{smatrixpp}
\begin{split}
  \langle \{\mathbf{p}_i\}| S |\{\mathbf{q}_j\} \rangle &= e^{i T  \left( \sum\limits_i \omega_{\vec p_i} + \sum\limits_j \omega_{\vec q_j} \right) }
\prod_{i} {\delta \over \delta \beta_{\vec{p}_i}} \, \prod_{j} {\delta \over \delta \beta_{-\vec q_j}} 
Z[\beta_0]\Big|_{\beta_0= 0}\\
\end{split}
\eeq
Using \eqref{normalization} (which tells us that $a({\vec k}) \rightarrow \frac{1}{\sqrt{2 \omega_k}} \beta({\vec k})$) we see that \eqref{sunitnew} can be rewritten as 
\begin{equation} \label{smatrixpp} 
	 \prod_{i}{\delta \over \delta \beta_{\vec{p}_i}^*} \,  \prod_{j} {\delta \over \delta \beta'_{-\vec q_j}}\, \exp\left(\int \frac{d^dk}{(2\pi)^d}   \, \frac{1}{2 \omega_k} {\delta \over \delta \bar\beta_{-\vec{k}}^*} \, {\delta \over \delta \bar\beta'_{\vec{k}}  } \right) \,\zpi^*[\beta,\bar\beta] \zpi[\beta',\bar\beta'] \bigg|_{\beta^{(')}, \bar \beta^{(')}= 0} = 2 \omega_{\{\vec p\}}(2 \pi)^d\delta (\{\vec p_i\}, \{\vec q_j\})
\end{equation}

\eqref{smatrixpp} (which applies to arbitrary choices of initial and final states) can be simplified as follows. 
We first make the algebraic observation that 
\begin{equation} \label{wicksm} 
\left( \prod_{i}{\delta \over \delta \beta_{\vec{p}_i}^*} \,  \prod_{j} {\delta \over \delta \beta'_{-\vec q_j}} \right) \exp\left(\int \frac{d^d p}{(2 \pi)^d}2 \omega_{\vec p} \beta^*_{\vec p}\beta'_{-\vec p}\right)\bigg|_{\beta^{(')}= 0}= 2 \omega_{\{\vec p\}}(2 \pi)^d\delta (\{\vec p_i\}, \{\vec q_j\})
\end{equation} 
(the straightforward proof of \eqref{wicksm} has some similarities to the derivation of the Wick Theorem from Gaussian integrals).
It follows that \eqref{smatrixpp} will be obeyed if and only if
\begin{equation} \label{smatrixpp1} 
 \exp\left(\int \frac{d^dk}{(2\pi)^d}   \, \frac{1}{2 \omega_k} {\delta \over \delta \bar\beta_{-\vec{k}}^*} \, {\delta \over \delta \bar\beta'_{\vec{k}}  } \right) \,\zpi^*[\beta,\bar\beta] \zpi[\beta',\bar\beta'] \bigg|_{\bar \beta^{(')}= 0} = \exp\left(\int \frac{d^d p}{(2 \pi)^d}2 \omega_{\vec p} \beta^*_{\vec p}\beta_{-\vec p}\right)
\end{equation}
(the various Taylor coefficients of \eqref{smatrixpp1} yield \eqref{smatrixpp} for various choices of scattering momenta). This is then the analog of the unitarity of the S-matrix in terms of the path integral with fixed boundary values. 

We now wish to trade the differential operators in the exponential of \eqref{smatrixpp1} for an integral involving a Kernel. Using manipulations similar to those in \S 3 of \cite{Gopakumar:2000zd}, it is easy to see that 
\begin{equation}\label{kernelon} 
 \exp\left( \frac{1}{a} {\delta \over \delta \bar{w}} {\delta \over \delta w } \right) f(w, {\bar w} ) \bigg|_{ {\bar w}= w= 0}
 =  \int \frac{ d^2w}{\pi} \exp\left( -a w {\bar w}  \right) f(w, {\bar w} )
 \end{equation} 
 Discretizing the the integral over ${\vec k}$ in \eqref{smatrixpp1}, and then applying \eqref{kernelon}
for each value of momentum separately, we find that: \footnote{If we discretize the integral over 
${\vec k}$ to a sum over ${\vec k}_i$, the measure on the LHS of the equation below gives us one factor of the 
infinitesimal volume element $\delta V_{k}$ in the numerator. However 
${\delta \over \delta \bar\beta'_{\vec{k}}  } \rightarrow \frac{1}{\delta V_{k}} {\delta \over \delta \bar\beta'_{\vec{k}_i}  }$. It follows that we effectively have one factor of  $\delta V_{k}$ in the denominator. Using \eqref{kernelon}, we then obtain a single factor of $\delta V_{k}$ in the numerator, the correct measure for an integral.} 

\begin{equation}\label{nonloc} 
\begin{split}
\exp\bigg(\int \frac{d^dk}{(2\pi)^d} &  \, \frac{1}{2 \omega_k}  {\delta \over \delta \bar\beta_{-\vec{k}}^*} \, {\delta \over \delta \bar\beta'_{\vec{k}}  } \,  \zpi^*[\beta,\bar\beta] \zpi[\beta',\bar\beta'] \bigg|_{\bar \beta^{(')}= 0} \bigg) =\\
&\int \mathcal{D}\bar\beta' \mathcal{D}\bar\beta \, \exp\left(-\int \frac{d^dk}{(2\pi)^d}   \, 2 \omega_k \bar\beta_{-\vec{k}}^*  \bar\beta'_{\vec{k}}  \right) \,\zpi^*[\beta,\bar\beta] \zpi[\beta',\bar\beta'] 
\end{split}
\end{equation}

Plugging this equality into \eqref{smatrixpp1} we obtain 
\begin{equation}\label{smatrixpp2} 
 \int \mathcal{D}\bar\beta' \mathcal{D}\bar\beta \, \exp\left(-\int \frac{d^dk}{(2\pi)^d}   \, 2 \omega_k \bar\beta_{-\vec{k}}^*  \bar\beta'_{\vec{k}}  \right) \,\zpi^*[\beta,\bar\beta] \zpi[\beta',\bar\beta'] = \exp\left(\int \frac{d^d p}{(2 \pi)^d}2 \omega_{\vec p} \beta^*_{\vec p}\beta_{-\vec p}\right)
\end{equation}
Using \eqref{bek}, the above equation can now be rewritten in terms of $\beta(x)$ rather than $\beta_{\vec{k}}$. We 
obtain 
\begin{equation}\label{unitsm}
\int \mathcal{D}\bar\beta^* \mathcal{D}\bar\beta'  e^{\int  d^d x'  d^d y' \, \bar{\wf}(x'-y')\bar\beta(x')^* \bar \beta' (y') } \,\zpi^*[\beta,\bar\beta] \zpi[\beta',\bar\beta']  = e^{\int d^d x  d^d y \,\bar{\wf}(x -y) \beta(x)^* \beta' (y)}
\end{equation}
where 
\beq
\bar{\wf}(x-y) = \int \frac{d^d p}{(2 \pi)^d} (2\omega_p) e^{i p (x-y)}
\eeq

The derivation presented in this subsection applies equally well to both the 
Dirchlet and the in out path integrals. In the rest of this section we will 
reexamine this equation separately for these two cases.

\subsection{Rederivation of the unitarity constraint for the in-out problem}
\label{rio}

In the case of the in-out problem, the equation \eqref{smatrixpp2} admits a simple interpretation in terms of the coherent quantization of \S \ref{coherent}. Starting with the matrix elements of equation $U^\dagger U=1$ between initial and final coherent states (here $U$ is the time evolution operator), we insert the coherent state resolution of identity 
between $U^\dagger$ and $U$ to obtain
\begin{equation} \label{cs}
\begin{split}
\langle \bar z_f | U^\dagger U| z_i  \rangle &=  \langle \bar z_f |  z_i  \rangle\\
\int \frac{d^2 z}{\pi} e^{-z \bar z}\langle \bar z_f | U^\dagger| z\rangle \langle z| U| z_i  \rangle &=  e^{\bar z_f z_i}\\
\end{split}
\end{equation}
Since the Hilbert Space of a scalar quantum field theory is the product of Hilbert Spaces (one for each value of the spatial momentum), \eqref{cs} has a simple generalization to such a quantum field theory. Noting that  $\beta_k$ and $z$'s are related by \eqref{betabarz}, we find 
\begin{equation} \label{ft}
\int  \mathcal{D}\bar\beta' \mathcal{D}\bar\beta \, e^{-\int \frac{d^dk}{(2\pi)^d}   \, 2 \omega_k \bar\beta_{-\vec{k}}^*  \bar\beta'_{\vec{k}}}\zpi^*[\beta,\bar\beta] \zpi[\beta',\bar\beta']  =  e^{\int \frac{d^d p}{(2 \pi)^d}2 \omega_{\vec p} \beta^*_{\vec p}\beta_{-\vec p}'}
\end{equation}
which is the same as \eqref{smatrixpp2} since  - as we have explained in \S \ref{coherent}  - the matrix elements of  $U$ evaluate precisely to the path integral as a function of positive and negative energy data. 

Note that the extremely simple derivation presented in this subsubsection is valid even at finite values of $T$.

\subsection{Comments on the Dirichlet Problem}\label{cdp}

In the case of the Dirichlet problem, one might, at first, be tempted to imitate 
the analysis of the previous subsection, but this time working in Schrodinger basis. Proceeding naively one `derives'
\begin{equation}\label{unitsch}
\int \mathcal{D}\bar\beta    \,\zpi^*[\beta,\bar\beta] \zpi[\beta',\bar\beta]  = e^{\int d^d x  d^d y \,\bar{\wf}(x -y) \beta(x)^* \beta' (y)}
\end{equation}
even at finite values of $T$. Note that \eqref{unitsch} differs from 
\eqref{unitsm}

While \eqref{unitsch} is true for the literal $\epsilon=0$ Lorentzian Dirichlet path integral,  this object does not capture S matrices (see \S \ref{bc}). The 
Lorentzian Dirichlet problem that does capture S matrices was specified in 
\S \ref{bc}, and \eqref{unitsch} does not apply to this problem \footnote{As $\omega \epsilon T $ is large, the effective time evolution operator computed by the almost Lorentzian Dirichlet path integral is significantly non unitary, so the naive assumptions that go into the derivation of \eqref{unitsm} do not apply.}. The reason that 
\eqref{unitsm} applies in this case is, presumably, related to the fact (see \S \ref{bc} ) that the leading large $T$ contribution to the almost Lorentzian Dirichlet problem is identical to that that of the in-out problem. 

For the reasons spelt out above it is clear that \eqref{unitsm} applies to the 
Dirichlet problem only in the large $T$ limit.

\section{Path integral as a functional of boundary values in position space: massive particles} \label{pimp} 

In \eqref{smatrix} we have derived the relationship between Taylor Series coefficients of the path integral
as a functional of boundary values, $G_{\rm bdry}$, and the S matrix. \eqref{smatrix} asserts that we 
obtain the S matrix by multiplying $G_{\rm bdry}$, by free solutions (of the modes whose S matrix we wish to compute) and integrating over all boundary points. More colloquially, the S matrix is the smeared path integral as a function of boundary values, with the smearing functions given by the free wave functions of the scattering 
states, restricted to the boundary. 

Atleast in appropriate situations (e.g. in theories with a mass gap)  the S matrix is well defined. The fact that we can recover this S matrix from appropriate smearings of  $G_{\rm bdry}$ in the limit $R\to \infty$ raises the following natural question: does $G_{\rm bdry}$ itself have a well behaved $R \to \infty$ limit, or does the smearing of $G_{\rm bdry}$ play a crucial role in ensuring the good behaviour  of the $S$ matrix?  

A second related question also poses itself. Recall that the $T$ matrix (the part of the $S$ matrix obtained after subtracting out identity) is expected to be an analytic function of  the momenta 
times a momentum conserving delta function. Does the analyticity 
of $T$ have its origins in the analyticity of $G_{\rm bdry}$ itself (as a function of positions) or does the smearing play a key role here?

Upto this point in this paper, all our discussions have been equally applicable to massless as well as massive theories; 
indeed nothing in our analysis so far has made a qualitative distinction between these two cases. However, it turns out that the structure of $G_{\rm bdry}$ as a function of boundary points is qualitatively different in the case of 
massive and massless fields. Accordingly, we consider these two cases separately: in this section we study the scattering of massive particles (i.e. we study scattering in a theory with a mass gap $m$). In order to study scattering we must, of course, ensure that the IR cut off of our spacetime, $T$, is much larger than $1/m$. It follows that in the limit of physical interest $m  T\gg 1$.  The existence 
of a large parameter simplifies the computation of Greens functions and the path integral as a function of 
boundary values. In practical terms the large parameter allows us to evaluate integrals via saddle points, yielding relatively simple and explicit results for the path integral as a function of boundary values. We will now explain how this works.

\subsection{Saddle point evaluation of the Greens function 
at long distances} 

 Let us suppose that the time ordered two point function of two scalars $\phi(x)$ admits the following Lehman decomposition in momentum space 
 \begin{equation}\label{lehmandecomp} 
 \left \langle \phi(p) \phi(p') \right \rangle  =  (2 \pi)^{D} \delta(p+p') \int d m^2  \frac{\rho(m^2)}{p^2+ m^2 -i \epsilon }
\end{equation} 
Let us also suppose that we have chosen the normalization of the field $\phi$ to ensure  that $\rho(m^2)$ takes the form 
\begin{equation}\label{rhom} 
\rho(m^2)= \delta(m^2-m_0^2) + \tilde{\rho}(m^2)
\end{equation} 
where the function $\tilde \rho(m^2)$ has no delta function type singularities and is assumed to be smooth apart from possible threshold type singularities (e.g. the function $\tilde \rho(m^2)$ may be nonzero only for $m^2 > 4 m_0^2$, and may have additional threshold singularities at 
$m^2= 16 m_0^2$, etc).

Under these assumptions, the $n$ point time ordered Greens function of the scalar fields $\phi$ in momentum space takes the form 
\begin{equation} \label{nptphi}
\langle \phi(p_1) \phi_(p_2) \ldots \phi(p_n) \rangle =
\prod_{i=1}^n \left( {1 \over p_i^2 + m_i^2 - i\epsilon} \right) \mathcal{ S}(p_i) + {G'}(p_i) 
\end{equation} 
where $\mathcal{ S}(p_i)$ contains the contributions from interaction and also contains overall momentum conserving $\delta$- function. The first term on the RHS of \eqref{nptphi} represents the contribution of the $\delta$ functions in \eqref{rhom} for all external states, while the second term on the RHS ($G'$) of  \eqref{nptphi} represents the contribution of ${\tilde \rho}(m^2)$, for atleast one of the external particles. 

Let us now use \eqref{nptphi} to evaluate the Greens function of 
our $\phi$ fields at or near the boundary of our spacetime (we will later explain how this quantity can be used to evaluate the 
quantity of main interest to us, namely $G_{\rm bdry}$). 

In order to evaluate the Greens function $\widetilde{G}(t_i,\vec{x}_i) =\langle \phi(x_1, t_1) \phi_(x_2, t_2) \ldots \phi(x_n, t_n) \rangle$ 
near the boundary of our spacetime, we Fourier Transform \eqref{nptphi}, and evaluate the Fourier Transform at parametrically large values of times and positions. We obtain:
\beq\label{gft}
\widetilde{G}(t_i,\vec{x}_i) = \int\frac{d^D p_i}{(2 \pi)^D}e^{- i p_i.x_i}\langle \phi(p_1) \phi_(p_2) \ldots \phi(p_n) \rangle
\eeq
For this purpose, it is sufficient to ignore the contribution of ${G'}$. \footnote{ The argument for this goes as follows. If we evaluate the two 
point function for the propagator \eqref{lehmandecomp} in position space at asymptotically large values of positions and times we find  
\begin{equation}\label{asympto}
\begin{split} 
  G(t_i,\vec{x}_i) \equiv \langle \phi(x) \phi(x')  \rangle &= \int d m^2 \frac{d^D p}{(2\pi)^D} \, e^{i p.(x-x')}\frac{\rho(m^2)}{p^2+ m^2 -i \epsilon }
 \approx \int d m^2\rho(m^2)  e^{im \sqrt{- (x-x')^2}} \\
& = e^{im_0\sqrt{- (x-x')^2}} + \int d m^2 \tilde {\rho} (m^2)  e^{im \sqrt{- (x-x')^2}} 
 \end{split} 
\end{equation} 
(in going from the first to the second expressions in \eqref{asympto} we have used the saddle point approximation).
Assuming $x$ and $x'$ are separated in a timelike manner (this will be the case of interest; see below), we find that the second expression on the 
second line of \eqref{asympto} is much smaller than the first (this is a consequence of cancellation due to the rapid oscillation of phases in the integral over $m^2$) and so can be ignored. It follows that in the appropriate asymptotic limit for positions, one can ignore the contribution of $\tilde \rho(m^2)$ to the two point function, and hence also ignore $G'$ (in \eqref{nptphi}) in comparison to the first term in \eqref{nptphi}. The physics behind this technical argument is the following. In order to specify a two particle state - in addition to its net momentum - one needs to specify its wave function for coordinate differences. The width of every such wave function grows like $\sqrt{t}$ at large times. It follows that the two particle state `dilutes out' in comparison to the one particle state, and so the late time behaviour is dominated by one particle states. Note, however, that this dilution is relatively mild (a power law) rather than the exponential damping one might, at first have expected from \eqref{asympto}. We believe that the mathematical explanation for this dilution lies in the fact that  $\tilde {\rho} (m^2)$ is non-analytic at the two particle threshold, and so has a Fourier transform that decays like a power law (rather than exponentially) at large momenta. } We use the exponential representation for the overall momentum conserving delta function in the S-matrix i.e.\footnote{$S(p_i)$ is not quite the S-matrix because the momentum arguments are not on-shell. But expanding this function in $p_i^0$ around the on-shell values gives the S-matrix as the $O(1)$ part.}
\beq\label{df}
\mathcal{S} (p_i) = S(p_i) \int \frac{d^D y}{(2 \pi)^D} e^{i p_i. y}  
\eeq 

Plugging \eqref{nptphi} and \eqref{df} in \eqref{gft} and taking large time $(t_i)$ limit, we obtain
\begin{equation} \label{invsmat} 
    \begin{split}
        \lim\limits_{t_i\rightarrow\infty} \widetilde G(t_i,\vec{x}_i) = \lim\limits_{t_i\rightarrow\infty}  \int \frac{d^D y}{(2 \pi)^D} \prod\limits_i  \left(\int \frac{d^D p_i}{(2 \pi)^D} {e^{-i p_i. (x_i - y)}\over p_i^2 + m_i^2 - i\epsilon}\right) S(p_i) 
    \end{split}
\end{equation}
 We can evaluate the $p_{i0}$ integral by closing the contour in the upper-half plane and picking up residue from the pole at $p_{i0} = - \omega_i + i \epsilon$ where $\omega_i = \sqrt{\vec{p}_i^2 + m_i^2}$. Assuming that the contributions of 
non-analyticities (e.g. from exchange poles) in $S(p_i)$ can be ignored (we will return to this point below) we find 
\begin{equation}\label{grn}
\begin{split}
        \lim\limits_{t_i\rightarrow\infty} \widetilde G(t_i,\vec{x}_i) = \lim\limits_{t_i\rightarrow\infty}\int \frac{d^D y}{(2 \pi)^D}\prod\limits_i\left( \int \frac{id^{D-1} 
        \vec{p_i}}{(2 \pi)^{D-1}} \frac{ e^{-i \omega_i (t_i-t) - i \vec{p}_i. (\vec{x}_i- \vec{y})}} {2 \omega_i}\right) S(p_i) 
    \end{split}   
 \end{equation}
where $y = (t , \vec{y})$ and the boundary insertions are at points $x_i = (t_i, \vec{x})$. As will be shown below, \eqref{grn} is essentially the inverse Fourrier transform of \eqref{smatrix}. \footnote{The way this works is as follows: the only differences between \eqref{grn} and \eqref{smatrix} are the factors of $2\omega_i$ sitting around in \eqref{grn}. But, we actually wanted to compute the boundary correlation function, while the object in \eqref{grn} is a bulk correlation function. The difference between these in the asymptotic limit, as will be explained below, is precisely the factors of $2\omega_i$. }

In the $t_i \rightarrow \infty$ limit, we can then evaluate the $\vec{p_i}$ integral using the saddle point approximation. Once again assuming that the contribution of any 
singularities in $S(p_i)$ can be ignored (again, we will return to this below) 
we obtain
\beq \label{psadpt} 
\vec{p_i} = -m_i \frac{\vec{x_i} - \vec{y}}{\sqrt{(t_i - t)^2 - (x_i -y)^2 }}
\eeq
Note that the saddle point \eqref{psadpt} lies on the contour of integration (for the variable ${\vec p}_i)$ only when the vector $x_i^\mu - y^\mu$ is timelike. When this vector is spacelike, on the other hand, 
we naively have two saddle points off the integration contour. Presumably, in this case, the reliable saddle point
is the one that gives the Euclidean answer (i.e. extremises $e^{-\sum_{i} m_i d_i}$ where $d_i$ 
represents real Euclidean distances. We will return to the study of these `Euclidean' saddles later. 
For now, we focus on the `Lorentzian' saddles, which exist only when $x_i^\mu - y^\mu$ is a timelike vector.

 After taking into account the one-loop determinant, the Greens function \eqref{grn} in the saddle point approximation is given by:
\beq \label{gffaraway}
  \lim\limits_{t_i\rightarrow\infty} \widetilde G(t_i,\vec{x}_i)  = \int \frac{d^D y}{(2 \pi)^D}\prod_i \left( \frac{i m_i ^{\frac{D-3}{2}}}{2 (2 \pi)^{\frac{D-1}{2}}}\frac{e^{-i m_i d_i }}{d_i^{\frac{D-1}{2}}} \right) S\left(\frac{m_i (\vec{x}_i - \vec{y})}{d_i}\right)
\eeq
where $d_i = \sqrt{(t_i -t)^2 - (\vec{x}_i- \vec{y})^2}$ is the distance between the boundary insertion points and $y$. We can think of $y$ as the point where scattering takes place in the bulk. 

As we have explained in Appendix \ref{apgrn} for the Dirchlet problem, and 
in Appendix \ref{digch} (see around \eqref{ndelz}) for the in out problem, 
in the limit  $m T \gg1$, the quantity $G_{\rm bdry}$ can be evaluated acting on the Greens function $G(t_i,\vec{x}_i)$ with the operator $2i n. \nabla$ for every external point. In this subsection we work, for concreteness,  the particular case of the  bulk theory with cutoff at constant time slices $T$ and $-T$. We refer to this spacetime as a `slab'. Using \eqref{gffaraway}, we conclude that  quantity $G_{\rm bdry}$,  with $n$ sources at the past boundary (i.e. at $-T$) and $m$ sources at the future boundary (i.e. at $T$) is given by:
\beq \label{cor}
\begin{split}
\frac{\delta^n Z[\beta_0]}{\delta \beta_0(y_1)\cdots \delta \beta_0(y_n)} =& \prod_{i = 1}^n (2i n. \nabla_i)  \widetilde G(t_i,\vec{x}_i)\bigg|_{t_i = - T} \quad \prod_{i=1}^m  (2i n. \nabla_j) \widetilde G(t_i,\vec{x}_i)\bigg|_{t_i =  T} \\
\approx & \int \frac{d^D y}{(2 \pi)^D}\prod_{i=1}^{n} \left( \left( \frac{m_i}{ 2 \pi}\right)^{\frac{D-1}{2}}\frac{(-T - t)}{{d^{in}_i}^{\frac{D+1}{2}}} e^{-i m_i d^{in}_i }  \right) \prod_{i=1}^{m} \left( \left( \frac{m_i}{ 2 \pi}\right)^{\frac{D-1}{2}}\frac{(T - t)}{{d^{out}_i}^{\frac{D+1}{2}}} e^{-i m_i d^{out}_i } \right) \\
& S\left(\frac{m_i (\vec{x}^{in}_i - \vec{y})}{d^{in}_i}, \frac{m_i (\vec{x}^{out}_i - \vec{y})}{d^{out}_i}\right)
\end{split}
\eeq
where $d^{in}$ and $d^{out}$ are the distances between the past insertions (at $t = -T$) or the future insertions (at $t = T$) and the bulk point $y=({\vec y}, t)$.  In going from the first to the second line of \eqref{cor} we
have acted the differential operators only on the exponents $e^{- i m_i d_i}$ (terms arising from the action 
of the derivative on other occurrences of $d_i$ are all subdominant compared to the term we have retained, in the limit $m T \gg 1$). 

In order to complete our evaluation of $G(t_i,\vec{x}_i)$, we now evaluate the integral over $y$. This integral can also be performed in the saddle point approximation, as we explore in detail in the subsequent subsections. Before turning to this, in the next subsection, we first present a consistency check of \eqref{gffaraway}

\subsection{Consistency of massive saddle points with the S matrix formula} \label{consist} 

In this subsection we present a consistency check of \eqref{gffaraway}.  Note that \eqref{cor} gives us a formula for the `boundary correlation function' in terms of the bulk S-matrix (as a function of incoming and outgoing momenta). On the other hand \eqref{smatrix} gives us the inverse relation: it gives us 
an expression for the S matrix in terms of a smearing of the `boundary correlation functions'. In order to verify the consistency of these two equations, we substitute \eqref{cor} in \eqref{smatrix}, to obtain
\beq
\begin{split} \label{smaton}
 S(p_i, q_i) =& \int \frac{d^{d+1} y}{(2 \pi)^D} \prod_{i= 1}^{n}\int d^dx_i \quad e^{i \omega_i T + i p_i x_i} \left( \frac{m_i}{ 2 \pi}\right)^{\frac{d}{2}}\frac{(-T - t)}{{d^{in}_i}^{\frac{d+2}{2}}} e^{-i m_i d^{in}_i }\\
 & \prod_{i= 1}^{m}\int d^d x'_i \quad e^{i \omega_i T - i q_i x'_i}\left( \frac{m_i}{ 2 \pi}\right)^{\frac{d}{2}}\frac{(T - t)}{{d^{out}_i}^{\frac{d+2}{2}}} e^{-i m_i d^{out}_i }\quad S\left(\frac{m_i (\vec{x}_i - \vec{y})}{d^{in}_i}, \frac{m_i (\vec{x}'_i - \vec{y})}{d^{out}_i}\right)
\end{split}
\eeq
where $d = D-1$ and we have used $\sqrt{h} = 1$ (this is true for the boundary of the slab spacetime we have specialized to)  and also substituted the plane wave expressions for the mode functions $f^*_{p_i}$ and $f_{q_i}$ (see \eqref{mode}). 

Note that the LHS and RHS of \eqref{smaton} are both given in terms of the S matrix. Since the S matrix is 
an arbitrary function, consistency demands that \eqref{smaton} holds as an identity. This is indeed the 
case as we now demonstrate, by explicitly evaluating the integrals on the RHS of \eqref{smaton}. 

As $mT \gg 1$,  the integrals over the boundary points $x_i$ and $x'_i$ can be performed  using the saddle point approximation. Once again we ignore the potential contributions of non-analyticities in $S$ to this saddle point integral (again we will see below why this is ok). Denoting the coordinates of the integration variable 
$y$ in \eqref{smaton} by $(t, {\vec y})$, we find that the  $x-$ and $x'_i$ saddle point is given by:
\beq \label{spuu}
\vec{x}_i = {\vec y}  - \left(\frac{T + t}{\omega_i}\right) \vec{p}_i,  \qquad \qquad \vec{x'}_i = y + \left(\frac{T - t}{\omega_i}\right) \vec{q}_i
\eeq
These saddle point equations \eqref{spuu} have a simple geometrical interpretation. They assert that the saddle point values of 
the boundary variables $x_i$ are obtained as follows. Starting 
at the bulk point $(t, {\vec y}$), shoot a particle out towards the 
future/past boundary with momentum ${\vec p_i}$. $x_i$ is the location of this projectile at the moment it reaches the boundary (i.e. at time $T$ for future insertions or time $-T$ for past insertions). 

After performing the $x$ and $x'$ integrals using the saddle point approximation, the RHS of \eqref{smaton} is given by:
\beq
\int \frac{d^{d+1} y}{(2 \pi)^D}  e^{i \left(\sum_i\omega^{out}_i - \sum_i \omega^{in}_i\right) t} e^{i \left(\sum_i \vec{p}_i - \sum_i \vec{q}_i\right).{\vec y} }S\left(\vec{p_i}, \vec{q_i}\right)
\eeq
The $y$- integral is now trivial to perform and yields the energy and momentum conserving delta functions i.e.
\beq
\delta\left(\sum_i\omega^{out}_i - \sum_i \omega^{in}_i\right) \delta^{d}\left(\sum_i \vec{p}_i - \sum_i \vec{q}_i\right) S\left(\vec{p_i}, \vec{q_i}\right) = 
\delta^{D} (\sum_i p_i^\mu  - \sum_i {q}_i^\mu )S\left(p_i^\mu, q_i^\mu \right)
\eeq
which matches the LHS including all factors of $2$ and $\pi$ etc.


\subsection{Saddle Point Evaluation of \eqref{cor} } \label{spe} 

We now return to the evaluation of the integral over $y$ in \eqref{cor}. Let us imagine that the S matrix 
${\cal S}$ that appears in \eqref{cor} a perturbative expansion in a coupling constant. Inserting 
this perturbative expansion into ${\cal S}$ on the RHS of \eqref{cor} gives us a perturbative expansion of 
$G_{\rm bdry}$. Individual terms in this expansion are obtained from the contribution of particular 
Feynman diagrams to the S matrix $S$. In this subsection we study the simplest of these terms: the contribution
to $G_{\rm bdry}$ of a tree level contact S matrix. In this case, the assumption we have made on several previous occasions - namely that non-analyticities in $S$ (apart from momentum conserving $\delta$ functions) do not contribute to various saddle point evaluations - is obviously true, just because $S$ itself is an analytic 
function of momenta. The integral over $y_i$ in \eqref{cor} can now be evaluated. In the saddle point approximation, the derivative of the sum of the large phases, $\sum_i m_i d_i$, has to vanish. In other words
$y$ is chosen to extremize the quantity 
\begin{equation} \label{extsl}
\sum_i m_i d_i
\end{equation}

Of course \eqref{extsl} applies in Lorentzian space. Let us momentarily, however, consider the Euclidean 
analogue of this equation (which infact arises when we consider the problem of evaluating action as a function of 
boundary values in Euclidean space). In this situation, in order to extremise of \eqref{extsl}, we must find the  bulk position $y$ that extremises the energy of $n$ constant tension `rubber bands' of energy per unit length -i.e. tension -  $m_i$.  One end of each of these bands is anchored at a fixed boundary point $x_i$. The other ends of each band are attached to each other, at a point $y$ that can be freely varied. The rubber bands explore various possibilities for $y$ and then settle on the value that extremises the total energy. The condition for this to happen is familiar from elementary undergraduate physics; it is simply the requirement that the forces exerted by each of the four rubber bands on its common meeting point cancel out, i.e. that 
\begin{equation}\label{forcecons} 
\sum_i m_i {\hat n}_i=0
\end{equation}
where the vector ${\hat n}_i$ is directed outwards from $y$ to the boundary point $x_i$.
Any configuration in which this happens is a saddle point for our problem. 

Returning to Lorentzian space the saddle point equation \eqref{forcecons} still holds, except that 
each of the ${\hat n}_i$ is now a (past or future directed) timelike unit vector in Lorentzian space (we assume that we are working with a Lorentzian saddle of \eqref{psadpt}). In this case the vector ${\hat n}_i$ 
parameterizes the direction of a particle trajectory - more precisely it is the `velocity $D$ vector (i.e.
it equals $(\gamma, v^i \gamma)$ ) associated with that trajectory. $m_i {\hat n}_i $ is the momentum 
$D$ vector associated with the trajectory, and \eqref{forcecons} simply asserts that the momentum of the scattering particle is conserved at the interaction vertex.  
\footnote{The relationship between Euclidean saddles (discussed in the previous paragraph) and Lorentzian saddles (discussed here) is subtle. Consider, for example, a configuration in which four insertion points are all located in the two-dimensional $R^{1,1}$ (which is a subspace of $R^{1,D-1}$) spanned by Cartesian coordinates $(t, x)$. Let the locations of the insertions be given by 
$(T, R)$, $(T, -R)$, $(-T, R))$, and $(-T, -R)$, with $R>T$. In this situation 
real contact saddle points certainly exist in Euclidean space but do not exist in Lorentzian space. The saddle points that are relevant to the considerations of this paper are always those that are real in Lorentzian space. That this is the case is clear on physical grounds. Mathematically, it is presumably a consequence of the 
discussion in subsection \ref{bc} which presumably determines the contour of integration that should be used for the saddle point approximation. }

It is useful to do some counting. Let us first focus on the neighborhood of the meeting point $y$. 
Let us suppose that $n$ particles meet at the interaction point (or $n$ constant tension rubber bands meet at 
this point in the Euclidean picture). Let us regard two `meetings' as the `locally equivalent' if the
particle momenta (forces in the Euclidean picture)  that emanate from the interaction point are  
the same (no matter where the meetings happen). It follows that the set of inequivalent meetings 
is parameterized by $n(D-1) -D$ parameters. The first term here gives the number of parameters in a collection of $n$ unit vectors, while the second term counts the number of momentum conservation (force balance in the Euclidean picture) equations. 
 
Of course, the particles (or Euclidean constant tension rubber bands) can meet anywhere 
in the $D$ dimensional bulk. This tells us that the total number of saddle points in our diagram 
equals $n(D-1)-D+D=n(D-1)$ parameters. But this is exactly the number of parameters 
in a collection of $n$ boundary points. It follows that there are at most discretely many saddle points corresponding to any choice of boundary points. Infact we believe 
that no collection of boundary points ever receives contributions from more than 
a single saddle (however we do not have a proof of this statement). 

The final answer for the contribution of a contact saddle to boundary Green's function is given by performing the integral over $y$ in \eqref{cor}. Working in the saddle point approximation (and  including also the determinant factor from the integral over $y$) we find 
\begin{equation}\label{corrsaddle}
\begin{split}
\frac{\delta^n Z[\beta_0]}{\delta \beta_0(x_1)\cdots \delta \beta_0(x_n)}
\approx & \frac{1}{(2 \pi)^{3D/2}\sqrt{\rm det \mathcal{A}}}\prod_{i=1}^{n} \left( \left( \frac{m_i}{ 2 \pi}\right)^{\frac{D-1}{2}}\frac{(-T - t)}{{d^{in}_i}^{\frac{D+1}{2}}} e^{-i m_i d^{in}_i }  \right) \prod_{i=1}^{m} \left( \left( \frac{m_i}{ 2 \pi}\right)^{\frac{D-1}{2}}\frac{(T - t)}{{d^{out}_i}^{\frac{D+1}{2}}} e^{-i m_i d^{out}_i } \right) \\
& S\left(\frac{m_i (\vec{x}^{in}_i - \vec{y})}{d^{in}_i}, \frac{m_i (\vec{x}^{out}_i - \vec{y})}{d^{out}_i}\right)
\end{split}
\end{equation} 
where the matrix $\mathcal{A}$ is given by:
\beq
\mathcal{A}_{\mu \nu} = \sum_{i=1}^{n + m}\frac{m_i}{d_i}\left(\frac{(x_i - y)_\mu (x_i -y)_\nu}{d_i^2} - \eta_{\mu\nu}\right) 
\eeq
where $x^\mu_i = (\pm T, \vec x _i)$, $d_i = d_i^{in} \,\rm{or} \, d_i^{out}$ depending on whether its a future/past insertion and $y^\mu = (t , \vec y)$ are chosen such that they satisfy \eqref{forcecons}. Notice that the matrix $\mathcal{A}_{\mu\nu}$ is a sum of matrices $A_i$ given by:
\beq\label{am}
A_i = \frac{m_i}{d_i}\left(\frac{(x_i - y)_\mu (x_i -y)_\nu}{d_i^2} - \eta_{\mu\nu}\right) 
\eeq
Hence the determinant of $\mathcal{A}$ is given by:
\beq
\rm{det}\left(\sum_{i =1}^{n+m} A_i\right) = \sum_{(k_1 \cdots k_D)}\rm{det}\mathcal{A}^{(k_1 \cdots k_D)}
\eeq
where the sum ranges over all $D$-tuples, $k_i$ ranges from $1$ to $n+m$ and  $A^{(k_1 \cdots k_n)}$ is the matrix with $i^{\rm {th}}$ row from the matrix $\mathcal{A}_{k_i}$ given in \eqref{am}.

\subsection{`Exchange' saddles} \label{es}

In obtaining the formula \eqref{cor} - and so our final formula for the boundary correlators \eqref{corrsaddle} -  we have ignored the possible non-analyticities in the $S$ matrix,  $S(p_i)$ in \eqref{invsmat} to the contour integral (that took us from \eqref{invsmat} to \eqref{grn}, and to the saddle point integral that took us from \eqref{grn} to \eqref{gffaraway}. In this subsection, we study a simple example of an S matrix with a non-analyticity - namely a tree level S matrix with pole exchange - and examine the effect of this non-analyticity on \eqref{cor} and \eqref{corrsaddle}. 

Consider the theory governed by the Euclidean Lagrangian
\begin{equation}\label{phipsi} 
\int d^Dx \left( \frac{1}{2} ( \partial \phi)^2 +
\frac{1}{2} m_\phi^2 \phi^2 + \frac{1}{2} ( \partial \psi)^2 +
\frac{1}{2} m_\psi^2 \psi^2 + g \phi^2 \psi\right)  
\end{equation} 
The tree level $\phi \phi \rightarrow \phi \phi$ scattering amplitude, in this theory, is given by 
\begin{equation} \label{tsa} 
S(p_1, p_2, p_3, p_4) = 
g 
\left( \frac{1}{m_\psi^2 +(p_1+p_2)^2 }+  \frac{1}{m_\psi^2 +(p_1+p_3)^2 } +\frac{1}{m_\psi^2 +(p_1+p_4)^2 } \right) 
\end{equation} 
Let us insert  this expression into \eqref{invsmat}. In going from  
\eqref{invsmat} to \eqref{grn} we performed the contour integral over the `energies' $p^0_i$, accounting for poles from the explicit denominators in \eqref{invsmat} but ignoring possible poles in the S matrix. In the example under study in this paper, we see from \eqref{tsa} that additional poles of this nature do exist, and must also be accounted for.
As we will see below, the consequence of this 
modification is that our final saddle point 
evaluation of the path integral will generically be a sum over saddle point contributions, only 
one of which is given by \eqref{corrsaddle}. 
Restated, while  \eqref{corrsaddle} (with 
\eqref{tsa} inserted into it) continues to contribute, it is only one of several possible saddle contributions. In the rest of this subsection, we explain how this comes about 
in the context of our simple example. In the subsequent subsection, we will examine the implications of this observation. 

Let us insert \eqref{tsa} into \eqref{invsmat}. 
It is useful to rewrite the resultant expression in the form 
\begin{equation} \label{invsmatexch} 
    \begin{split}
        \lim\limits_{t_i\rightarrow\infty} \widetilde G(t_i,\vec{x}_i) =\lim_{t_i\rightarrow\infty} 3 g   \prod_{i=1}^{4}&\left( \int \frac{d^D p_i}{(2 \pi)^D}\right) \frac{d^D p}{(2 \pi)^D} \int \frac{d^D y_1}{(2 \pi)^D} \frac{d^D y_2}{(2 \pi)^D} \frac{e^{-i p (y_1 -y_2)}}{p^2+m_\psi^2 -i \epsilon}\\
        &\prod_{i =1}^{2}\frac{e^{-i p_i.(x_i-y_1)}}{p_i^2 + m_{\phi}^2 - i\epsilon}\prod_{i =3}^{4}\frac{e^{-i p_i.(x_i-y_2)}}{p_i^2 + m_{\phi}^2 - i\epsilon}  
    \end{split}
\end{equation}
We now use contour methods to perform the integral over $p_{0i}$ (for $1\ldots 4$) and $p_0$
to obtain 
\begin{equation}\label{grnexch}
    \begin{split}
        \lim\limits_{t_i\rightarrow\infty} \widetilde G(t_i,\vec{x}_i) =\lim_{t_i\rightarrow\infty} 3 g   \prod_{i=1}^{4}&\left( \int \frac{id^{D-1} \vec{p}_i}{(2 \pi)^{D-1}}\right) \frac{id^{D-1} \vec{p}}{(2 \pi)^{D-1}} \int \frac{d^D y_1}{(2 \pi)^D} \frac{d^D y_2}{(2 \pi)^D} \frac{e^{-i \omega_p (t-t')-i \vec{p}(\vec{y}_1 -\vec{y}_2)}}{2\omega_p}\\
        &\prod_{i =1}^{2}\frac{e^{-i \omega_i (t_i -t)-i \vec{p}_i.(\vec{x}_i-\vec{y}_1)}}{2 \omega_i}\prod_{i =3}^{4}\frac{e^{-i \omega_i (t_i -t')-i \vec{p}_i.(\vec{x}_i-\vec{y}_2)}}{2 \omega_i}  
    \end{split}
\end{equation} 
where we have paramterized bulk points $y_1$ and $y_2$ as $(t, \vec{y_i})$ and $(t', \vec{y_2})$. As under \eqref{grn}, we now evaluated the integrals over ${\vec p_i}$ (for $1\ldots 4$)
and ${\vec p}$ in the saddle point approximation. The resultant saddle point equations are 
\beq \label{psadptexch} \begin{split} 
&\vec{p_i} = -m_\phi \frac{\vec{x_i} - \vec{y_1}}{\sqrt{(t_i - t)^2 - (x_i -y_1)^2 }} 
~~~~~~~~~{\rm For ~i=1, 2}\\
&\vec{p_i} = -m_\phi \frac{\vec{x_i} - \vec{y_2}}{\sqrt{(t_i - t')^2 - (x_i -y_2)^2 }} 
~~~~~~~~~{\rm For ~i=3, 4}\\
&\vec{p} = -m_\psi \frac{\vec{y_1} - \vec{y_2}}{\sqrt{(t - t')^2 - (y_1 -y_2)^2 }} 
~~~~~~~~\\
\end{split} 
\eeq
The analogue of \eqref{gffaraway} is 
\beq \label{gffarawayexch}
\begin{split}
  \lim\limits_{t_i\rightarrow\infty} \widetilde G(t_i,\vec{x}_i)  = \int \frac{d^D y_1}{(2 \pi)^D}\int \frac{d^D y_2}{(2 \pi)^D}&\left(\frac{i m_\phi ^{2(D-3)} m_\psi^{\frac{D-3}{2}}}{32 (2 \pi)^{\frac{5(D-1)}{2}}}\right)\prod_{i=1}^{2} \left( \frac{e^{-i m_\phi d_i(y_1) }}{\left(d_i(y_1)\right)^{\frac{D-1}{2}}} \right)\prod_{i=3}^{4} \left( \frac{e^{-i m_\phi d_i(y_2) }}{\left(d_i(y_2)\right)^{\frac{D-1}{2}}} \right)\\
  &\left( \frac{e^{-i m_\psi d(y_1, y_2) }}{\left(d(y_1,y_2)\right)^{\frac{D-1}{2}}} \right) S\left(\frac{m_\phi (\vec{x}_i - \vec{y_1})}{d_i(y_1)}, \frac{m_\phi (\vec{x}_i - \vec{y_2})}{d_i(y_2)}\right)
\end{split}
\eeq
where $d_i(y_1)$ ($d_i(y_2)$) represent the distance between points $x_i$ and $y_1 (y_2)$ and $d(y_1,y_2)$ is the distance between two bulk points $y_1$ and $y_2$. Note that \eqref{gffarawayexch} involves an integral over two bulk points, $y_1$ and $y_2$, 
in contrast to \eqref{gffaraway} which involves an integral over only a single bulk point. 
This difference carries through to the analogue of \eqref{cor}, which takes the form 
\begin{equation}\label{corexch} 
\begin{split}
\frac{\delta^n Z[\beta_0]}{\delta \beta_0(x_1)\cdots \delta \beta_0(x_4)} \approx & \int \frac{d^D y_1}{(2 \pi)^D}\int \frac{d^D y_2}{(2 \pi)^D}\prod_{i=1}^{2} \left( -\left( \frac{m_i}{ 2 \pi}\right)^{\frac{D-1}{2}}\frac{(T + t) e^{-i m_\phi d_i(y_1)}}{{d_i(y_1)}^{\frac{D+1}{2}}}   \right) \left( \frac{m_\psi^{\frac{D-3}{2}}e^{-i m_\psi d(y_1, y_2) }}{2 (2\pi)^{\frac{D-1}{2}}\left(d(y_1,y_2)\right)^{\frac{D-1}{2}}} \right)\\
&  \prod_{i=3}^{4} \left( \left( \frac{m_\phi}{ 2 \pi}\right)^{\frac{D-1}{2}}\frac{(T - t') e^{-i m_\phi d_i(y_2)}}{{d_i(y_2)}^{\frac{D+1}{2}}} \right)  S\left(\frac{m_\phi (\vec{x}_i - \vec{y_1})}{d_i(y_1)}, \frac{m_\phi (\vec{x}_i - \vec{y_2})}{d_i(y_2)}\right)
\end{split}
\end{equation} 
We can similarly find the analog of \eqref{corrsaddle} by integrating over $y_1$ and 
$y_2$. These integrals can be performed by using the saddle point approximation, with the large parameter again being the IR cutoff $T$. The saddle points can be found by extremizing the energy functional
\begin{equation}\label{enfunct}
E(y_1, y_2) = m_\phi|x_1-y_1| +m_\phi|x_2-y_1| +m_\phi|x_3-y_2| +m_\phi|x_4-y_2|+ m_\psi|y_1-y_2|
\end{equation}
Hence $y_1$ and $y_2$ obey the equations : 
\begin{equation}\label{spexch} 
\begin{split}
    \frac{m_\phi(x_1-y_1)^\mu}{|x_1 -y_1|} + \frac{m_\phi(x_2-y_1)^\mu}{|x_2 -y_1|} + \frac{m_\psi(y_1-y_2)^\mu}{|y_1 -y_2|} &=0\\
    \frac{m_\phi(x_3-y_2)^\mu}{|x_3 -y_2|} + \frac{m_\phi(x_4-y_2)^\mu}{|x_4 -y_2|} - \frac{m_\psi(y_1-y_2)^\mu}{|y_1 -y_2|} &=0
\end{split}
\end{equation} 
where $\mu$ runs from $0 \cdots D-1$.
Let us take note of several points 
\begin{enumerate} 
\item As we have already noted above, it was always clear that one solution to the saddle point equations is given by \eqref{corrsaddle}. 
This follows from the analysis of section \ref{spe} (which correctly captures the contribution of the saddle it focuses on, even in the example discussed in this subsection, even though it incorrectly ignores the contributions of other saddles). In the language of this subsection, comes from the contribution of `a saddle at $y_1=y_2$. As this is a point of non-analyticity of the integrand in \eqref{corexch}, this saddle is complicated to analyze 
in the current formalism. As we already understand the contribution of this saddle, this is no great loss.  In the rest of this subsection, we will focus our attention on saddles with $y_1 \neq y_2$.  
\item The saddle point equations 
\eqref{spexch} have a physical interpretation that is very similar to the one described under 
\eqref{forcecons}. In Lorentzian space, the saddle point described the (straight) world line of two particles of mass $m_\phi$ that arise at the boundary, propagate till they meet at the bulk point $y_1$, merge at $y_1$ into a particle of mass $m_\psi$, propagate to $y_2$ where this particle once again bifurcates into two particles, each of mass $m_\phi$, that head to the boundary. The merging and bifurcations both 
respect $D$ momentum conservation. 
\footnote{ In Euclidean space we 
have two rubber bands of tension $m_\phi$ emerging from the boundary,  meeting an interaction point $y_1$, merging into a rubber band of tension $m_\psi$ and then, at 
point $y_2$, bifurcating again to head to the boundary. The rubber bands are always straight and obey force conservation at points of merging. } Note that both the external particles and the internal particle propagate over distances of order the IR cut off $T$ in this saddle point. Physically, this is possible only because ``all particles propagate onshell'' in the saddle point  
(that this is the case is clear from \eqref{psadptexch}), and because all particles are stable (an unstable particle with a finite lifetime would decay long before it was able to propagate a distance $T$). 
\item It follows from the previous point 
that nontrivial real Lorentzian saddles with $y_1 \neq y_2$ exist only when 
\begin{equation}\label{condofexist} 
m_\psi \geq 2 m_\phi
\end{equation} 
\footnote{On the other hand an examination of the force conservation condition will convince the reader that a nontrivial real Euclidean 
saddle exists only when $m_\psi \leq 2 m_\phi$: note this is the 
converse of \eqref{condofexist}. Euclidean and Lorentzian saddles are distinct (they are not analytic continuations of each other) and they do not coexist. See 
Appendix \ref{esap} for a discussion.}
This follows from the fact that 
an onshell three point S matrix between two particles of mass $m_\phi$ and a single particle of mass $m_\psi$ exists only when \eqref{condofexist} is obeyed.
\item It is not difficult to repeat
the counting of parameters that 
label different exchange saddle point diagrams. Imitating the analysis of subsection \ref{spe}, we first count the parameters that characterize the full  `exchange interaction vertex' (i.e. the full effective four point interaction vertex) with its $`y_1'$ vertex
located at a given point, and then add to this count the $D$ parameters generated by spacetime translations (i.e. add the $D$ parameters $y_1^M$).  The number of inequivalent effective interaction vertices is parameterized by four vectors $k_1$ each of definite Lorentzian norm, that in addition to the condition $\sum_i k_i=0$ that we encountered in subsubsection \ref{spe}, also obey the conditions $(k_1+k_2)^2=m^2$ \footnote{This condition, plus the relation $\sum_i k_i=0$ automatically guarantees $(k_3+k_4)^2=m^2$). } together with 
one additional number - the length of the vector $y_1-y_2$, which is left undetermined by the saddle point equations. It follows that the interaction vertex is parameterized by $4D-4 -D-1+1=4(D-1)-D$ numbers, 
precisely as in subsection \ref{spe}. Adding in the $D$ parameters for the location of the 
interaction, we find that the total number of parameters in such saddles equals $4(D-1)$. This is exactly the same number of parameters as we found in subsection \ref{spe}. 
Although we have performed the counting for 4 particle scattering,
the fact that there are as many parameters in the exchange diagram as in the `contact' diagram of 
subsection \ref{spe} is easily verified to hold for the scattering of any number of particles. 
\item  Recall that in subsection 
\ref{consist} we have demonstrated that inserting the contribution of the saddle point of subsection \ref{spe} into the RHS of \eqref{smatrix}, exactly reproduces the LHS of that equation. The reader may wonder whether the analysis of this subsection turns the success of 
subsection \ref{consist} into a paradox. Would not the additional contributions of the saddle points of this subsection invalidate the previous agreement? A little further thought will convince the reader, however, that this is not the case. 
The saddle points of this section do not contribute at generic values of external momenta, but only for those momenta that obey the equation 
\begin{equation}\label{pop}
(p_1+p_2)^2 =-m_\psi^2
\end{equation} 
These are precisely those values of momenta at which the tree level S matrix is anyway ill-defined (because of the intermediate tree level exchange pole). Consequently, the agreement of subsection \ref{consist} persists at all values of external momenta for which the S matrix is well defined. 
\item The skeptical reader may continue to find herself unconvinced by the previous point. She might argue as follows. The fact that the 
S matrix blows up when \eqref{pop} is obeyed is an artifact of the tree approximation. In the case that $m_\psi> 2 m_\phi$ (recall this condition is needed in order for the new saddles to exist) the pole corresponding to the intermediate $\psi$ particle will develop an 
imaginary part and so move off the 
real axis. In this situation the 
amplitude will be well defined even 
when \eqref{pop} is obeyed, reviving the possibility of a contradiction. 
We turn to an analysis of this point in the next subsection. 
\end{enumerate} 

\subsection{Additional Saddles and Unstable Particles} \label{unst}

As we have seen in the previous subsection in the context of the Lagrangian \eqref{phipsi}, order by order in perturbation theory, the integral in \eqref{invsmat} receives contributions from additional saddle points (other than the universal saddle described in subsection \ref{spe}). Even though the new saddles contribute only at special 
values of scattering momenta, once we move to position space (i.e. to the computation of the path integral as  a function of boundary values) the new saddles contribute at generic boundary locations, and may also yield the dominant contribution over 
a (generic) range of boundary positions. 

We will now argue that the conclusion of the previous paragraph is an artifact of working order 
by order in perturbation theory. In other words, we will argue that the conclusions of the previous paragraph are valid if we first take $g \to 0$ and then take $m T \to \infty$. For physical purposes, 
however, we should first take $mT \to \infty$ and then expand the resultant answer in a perturbation 
in the coupling constant $g$ (i.e. then take $g \to 0$). In other words, we should ensure that 
$\frac{1}{mT}$ is the smallest parameter in the problem; much smaller than any dynamical couplings in the problem \footnote{When this is not the case, there is not enough time for the scattering process under study to be properly completed before the asymptotic particles hit the IR 
cut off. For instance, if one of the particles involved in the scattering process is unstable, with a lifetime that goes to infinity as $g \to 0$ then the particle will not have the time to decay before it meets the IR cut-off.}. In this physical situation the conclusions of the previous paragraph change at the qualitative level. 

As we have remarked above, the new saddle point of the previous subsection exists only when 
$m_\psi> 2 m_\phi$. In this situation, however, 
the $\psi$ particle is no longer stable because 
it can decay into two $\phi$ particles. The interaction that makes this decay possible is precisely the term $g \phi^2 \psi$ in \eqref{phipsi}
that also made the exchange diagram of the previous subsection possible. At the technical level, the instability of the $\psi$ particle can be seen from the one loop contribution to the $\psi$ propagator, which, for example in $D=4$ is given by:
\beq\label{onelooppr}
G_{\psi}(p^2) = \frac{1}{p^2 + m_\psi^2 - \Pi(p^2)}
\eeq
where 
\beq\label{imgp}
\Pi(p^2) = \frac{g^2}{2(4 \pi)^3}\left(c_1 p^2+c_2 m_{\phi }^2+ p^2 r^3 \log \left(\frac{r+1}{r-1}\right)\right)
\eeq
where
\beq
c_1 =  3-\pi  \sqrt{3}, \qquad \qquad c_2 = 3-2 \pi  \sqrt{3}, \qquad \qquad r= \sqrt{\frac{4 m_{\phi }^2}{p^2 - i\epsilon}+1}.
\eeq
From \eqref{imgp} one can note that the loop corrections $\Pi(p^2)$ has an imaginary part (due to the $\log$ piece in \eqref{imgp}). Finally, the intermediate particle $\psi$ decays, and the decay rate is given by (eq $4.24$ in \cite{Xianyu:2016fd}):
\beq
\Gamma = \frac{g^2\pi}{6(4 \pi)^3} m_\psi\left(1- \frac{4 m_{\phi }^2}{m_\psi^2}\right)^{3/2}
\eeq

As we have explained above, we should first take the limit $m T \to \infty$ and only then take the limit $g \to 0$. In other words, we should take the limit  $m T \to \infty$ at a fixed value of the coupling (no matter how small). This procedure changes the analysis of the previous subsection in the following important manner: the tree level $\psi$ and $\phi$ propagators that appear in \eqref{invsmatexch} 
should be replaced by the exact quantum effective 
propagator. While this exact propagator might be rather complicated, the only feature of the propagator that played an important role in the analysis that followed \eqref{invsmatexch} is the location of its poles. As we have seen above, in the case that $m_\psi> 2 m_\phi$ (so that the exchange saddle of the previous subsection exists at all) 
the quantum effective $\psi$ propagator changes, and the mass $m_\psi$ effectively becomes imaginary. 
It follows that the third equation in \eqref{psadptexch} no longer has a solution on the contour of integration. Consequently, there are now no real exchange saddle 
point solutions. 

The mathematical discussion of the previous paragraph has a clear physical translation. 
The imaginary contribution to $m_\psi$ captures 
the physical fact that the $\psi$ 
particle decays (over a time scale of order $1/m_\psi g^2$) and so cannot propagate over distances of order $T$ (recall that $m T \gg 1/g^2$). However, the exchange saddle precisely 
corresponds to a configuration in which $\psi$ propagates over a distance of order $T$. On physical grounds this process should be exponentially 
suppressed like $e^{- a \frac{m_\psi T}{g^2}} $
(here $a$ is a positive number of order unity). 
This conclusion can also, presumably, be reached 
by modifying the saddle point analysis of subsection \ref{es} to account for the fact that $m_\psi$ is 
imaginary. 

Although the discussion of this section has been in the context of the particular example of (tree level exchange) of the previous subsection, it is easy to argue that the final conclusion is true more generally (It holds for loop diagrams as well). Despite first appearances, 
the saddle point of section \ref{spe} is the only 
one that contributes to the evaluation of \eqref{invsmat}. It follows, in other words, that the path integral as a function of boundary values
is accurately evaluated by \eqref{corrsaddle}, and 
so determined in terms of the S matrix in a relatively simple and direct matter. We examine this point further in the next subsection. 

\subsection{`Holographic Renormalization'} \label{hr} 

 In this subsection, we will describe an algorithm that can, in principle, be used to evaluate the path integral as a function of boundary values given the S matrix using \eqref{corrsaddle}. 
 
 Consider a momentum conserving scattering process involving momenta
 external particles with masses $m_i$ and momenta 
${\vec p}_i$. Let us imagine that the scattering occurs at the bulk point with coordinates $y_\mu$. 
The equations \eqref{spuu} then determine 
the of boundary points ${\vec x}_i(y^\mu, {\vec p}_i)$. As we have explained above,  $\{ {\vec x}_i \}$ and $\{ y^\mu, {\vec p}_i \}$ both contain 
$n(D-1)$ variables (where $n$ is the number of scattering particles, or, equivalently, the number of boundary points). It follows that the functional relations ${\vec x}_i(y^\mu, {\vec p}_i)$ can be inverted at least locally. This inversion yields 
the functions 
\begin{equation}\label{invfunct}
y^\mu( {\vec x}_i), ~~~~{\vec p}^j({\vec x}^i)
\end{equation}

\eqref{invfunct} associates a collection of onshell momenta and a bulk point with every collection of boundary points. In order to evaluate the path integral as a function of boundary values, we are 
instructed to 
\begin{itemize} 
\item First evaluate the S matrix with the associated momenta. 
\item Next multiply this S matrix by the complicated `counterterm' 
\begin{equation}\label{fscounter}
\frac{1}{(2 \pi)^{3D/2}\sqrt{\rm det \mathcal{A}}}\prod_{i=1}^{n} \left( \left( \frac{m_i}{ 2 \pi}\right)^{\frac{D-1}{2}}\frac{(-T - t)}{{d^{in}_i}^{\frac{D+1}{2}}} e^{-i m_i d^{in}_i }  \right) \prod_{i=1}^{m} \left( \left( \frac{m_i}{ 2 \pi}\right)^{\frac{D-1}{2}}\frac{(T - t)}{{d^{out}_i}^{\frac{D+1}{2}}} e^{-i m_i d^{out}_i } \right)
\end{equation}
(where $d_i$ is the timelike distance between the $i^{th}$ boundary point at $y^\mu$ and all other quantities are defined below \eqref{corrsaddle}) 
\end{itemize} 
It follows from \eqref{corrsaddle} 
that this procedure correctly reproduces the 
`path integral as a function of boundary values'. 

While the S matrix is simple and beautiful the 
factor listed in \eqref{fscounter} is less elegant. This both oscillates rapidly as well as decays in a power law fashion as $R \to \infty$. In the language of the AdS/CFT correspondence, it is natural to think of the term in \eqref{fscounter} as a factor that should be stripped away to yield the `renormalized' path integral which then equals the S matrix. 
Notice that, unlike in $AdS$ space, the renormalization that removes the factor \eqref{fscounter} is not a product of local factors, one for each insertion. It instead depends on 
all insertion points together. There is an analog of this in momentum space as well. Recall from equation \eqref{smatrix} that in obtaining the S-matrix from the Fourier transform of the boundary correlation function, one must strip off the external wavefunctions. These wavefunctions can be thought of as the momentum space analog of the renormalization factors desribed above, and behave like $e^{i\omega_{\vec{k}}T}$ in the asymptotic limit. Crucially, note that they depend on the momentum, or more precisely on $|\vec{k}|$. In contrast, the renormalization factors in AdS do not depend on the momentum. 

Of course, the 
`renormalization' prescription suggested here 
is simply a definition. Very optimistically one might hope that - in the situation that 
the scattering particles include gravitons - there exists an independent mathematical structure on the boundary of our spacetime that generates the partition function \eqref{corrsaddle}, and that 
the renormalization procedure suggested above is natural within this as yet undetermined structure. 
The optimistic reader may hope that this fantasy 
is borne out in future work. However, the above discussion suggests that this boundary structure will involve non-locality.
 
\section{Path Integral as a functional of boundary insertions in position space: Massless Scattering}  \label{sec:massless}

We now turn to the study of the scattering of massless particles. The `inverse LSZ formula' \eqref{invsmat} (with $m_i$ set to zero) and \eqref{grn} continues to hold, and gives us a clear formula for bulk time ordered Greens functions in terms of the S matrix.  However \eqref{gffaraway}, which was obtained using the saddle point approximation justified by the largeness of $m_i R$, no longer holds. Moreover the arguments of Appendix \ref{apgrn} - that guarantee that $G_{\rm bdry}$ is obtained from the bulk Greens function $G$ by acting on the latter by the differential operator $2 {\hat n}. \nabla$ at each external point - no longer hold when external particles are massless.

Despite these apparent difficulties, in this section, we will 
explain that we can still make many precise statements about the position space bulk Greens function $G$, the position space boundary quantity $G_{\rm bdry}$ and the relationship of these quantities to the S matrix even when the scattering particles are massless. In particular, in this section, we will argue that
\begin{itemize} 
\item The bulk Greens function $G$
 is a meromorphic function of insertion locations $x_i$, whose singularities occur whenever, first,  there exists a bulk point  $y$, which is null separated from a collection of some subset of these $x_i$, and second, when the null rays from these $x_i$ to $y$ are so oriented that they can scatter, in a manner that 
 conserves momentum. 
\item Unlike in the massive case, the above condition has no solutions when the insertion locations of the Greens function $G$ are chosen generically. These equations 
have solutions only on a space of codimension $c \geq 1$ in the space of locations of the insertions. 

\item While for general boundary surfaces the relationship between $G_{\rm bdry}$ and $G$ is complicated, in the neighborhood of the singular sub-manifold, $G_{\rm bdry}$ is
obtained from $G$ by the action of $2 {\hat n}. \nabla$ for 
every external point, as in the previous subsection.
\item The bulk S matrix is encoded in the precise nature of the `bulk point singularities' described above. In the special case that $c=1$, the dependence of the bulk S matrix on angles and ratios of energies of scattering particles is captured by the 
`residue' of the singularity, while the dependence of the S matrix on the overall energy scale is captured by the `degree' of the singularity. When $c\geq 1$ the singularities also encode the $S$ matrix, but in a slightly more involved way (see below for details).
\end{itemize} 

\subsection{Landau Singularities in $G$}\label{singular} 

As we have mentioned above, in the absence of the large parameter $mR$, the saddle point equations of the previous 
paragraph does not apply to the study of massless scattering. 
Despite this fact, we will see in this subsection that the equations similar to the Lorentzian analogues of \eqref{forcecons} continue to have physical significance even in the case of massless scattering. These equations turn out to be the Landau equations that determine the locations of singularities in the bulk correlator $G$.

To see how this works, we  follow the discussion in \S 3 of \cite{maldacena2017looking}  (see also \cite{Chandorkar:2021viw}). To start with let us consider the contribution of 
a tree level contact diagram to the Lorentzian bulk correlator $G(x_1, x_2, \ldots x_n)$.  Such a contribution takes the form 
\begin{equation}\label{treecontact} 
\int d^Dy  \prod_{i=1}^n G(x_i-y) 
\end{equation} 
where $x_i$ are boundary points and $y$ is a bulk point. In $D$ spacetime dimensions, the time ordered two point Greens function is extremely simple; it is given by 
\begin{equation}\label{togf} 
G(x, y) = \frac{1}{\left( (x-y)^2-i\epsilon \right)^\frac{D-2}{2}}
\end{equation} 
The integral \eqref{treecontact} develops a singularity if one or more of the factors in \eqref{togf} go singular in such a way that it is impossible to deform the integral over $y$ (in the complex plane) to move away from  the all the singularities without crossing any one them in the process. 

Let us suppose that our integration contour $y$ passes through a point at which  the first $m$ factors (with $m \leq n$) in \eqref{treecontact} go singular, i.e. that  
\begin{equation}\label{xys}
(x_i-y)^2=0, ~~~{\rm for}~~i=1 \ldots m.
\end{equation}
If it is possible to find a small deformation $i\delta y$ such that 
\begin{equation}\label{defmass}
(x_i-y). \delta y >0 ~~~ i=1\ldots m
\end{equation} 
then it is possible to deform our contour away from the singular point in an allowed manner, and the contribution to the integral from this point is nonsingular. If, on the other hand, it is not possible to find such a deformation, then the integral in the neighborhood of this point
generically yields a singular contribution. 

Let us now suppose that our boundary points are such that 
\begin{equation}\label{momcons} 
\sum_{i=1}^m \omega_i (x_i-y)=0
\end{equation} 
where $\omega_i$ are all positive (or all negative) real numbers. In such a situation it is clearly impossible to find any $\delta y$ that satisfies all the equations \eqref{defmass} (as can be seen by dotting \eqref{momcons} with $\delta y$) \footnote{A very similar argument - in a closely related context -  was given in \S 3 of \cite{maldacena2017looking}}. It follows that when \eqref{momcons}
is obeyed, it is impossible to deform the $y$ contour away from the singular point, and our boundary correlators generically develop a 
singularity in the variables $x_i$. 

Notice that the relationship \eqref{momcons} has a simple interpretation. 
This equation is obeyed whenever it is possible to find a set of 
$m$ spacetime momenta, $k_i$, that are proportional to $(x_i-y)$ (with 
positive proportionality constants) such that the sum of the momenta vanishes. In other words \eqref{momcons} is obeyed whenever it is possible 
to perform a scattering experiment at the location $y$ with the particle sources and detectors located at $x_i$ (the momenta of all particles 
in such an experiment are forced to be proportional to $x_i-y$; recall that all particles participating in this experiment are massless; this follows from the condition that $(x_i-y)^2=0$, the relationship that tells us that the propagator was singular at the point $y$). 

Notice that the conditions we have derived above do not involve the 
remaining positions - i.e. do not involve $x_i$ for $i=m+1 \ldots n$ -  in any manner. These positions can take any value whatsoever. 

Notice that the equation \eqref{momcons} is very similar to the massive 
saddle point equation \eqref{forcecons}. There are, however, two important differences between \eqref{momcons} and \eqref{forcecons}. First, while \eqref{forcecons} necessarily involves all $n$ particles, \eqref{momcons} works for any subset of $m$ particles . The second difference is that the proportionality constants in \eqref{forcecons} are fixed (they are the masses of the scattering particles) while the proportionality constants in \eqref{momcons} are free (though positive) parameters. 
Indeed there is no natural way of choosing one of these proportionality 
constants as special, as all the vectors that appear in 
\eqref{momcons} are null (in contrast, the vectors that appear in the 
Lorentzian version of \eqref{forcecons}, which are all timelike). 

\subsection{Counting and Geometry of singularities} \label{cod}

As we have seen above, the singularities in $n$ point massless correlators 
occur whenever there exists a bulk point 
$y$, and an onshell scattering experiment (involving only massless particles at $y$), s.t. the worldlines of the scattered particles reach the chosen boundary points $x_i$.

The total number of parameters needed to specify both all the momenta and the bulk location of an  $m$ particle scattering is $m(D-1)$ (we have $m(D-1)-D$ parameters in the scattering momenta, and $D$ parameters in the scattering location). This is the same as the total number of parameters we need to label $m$ boundary points. This part of the counting is identical to the massive case. 

In the case of massive scattering, the agreement between these two different numbers tells us that the boundary correlator - for every choice of boundary locations - is determined by a particular S matrix (see \eqref{corrsaddle}). This conclusion followed because two distinct scattering processes (processes with either distinct momenta or with distinct bulk scattering locations or both) always `ended up' at distinct boundary locations.  The last conclusion, in turn, followed because massive particles with distinct momenta travel at different speeds (and so at different spacetime `angles').

In the case of massless scattering, however, a particle with momentum $k_\mu$ travels in exactly the same direction as a particle with momentum $\alpha k_\mu$. Consider two scattering processes that take place at the same bulk point. The first process has momenta $(k^i_\mu)$. The second has momenta $\alpha_i k_\mu^i$. Both these processes  `end up' at exactly the same boundary points. Conservation of momentum for the first scattering process tells us that  
\begin{equation} \label{Pparor}
\sum_i  k_i=0
\end{equation} 
whereas the same requirement for the second scattering process implies
\begin{equation} \label{Ppar}
\sum_{i=1}^m \alpha_i k_i=0
\end{equation} 

We now wish to address the following question: given that \eqref{Pparor} holds, what is the dimensionality of the vector space of coefficients $\alpha_i$ for which \eqref{Ppar} holds? The answer to this question is easily found. Note that \eqref{Ppar} is a set of $D$ homogeneous linear equations for $m$ variables. The total number of parameters in the general solution to this equation equals 
\begin{equation}\label{Pform}\begin{split} 
c= (m-1)-D +1 = m-D ~~~&{\rm if ~~~~ m-1>D} \\
c=1  ~~~~~~~&{\rm if ~~~~ m-1 \leq D}\\
\end{split} 
\end{equation} 
(we always have atleast one parameter because all $\alpha_i$ equal to the same arbitrary number is always a solution).
\footnote{One careful way to argue for \eqref{Pform} goes as follows. We can use \eqref{Pparor} to solve for $k_m$ in terms of $k_i$ for $i < m$. Inserting this solution into \eqref{Ppar} turns it into
\begin{equation} \label{Pparn}
\sum_{i=1}^{m-1} \alpha_i' k_i=0
\end{equation} 
where $\alpha_i'= \alpha_i-\alpha_m$. The momenta that appear in \eqref{Pparn} are now unconstrained. \eqref{Pparn} may thus be thought of as $D$ homogeneous equations for the $m-1$ variables $\alpha'_i$. Consequently, we have only the unique solution $\alpha_i'=0$ when $m-1 \leq D$, but have an $m-1-D$ dimensional space of solutions for $\alpha_i'$ when $m-1 >D$, i.e. when $m > D+1$. Now given any solution for $\alpha_i'$, we can generate an extra one parameter set of solutions for $\alpha_i$ by setting $\alpha_i=\alpha_i'+ \theta$, and $\alpha_m=\theta$ for any $\theta$. Note, in particular, that this symmetry can be used to make all $\alpha_i$ positive, as is required for 
the pinch (existence of a Landau Singularity).
}

We have just argued that a $c$ parameter set of distinct scatterings, occurring at any particular bulk point, all end up at the 
same configuration of boundary points. We now ask a related but distinct  question. Given particular fixed locations of boundary points, do they receive (Landau singular) contributions from one (or a discrete collection) of bulk points, or from a manifold of bulk points with some dimensionality? The answer to this question can also be obtained from a simple counting argument, as follows. The lightcones of each of the $m$ insertion points are spacetime surfaces of codimension one. The nature of the intersection manifold of all of these lightcones is investigated in Appendix \ref{lightconeapp}. The situation is qualitatively different depending on whether $c=1$ or $c>1$. 
In the case $c=1$, and assuming that the scattering hypersurface - the surface spanned by the vectors between insertion points - is an $R^{m-2,1}$ (see the Appendix for details), we demonstrate in the Appendix that there are three possibilities for this intersection manifold, depending on the details of the insertion points. For one open set of insertion points, the intersection manifold 
is simply empty. For another open set of insertion points, 
the intersection manifold is a spatial sphere of dimension 
$D-m$, but in this case momentum conservation can never be satisfied. On boundary between these two cases (i.e. as the sphere described above shrinks to zero), the intersection manifold is a single point that lies on the 
scattering hypersurface. This is the only case in which momentum conservation can be satisfied. Note that reaching this case involves a one parameter tuning. 

When, on the other hand, $c>1$, the intersection of lightcones is generically empty. Obtaining such an intersection requires an $m-D$ parameter tuning (this is intuitive from the fact that $m$ codimension one manifolds only have a common intersection in a $D$ dimensional space once we perform $m-D$ tunings). When such an intersection point exists, it is generically unique (see Appendix \ref{lightconeapp}). When the lightcones intersect 
there is no barrier to finding momentum 
conserving configurations, because the the demand that \eqref{Pparor} is obeyed constitutes $D$ conditions on $m$ variables, and $m>D$.

Let us summarize. We have argued that the 
quantity $c$, defined in \eqref{Pform}, is significant in two ways. 
 \begin{enumerate}
\item First, it computes the codimension
of the singular submanifold in the $m(D-1)$ dimensional space of all possible boundary 
points. Any point on this singular manifold receives contributions from scattering 
occurring at a particular bulk point. 
\item Second, consider scattering that occurs at a particular bulk point. 
The total number of parameters in the space of scattering momenta that are both onshell and conserve momentum is $m(D-1)-D$. This space is foliated into an $m(D-1)-D-c$ parameter set of  `leaves' of dimensionality $c$, such that all momenta 
that lie in the same leaf end up at the same boundary points, while momenta that lie in distinct leaves end up in different boundary points. 
\item The relationship between points [1] and [2] is the following. 
We have seen from point $\mathcal{Q}$2 that scattering at any given bulk location reaches a $m(D-1)-D-c$ set of boundary points. Since the map between bulk and boundary is 
one to one, upon varying over the full 
$D$ parameter set of bulk points, we end up reaching a $m(D-1)-c$ parameter set of boundary points, or codimension $c$ submanifold of the boundary. But this is 
precisely the scattering submanifold of the boundary, as asserted in point [1].
\end{enumerate} 

Let us re-emphasize that every point on  the scattering manifold - which is of codimension $c$ - receives contributions from a $c$ parameter set of S matrices (all scatterings take place at exactly one bulk point).

\subsection{Condition for singularities in terms of the matrix of squared distances}\label{Nmatrix}

Consider a bulk $n$ point function at insertion points 
$x_i^\mu$ with $i=1 \ldots n$. 
Let us define the symmetric matrix $M$ 
so that 
\begin{equation}\label{matelem}
M_{ij}= (x_i-x_j)^2
\end{equation}
By definition, the matrix $M$ (which we refer to as the matrix of squared distances)  is Lorentz invariant. In this subsection, we will characterize the singular surface for correlators in terms of the matrix $M$.

As we have seen in the previous subsection, our correlation function is singular whenever \eqref{xys} and 
\eqref{momcons} are satisfied. 
Let us suppose \eqref{momcons} is obeyed, for some $m$ of the insertion points. Let us now consider the $m \times m$ restriction of the matrix $M$, 
with $i$ and $j$ restricted to these special insertion points. 
corresponding (in an obvious manner) to these selected insertions. Let us call this $m \times m$ symmetric matrix $N$. Let us assume that there exists a bulk point $y$ that obeys \eqref{xys}. It follows that
\begin{equation}\label{nmc}
N_{ij}= (x_i-x_j)^2= 
\left( (x_i-y)-(x_j-y) \right)^2 =2(x_i-y). (x_j-y)
\end{equation} 
where we have used that 
$(x_i-y)^2=0$ for all $i$ (see \eqref{xys}). Multiplying \eqref{nmc} with $\omega_i$, and summing over $i$ and using \eqref{momcons}, we conclude that for each $j=1$ to $m$,
\begin{equation} \label{zeroeig}
\sum_{i=1}^m \omega_i N_{ij} = 0
 \end{equation} 
In other words, it follows that 
at least one of the eigenvalues of $N$
is zero, and all entries of the corresponding eigenvector are of the same sign. In particular, it follows that the determinant of $N$ vanishes.

We have found that our correlation functions are singular whenever $N$ has a zero eigenvalue with all entries of the corresponding eigenvector being of the same sign. In fact, the converse of this statement is also true.  Whenever \eqref{zeroeig} holds either 
$\sum_i \omega_i=0$ \footnote{As we see in \S \ref{cgr1}, for $m>D+1$ the matrix $N$ generically has such \emph{non-scattering} zero eigenvalues with $\sum_i \omega_i=0$} or, as we show in Appendix F, there exists a bulk scattering point $y$ such that bulk scattering conserves momentum (i.e. \eqref{momcons} holds). In particular, it follows that whenever \eqref{zeroeig} holds with all $\omega_i$ of the same sign (so that $\sum_i \omega_i \neq 0)$), bulk scattering happens, leading to a singularity in the correlator.


\subsection{Properties of the matrix $N$}\label{nprop}
We now elaborate on the properties of the matrix $N$. First, let us characterize the nature of its zero eigenvalues. We observe that the eigenvalue equation \eqref{zeroeig} is satisfied differently depending on whether $\sum_i \omega_i=0$ or $\sum_i \omega_i\neq 0$. To see this, consider the difference of two of the equations in \eqref{zeroeig},

\begin{equation} \label{diffeig}
\begin{split}
    \sum_i \omega_i (N_{ij}-N_{ik})&=0 \\
    \implies 2 \sum_i \omega_i x_i \cdot (x_j-x_k)&=(x_j^2-x_k^2)\sum_i \omega_i
\end{split}
\end{equation}
For a generic set of insertions, the vectors $(x_j-x_k)$ span a $(m-1)$ dimensional space. Eq. \eqref{diffeig} provides the components of the $d$-dimensional vector $\sum_i \omega_i x_i^{\mu}$ in this basis. If $m>D+1$, then imposing \eqref{diffeig} completely fixes the vector $\sum_i \omega_i x_i^{\mu}$, and if $m\leq D+1$, then the vector is only fixed partially, i.e. its components are fixed along those directions which span the space formed by the insertions. For $m>D+1$, we will see in \ref{cgr1}, that the matrix $N$ generically has zero eigenvalues with $\sum_i \omega_i=0$ which do not correspond to scattering. If $\sum_i \omega_i\neq 0$, then the insertions are in a scattering configuration, as shown in Appendix \ref{reverse}. In this case, we notice that \eqref{diffeig} pits the condition for finding a bulk point null separated from all insertions (the RHS of the equation) versus the condition that momenta can be conserved at a bulk point (the LHS of the equation). In other words, if the former condition is satisfied then \eqref{diffeig} imposes the latter, and if the latter condition is satisfied, then it imposes the former. So the number of conditions on the set of insertions that are needed to satisfy \eqref{diffeig} is the minimum number of tunings needed to satisfy either of these two conditions separately. This minimum number depends on whether $m\leq D+1$ or $m>D+1$. Subsequently, the matrix $N_{ij}$ has significantly distinct properties depending on whether $c=1$ or $c>1$ (i.e. whether $m\leq D+1$ or $m>D+1$). We study these two cases separately. 

\subsubsection{$c=1$ i.e. $m\leq D+1$}

In this case, the matrix $N$ is built out of the positions of the $m$ insertions modulo Lorentz transformations. The number of variables that specify 
locations of $m$ insertions modulo Lorentz transformations equals 
$$0+1+2 \ldots m-1= \frac{m(m-1)}{2}$$
which is identical to the number of independent elements 
of the matrix $N$. As we vary over all possible insertion locations, our $N$ matrices thus fill out an open set in the space of symmetric matrices with zero diagonal elements. It follows that our matrix $N$ generically has nonzero determinant.  

We have argued above that the Greens functions develop singularities on a submanifold in the space of external boundary insertions. These singularities occur whenever 
there exists a bulk point null separated from all insertions s.t. \eqref{momcons} is obeyed. As we have argued above, this implies \eqref{zeroeig}, and so 
implies that  $det N=0$ \footnote{Whenever any two insertions are coincident, two rows (and two columns) of $N$ become the same, and so $det N$ trivially vanishes. We are not interested in these coincident singularities, hence we will assume that all our insertions are separated from each other.}. In this case, consequently, we have a simple algebraic characterization of the singular manifold. It is an open subset
\footnote{The singular manifold is an open subset of this manifold - rather than the full manifold itself - because vanishing of the determinant does not guarantee the positivity of the entries of the eigenvector.}
on the codimension one submanifold in the space of insertions: the submanifold in question is given by the equation $det N=0$

Note that each matrix element of $N$ - and therefore the matrix $N$ - is invariant under Poincare and Lorentz transformations. Moreover, every element of $N_{ij}$ also has weight two under scalings.  In Appendix \ref{cis} we demonstrate that the individual matrix elements of $N_{ij}$  also transform 
homogeneously under special conformal transformations. As a consequence, the condition that 
$det N$ vanishes is conformally invariant, hence can only depend on conformal crossratios.  

\subsubsection{$c>1$ i.e. $m>D+1$} \label{cgr1}

In this case the total number of variables that specify the location of $m$ particles upto Lorentz transformations equals 
$$ \frac{(D+1)D}{2} + (m-D-1)D= \frac{D(2m -D-1)}{2}. $$
This number is smaller than $m(m-1)/2$, the total number 
of independent matrix elements in the matrix $N$. 
For this reason, the matrix $N$ is not generic in this case. In particular, we will now demonstrate that the rank of the matrix $N$ is always less or equal to the minimum of $m$ and $D+2$. As $m>D+1$, this minimum is 
simply $D+2$. When $m=D+2$ we expect the determinant of $N$ to be generically nonzero (as in the case $c=1$). For every larger value of $m$, $det N$ vanishes (and, infact, the matrix $N$ always has atleast $m-D-2$ zero eigenvalues). 

{\it The rank of $N$}

The rank of $N$ may be estimated as follows. We start with the following decomposition of the $m\times m$ matrix $N$,
\begin{equation} \label{matdecomp}
\begin{split}
     N &=\begin{pmatrix}
        x_1^2 & x_1^2 & \ldots  & x_1^2\\
        x_2^2 & x_2^2 & \ldots  & x_2^2\\
        \ldots & \ldots & \ldots & \ldots\\
        x_m^2 & x_m^2 & \ldots  & x_m^2\\
    \end{pmatrix} 
    +\begin{pmatrix}
        x_1^2 & x_2^2 & \ldots  & x_m^2\\
        x_1^2 & x_2^2 & \ldots  & x_m^2\\
        \ldots & \ldots & \ldots & \ldots\\
        x_1^2 & x_2^2 & \ldots  & x_m^2\\
    \end{pmatrix} 
    -2 \begin{pmatrix}
        x_1^2 & x_1 \cdot x_2 & \ldots  & x_1 \cdot x_m\\
        x_1 \cdot x_2 & x_2^2 & \ldots  & x_2 \cdot x_m\\\
        \ldots & \ldots & \ldots & \ldots\\
        x_1 \cdot x_m & x_2 \cdot x_m & \ldots  & x_m^2\\
    \end{pmatrix} \\
    &=1^T \cdot Y +Y^T \cdot 1 -2 X^T \cdot \eta \cdot X
\end{split}
\end{equation}
where $1$ is the coloumn vector of ones, $Y$ is a column vector of size $m$ with $i^{th}$ entry $x_i^2$, $\eta$ is the Minkowski metric, and $X$ is a $D \times m$ matrix whose $ij^{th}$ entry is given by $x_j^{(i)}$, the $i^{th}$ coordinate of the insertion point $x_j$,
\begin{equation}
    X= \begin{pmatrix}
        x_1^{(1)} & x_2^{(1)}  & \ldots  &  x_m^{(1)}\\
       x_1^{(2)} & x_2^{(2)}  & \ldots  &  x_m^{(2)}\\ 
        \ldots & \ldots & \ldots & \ldots\\
        x_1^{(D)} & x_2^{(D)}  & \ldots  &  x_m^{(D)}\\
    \end{pmatrix} 
\end{equation} 
We now use the fact that the rank of a sum of matrices is upper bounded by the sum of ranks of the individual matrices. The rank of the first two factors in the decomposition \eqref{matdecomp} is one and the rank of the last factor is $min(m,D)$. Hence, given that $N$ is a $m\times m$ matrix, we find that the rank of $N$ is $min(m,D+2)$. Thus when $m>D+2$, generically $N$ has $m-D-2=c-2$ zero eigenvalues, where $c$ is the codimension of singularity given in \eqref{Pform}.

{\it Nature of the zero eigenvalues}

We will now explore the nature of the generic zero eigenvalues of $N$, and demonstrate that these eigenvalues never lead to singularities of boundary correlators. We will also point out that a $c$ parameter 
tuning of the matrix $N$ leads it to develop two 
additional zero eigenvalues (over and above the $c-2$ generic ones), giving rise to scattering type singularities in correlators. 

The equation \eqref{zeroeig} can be expanded as
\begin{equation} \label{zev}
    \sum_j \omega_j N_{ij} = \left(\sum_j \omega_j\right) x_i^2 + \sum_j x_j^2 \omega_j - 2 x_{i}. \sum_j x_{j} \omega_j = 0
\end{equation}
If $\sum_j \omega_j \neq 0$, then \eqref{zev}
can only be satisfied for a generic set of insertions if each of the  terms in that equation vanishes individually. This is because, satisfying \eqref{diffeig} for $m>D+1$ requires every component of $\sum_j x_{j}^{\mu} \omega_j$ along each of the basis vectors $(x_i-x_k)$ be zero, which means that $\sum_j x_{j}^{\mu} \omega_j=0$. 
From \eqref{zev}, we then have that
\begin{equation}\label{zev1}
    \sum_j \omega_j = 0 , \qquad  \sum_j x_j^2 \omega_j =0 , \qquad \sum_j x_{j}^{\mu} \omega_j = 0.
\end{equation}
\eqref{zev1} constitutes $D+2$ conditions on $m$ entries of the eigenvectors $\omega$. Hence generically the above system has $m-D-2$ subspace of solutions. We have thus found the $m-D-2$ dimensional space of generic zero eigenvalues anticipated above. 

Note that the first conditions in \eqref{zev1} require the sum of the entries of eigenvector to vanish. Clearly, this condition cannot be satisfied when $\omega_j >0$ for all $j$. It follows that these generic zero eigenvectors of the matrix $N_{ij}$ do not lead to singularities in the correlation function, as might have been anticipated on physical grounds. 

We have already demonstrated in Appendix \ref{reverse} that if $\sum_i \omega_i \neq 0$, then \eqref{zev} implies that there exists a bulk point $y$ such that all insertions are null separated from $y$.  In other words, if the lightcones emerging from $m$ boundary points can intersect at a bulk point ($y$) then \eqref{zev} can be satisfied. As explained in \S \ref{cod}, for $m> D$, generically the lightcones do not intersect and we need $m-D$ tunings to find such an intersection. Once such tunings are performed, one can always find momentum conserving configurations \footnote{Infact the space of 
solutions of this constraint is a $m-D$ dimensional vector space, as remarked in subsection \ref{cod}.}. As 
the condition $\sum_i \omega_i \neq 0$ in this case, it is possible for all $\omega_i$ to be positive. 
Consequently, the codimension $m-D$ surface in the space of all boundary insertions, that obey \eqref{zev} with positive $\omega_i$ can 
(and do) give rise to scattering type singularities in correlators.

\subsection{The Geometry of scattering manifolds for $c>1$} \label{geom}

We have seen above that the scattering of $m$ particles in $D$ dimensions happens on a scattering manifold of 
codimension $c=m-D$ when $m>D+1$. In this section we will 
explain that this scattering manifold can be thought of as the intersection of the scattering of 
$m-D$ scattering subsheets, each of these sheets relating to the $c=1$ scattering of $D+1$ particles in $D$ dimensions. The sheets in question may be chosen to be permutations of scattering of particles 
\begin{equation}\label{scatsheetbuild}
(1,2,3,...D+1), ~~~(2, 3,4 ... D+2) \ldots (m-D, m-D+2 \ldots m)
\end{equation}
Each of the scattering sheets listed in \eqref{scatsheetbuild} is of codimension one, and so the 
intersection of these $m-D$ sheets yields the codimension $m-D$ scattering sheet under study. 

In order to avoid notational clutter, we explain the point described above in the context of a simple example, with the hope that the generalization will be clear to the reader.

Consider, as an example, the scattering of 
5 particles in 3 dimensions. In this case, 
$c$ defined in \eqref{Pform} equals 2. This 
means that, the singular manifold is codimension two and that, on this manifold, the $N$ matrix has two linearly independent zero eigenvectors. Away from the singular manifold, the determinant of $N$ is nonzero. 

Let us work with a generic point on the singular manifold (i.e. with a set of boundary insertions that are tuned to receive contributions from physical 5 particle scattering in the bulk, but are not further tuned in any way). As we have argued above, such configurations receive contributions from a two parameter set of 
physical scatterings. It follows that 
it is possible to choose a basis  for the 
two dimensional space of zero eigenvalue eigenvectors that take the form 
\begin{equation} \label{zeroform} 
\omega^{(a)}_i, ~~~ a=1\ldots 2, ~~~i=1 \ldots 5  ~~~~ {\rm s.t.} ~~~ \omega^{(a)}_i >0 \
\forall ~i ~{\rm and}~ \forall ~a
\end{equation} 
(here $a$ is a `which eigenvector' label
while $i$ is a `which row in the eigenvector' label). 

Of course the most general eigenvector 
in the zero eigenspace takes the form 
\begin{equation}\label{mostgenev} 
\sum_a \zeta_a \omega^{(a)}
\end{equation} 
Each eigenvector of the form \eqref{mostgenev} that obeys 
\begin{equation} \label{chambers} 
\left( \sum_a \zeta_a \omega^{(a)}_i \right) >0 ~~~\forall~ i 
\end{equation} 
corresponds to a different physical scattering configuration. 

\eqref{chambers} is clearly obeyed when
$\zeta_1$ and $\zeta_2$ are both positive. 
However, it will also generically  be obeyed when one of these variables (say $\zeta_1$) turns negative provided $|\zeta_1|$ is small enough. More precisely, this condition is obeyed provided 
\begin{equation} \label{condonzeta}
\frac{ |\zeta_1|}{\zeta_2} 
<{\rm min}_{i} ~\frac{ \omega^{(2)}_i}{\omega^{(1)}_i } 
\end{equation}
When the inequality in \eqref{condonzeta}
is saturated, something special happens. 
In this case the five particle scattering, 
degenerates: the momentum of one of the particles (the particle for which 
$\frac{ \omega^{(2)}_i}{\omega^{(1)}_i }$ is minimum, let us call this $\mu_1$) goes to zero (i.e. this particle becomes soft). At precisely the degeneration point this particle has zero momentum, and hence drops out of scattering kinematics. At this point scattering kinematics is precisely that of the scattering of four particles (all particles except $\mu_1$). It follows that this point on the 5 particle scattering manifold also lies on the four particle scattering manifold obtained by excluding $\mu_1$. 

In a manner similar to the discussion above, we can now keep $\zeta_1$ positive, but 
take $\zeta_2$ negative. In this case we obtain a legal five particle scattering process when 
\begin{equation} \label{condonzetan}
\frac{ |\zeta_2|}{\zeta_1} 
<{\rm min}_{i} ~\frac{ \omega^{(1)}_i}{\omega^{(2)}_i } 
\end{equation}
Let us call the $i$ for which 
$\frac{ \omega^{(1)}_\mu}{\omega^{(2)}_\mu }$ is minimum $\mu_2$. As in the paragraph 
above, the particle $\mu_2$ becomes soft 
when \eqref{condonzetan} is near to being saturated. When the inequality is saturated, the kinematics of the 5 particle scattering process reduces to that of the four particle scattering obtained  by removing the particle $\mu_2$. Once again we conclude that this point on the five particle scattering manifold also lies on this 
four particle scattering manifold. 

The analysis above applies for any generic point on the five particle scattering surface, for some choice of $\mu_1$ and $\mu_2$. It follows that, generically, every point on the five particle scattering surface lies on the intersection of exactly two four particle scattering planes. We see that generic points on the five particle scattering surface can be obtained as the intersection of exactly two distinct four particle scattering surfaces. This point of view explains why the five particle scattering sheet is codimension two in the space of boundary configurations: it is obtained from the intersection of two codimension one surfaces. 

Now there are 5 distinct four particle scattering sheets (those obtained by eliminating each of the five scattering particles in turn). The five particle 
scattering sheet may be decomposed into 
regions or cells. The cell $C_{\mu\nu}$ 
is the part of the five particle scattering sheet which is obtained as the intersection of the four particle scattering sheet omitting the particle $\mu$ \footnote{The insertion of the particle  $\mu$ can then lie anywhere.That is why this configuration is codimension one.} , and the four particle scattering sheet omitting the particle $\nu$ \footnote{Again, the particle $\nu$ can lie, anywhere, yielding a codimension one configuration}. The cell $C_{\mu \nu }$ borders the cell $C_{\mu \rho}$ $(\nu \neq \rho)$. The boundary between these cells is codimension one on the five particle scattering sheet, or codimension 3 in the space of all external configurations. This boundary is obtained as the common intersection of  the four particle scattering sheet excluding $\mu$, the four particle scattering sheet excluding $\nu$ and the four particle scattering sheet excluding $\rho$. The fact that it is the intersection of three codimension one surfaces explains why this boundary is codimension $3$ in the space of configurations. 

In the case under study, the matrix $N$ has no zero eigenvalues as generic values of the locations of boundary insertions. When any of the four particles (lets say particles 1, 2, 3, 4 ) lie on the four particle scattering manifold, then our matrix $N$ has a zero eigenvector for
any value of the position of the 5th particle (the eigenvector in question has a zero entry in the fifth column). This condition is met on a codimension one surface in the space of boundary insertions of five particles. Similarly, when another set of four particles, 
lets say $(2, 3, 4, 5)$, lie on their scattering manifold
then we have a zero eigenvector for any value of the boundary location of the first particle (the eigenvector 
in question will have zero entry in the first column). 
This condition is also met on a codimension one surface in the space of boundary insertions of 5 particles. When 
both these (or two similar such) conditions are met 
then the zero eigenspace is two dimensional. At this point
we can take linear combinations of the two zero eigenvectors above, to obtain an eigenvector whose column has no zero values. Such an eigenvector describes genuine 5 particle scattering. Clearly this condition happens on a space that is codimension two in the space of 5 boundary insertions.

In the special example studied above a configuration whose $N$ matrix has a zero eigenvalue, whose eigenvector components obey  
\begin{equation}\label{eveo}
\omega_i >0, ~~~\forall ~i
\end{equation}
corresponds to genuine five particle scattering. Notice, however, that the slightly modified condition
\begin{equation}\label{eveom}
\omega_i \geq 0, ~~~\forall ~i
\end{equation}
is much less restrictive. This condition is met even on the four particle scattering manifold (we simply set 
$\omega$ to zero for the fifth particle). Consequently 
\eqref{eveom} is met on a codimension one manifold in the space of boundary locations. As explained above, the intersection of two of these submanifolds 
yields an $N$ matrix with a two dimensional space of zero eigenvalues. This codimension two location corresponds to genuine 5 particle scattering. 

While we have phrased the discussion of this subsection in terms of a simple example, the generalization of the beautiful interlinked geometry of scattering sheets to  higher number of external insertions, $m$ and higher dimensions, $D$ is hopefully clear. 

\subsection{Examples of Singularities in $G$}\label{egsing}

In subsection \ref{cod}, we saw that the co-dimension of Landau singularity for an $m$ point Greens function can either be one or $m-D$ depending on the spacetime dimension $D$. We will discuss the two cases, namely, $c=1$ and $c > 1$, separately. 

\subsubsection{Co-dimension one singularity $(c=1)$} \label{ceq1}
As an example of co-dimension one singularity, we consider tree level four-point Greens function for massless scalars in $D \geq 3$, and assume that 
the scalars in question interact with the contact 
Lagrangian 
\begin{equation}\label{contactlag} 
\frac{g}{2^{n+m}} \int d^D x \sum a_{m, n} \left( \partial_{\mu_1} \ldots 
\partial_{\mu_m} \partial_{\nu_1} \ldots 
\partial_{\nu_n} \phi \right) 
\left( \partial^{\mu_1}\ldots \partial^{\mu_m} \phi \right)  \left( \partial^{\nu_1} \ldots \partial^{\nu_n} \phi\right) 
\phi 
\end{equation} 
Let us define the function 
\begin{equation}\label{hub}
T(a, b) = g \sum_{m, n} a_{m, n}  a^m b^n
\end{equation} 
The interaction \eqref{contactlag} has been chosen to ensure that the tree level S matrix that follows from this Lagrangian equals 
$$ T(s, t)+ T(s, u) + T(t, u).$$

The bulk Greens function is given by:
\beq\label{bsg} \begin{split} 
G =  g  &\int ~ d^D y 
\sum a_{m, n} \left( \partial_{\mu_1} \ldots 
\partial_{\mu_m} \partial_{\nu_1} \ldots 
\partial_{\nu_n} D(x_1, y) \right) 
\left( \partial^{\mu_1}\ldots \partial^{\mu_m} 
D(x_2, y)\right)  \left( \partial^{\nu_1} \ldots \partial^{\nu_n} D(x_3, y)\right) 
D(x_4, y) \\
&+ (2 \leftrightarrow 3) + (2 \leftrightarrow 4)
\end{split} 
\eeq
where $y$ is the bulk point and $x_i$ are the four boundary points parameterized above, the 
derivatives $\partial_\mu$ all act on $y$ while the derivatives $\nabla$ act on the various $x_i$.

The bulk Greens function in $D \geq 3$ is given by:
\beq \label{bulkgr}
D(x,y) =\frac{\Gamma \left(\frac{D-2}{2}\right)}{4 \pi^{\frac{D}{2}}} \frac{1}{((x-y)^2)^{\frac{D-2}{2}}- I \epsilon} 
\eeq
The bulk Greens function $G$ can be represented as follows using Schwinger parametrization.
\beq\label{sg}
D(x,y) = \frac{e^{-\frac{ i \pi (D-2)}{4} }}{4 \pi^{\frac{D}{2}}} \int_0^\infty ~ d \omega ~ \omega^{\frac{D-2}{2}-1}e^{i\omega((x-y)^2 -i \epsilon)} 
\eeq
One advantage of this representation is that 
it makes the taking of derivatives very simple.

Substituting \eqref{sg} in \eqref{bsg}, we obtain:
\beq\label{bsg1}\begin{split} 
G &=   A \int ~ d^{D}y~ \left(\prod_{i = 1}^{4} d\omega_i~ \omega_i^{\frac{D-2}{2}-1}\right) e^{i\sum_i\omega_i((x_i-y)^2-i \epsilon)}~
\left( T(s, t)+ T(t, u)+ T(u, s) \right) \\
& s= 2 \omega_1 \omega_2 (x_1-y).(x_2-y), ~~~~t= 2 \omega_1 \omega_3 (x_1-y).(x_3-y)~~~~u= 2\omega_1 \omega_4 (x_1-y).(x_4-y) \\
& \rm where\\
&\qquad A= \frac{e^{-i \pi (D-2) }}{4^4\pi^{2D}}\\
\end{split} 
\eeq

Recall that our the bulk correlator develops a singularity whenever one or more eigenvalue of the matrix $N_{ij}$ vanishes. In this subsection we specialize to situations in which no more than one eigenvalue of the matrix $N_{ij}$ vanish simultaneously. We name this small (and 
potentially vanishing) eigenvalue $\lambda_0$. Our 
correlator is singular when $\lambda_0$ vanishes; as this is one condition, it happens on a codimension one submanifold in the manifold of all external insertions. We will refer to this submanifold as the singular submanifold, 
because we expect our correlator to develop a singularity 
on this submanifold. In this subsection we determine the precise form of this singularity, and also determine a precise formula for the `residue' of this singularity
in terms of an appropriate S matrix. 

Let us suppose that our external insertions are near (but not precisely on) the singular manifold, so that $\lambda_0$ is small compared to all other eigenvalues $\lambda_i$, which, in turn, are each of order $R^2$. 
When this happens, our insertion points $\{ x_i \}$ are necessarily located near some point ${\tilde x}_i$  on the singular manifold. Of course the precise choice of ${\tilde x}_i$ can be made in many ways; this ambiguity will not affect the final answer for
our analysis, and so can be resolved in any convenient manner. \footnote{ One concrete way of establishing a map between points near the singular manifold and points on the singular manifold is to use a congruence of geodesics that intersects the singular manifold. Concretely, let $\alpha(\{x_i \})$ denote the lowest (in modulus) eigenvalue of the matrix $N_{ij}$. The $n(D-1)$ dimensional space of external insertions can be foliated into codimension one surfaces of 
constant $\alpha$. The one form field $n_{i\mu}(\{x_i \})=\partial_{x_i^\mu} \alpha $ is `normal' to this foliation in the following sense: a small deviation that $\delta x_i^\mu$ that obeys the equation $\delta x. n=0$ describes a deformation along a foliation direction (i.e. 
along a direction that keeps the value of $\alpha$
fixed). We may (arbitrarily) define the metric 
\begin{equation}\label{metx}
ds^2=\sum_i \eta_{\mu\nu}\delta x_i^\mu \delta x_i ^\nu
\end{equation} 
on the space of external insertions. With all these definitions, we can associate a point $\{x_i^\mu \}$ in the space of insertions, to the  singular point ${\tilde x}$ on the manifold, if a  geodesic shot normal from ${\tilde x}$ intersects the point $\{x_i^\mu \}$.} 

Insertions at the point ${\tilde x}$ yield a singular  correlator precisely because, at these values of external insertions, there exists a bulk point ${\tilde y}$ such that \eqref{momcons} is satisfied for some values of ${\tilde \omega}_i$. Note that the overall scale of ${\tilde \omega}_i$ is ambiguous. 
Let us now return to the formula \eqref{bsg1}. 
Motivated by the discussion above, we make the 
change of variables $y \rightarrow y+{\tilde y}$. The exponent in the resulting integral (ignoring the $i \epsilon$ for now) becomes 

\begin{equation}
\begin{split}
&i\sum_i\omega_i(x_i-y)^2=i \left(\sum_i \omega_i \right) \left(  y -
\frac{\sum_{i}( \omega_i (x_i-{\tilde y}) )}{\sum_i \omega_i} \right)^2 
- i\frac{ \left( \sum_i \omega_i(x_i-{\tilde y}) \right)^2 }{
\sum_i \omega_i}+ i\sum_i \omega_i (x_i-\tilde{y})^2
\end{split}
\end{equation}
The last two terms can be massaged to see that they depend only on ${x_i}$ as
\begin{equation} \label{expN}
\begin{split}
- i\frac{ \left( \sum_i \omega_i(x_i-{\tilde y}) \right)^2 }{
\sum_i \omega_i}+ i\sum_i \omega_i (x_i-\tilde{y})^2 &=
i \frac{1}{\sum_i \omega_i} \sum_{\substack{i,j \\ i \le j}} \omega_i \omega_j (x_i-x_j)^2\\
&=i \frac{1}{\sum_i \omega_i} \sum_{\substack{i,j \\ i \le j}} \omega_i \omega_j N_{ij}
\end{split}
\end{equation}
It follows that the the exponent in \eqref{bsg1} is given by 
\begin{equation}\label{expbsgo}
i \left(\sum_i \omega_i \right) \left(  y -
\frac{\sum_{i}( \omega_i (x_i-{\tilde y}) )}{\sum_i \omega_i} \right)^2  + i \frac{1}{2\sum_i \omega_i} \sum\limits_{i,j} \omega_i \omega_j N_{ij}
\end{equation} 
where, in the last line, we have dropped the limit $i \leq j$ in the sum $\omega_i \omega_j N_{ij}$, and compensated this by including a factor of $1/2$. We need to perform the integral over $y$ and $\omega_i$. 
Let us first study the integral over $y$. The exponent 
in \eqref{expbsgo} is a Gaussian with centre $\frac{\sum_{i}( \omega_i (x_i-{\tilde y}) )}{\sum_i \omega_i}$ and width $\propto \frac{1}{\sqrt{\sum_i \omega_i}}$. We will see below (once we do the integral over $\omega_i$) that the integral is dominated by values of $\omega_i$ such that the center of the Gaussian 
is of order $\frac{\sqrt{\lambda_0}}{R}$ and the width of the Gaussian is $\sqrt{\lambda_0}$. The important point is that  both the center and the width are parametrically small as $\lambda_0 \rightarrow 0$. It follows that the coefficient of the leading singularity as $\lambda_0 \rightarrow 0$ may accurately be captured by 
setting $y'=0$ (i.e. by replacing $y$ by $\tilde{y}$) (and also replacing $x_i$ by $\tilde{x_i}$) in the expressions of Mandlestam variables appearing in $T$ matrix (the polynomial prefactors of the Gaussian integral in $y$). After we make these replacements, the Gaussian integral over  $y$ can be performed, leaving us with 
\beq\label{bsg2}\begin{split} 
G &=   A (i\pi)^{D/2}\int ~ \left(\prod_{i = 1}^{4} d\omega_i~ \frac{\omega_i^{\frac{D-2}{2}-1}}{{\sum_i \omega_i}^{D/2}}\right) e^{i\frac{1}{2\sum_i \omega_i} \sum \omega_i \omega_j N_{ij}} ~
\left( T(s, t)+ T(t, u)+ T(u, s) \right) \\
& s= 2 \omega_1 \omega_2 (\tilde{x}_1-\tilde{y}).(\tilde{x}_2-\tilde{y}), ~~~~t= 2 \omega_1 \omega_3 (\tilde{x}_1-\tilde{y}).(\tilde{x}_3-\tilde{y})~~~~u= 2\omega_1 \omega_4 (\tilde{x}_1-\tilde{y}).(\tilde{x}_4-\tilde{y})\\
\end{split} 
\eeq

Now, we turn to the $\omega_i$ integrals. Let the matrix $N_{ij}$ have eigenvalues  $\{\lambda_i\}$ and corresponding unit normalized eigenvectors $\{\vec{\zeta_i}\}$.
Let  $\vec{\omega}$ be the vector with entries $\{\omega_i\}$. We expand $\vec{\omega}$ in the basis of eigenvectors of $N_{ij}$ as $\vec{\omega}=\sum_{i=0}^{n-1} a_i \vec{\zeta_i} $ for some constants $a_i$. The integral over $\omega_i$ can now be expressed as integral over $a_i$. It is easy to convince oneself that the Jacobian for this variable change is unity. In these new variables, the exponent in \eqref{bsg2} takes the form 
\beq \label{neweq} 
\begin{split}
  &i \frac{1}{2\sum_{\mu,i} \zeta_i^\mu a_i } \left(a_0^2 \lambda_0 +\sum_{i=1}^{n-1} a_i^2 \lambda_i \right) \\
  \end{split}
  \end{equation} 
  where the index $\mu$ labels each entry of the eigenvector $\vec{\zeta_i}$. 
  
 Recall we have assumed that smallest eigenvalue $\lambda_0$ is much smaller than all the other eigenvalues. Consequently, it follows from \eqref{neweq} that in the integral over $a_0$ and $a_i$, generically 
 $a_o \gg a_i$ \footnote{More precisely, $\frac{a_0}{a_i} = {\cal O}(\sqrt{\frac{R}{\lambda_0}})$.}.
 It follows that the sum in the denominator of \eqref{neweq} is dominated by $i=0$. For the purpose of determining the leading singularity as $\lambda_0 \rightarrow 0$, consequently, the exponent in \eqref{neweq} may be replaced by 
  \begin{equation} \label{neeqtwo} 
 \begin{split}       
  &\approx i \frac{1}{2\sum_{\mu} \zeta_0^\mu a_0 } \left(a_0^2 \lambda_0 +\sum_{i=1}^{n-1} a_i^2 \lambda_i \right)
\end{split}
\eeq 
Moreover, for the same reason,  we can also replace $\omega_i$ in the Mandlestam variables (which appear in the polynomial prefactors in \eqref{bsg2}) by $\vec\omega \sim a_0 \vec\zeta_0$. Given that all Schwinger parameters $\omega_i$ are positive, this replacement is only possible if all entries of ${\vec \zeta}_0$ have the same sign. \footnote{This is why the singularity, that we will find later in this section when $\lambda_0 \to 0$, actually occurs only when all entries of its eigenvector 
have the same sign.} Let us assume that the sign of these 
eigenvector entries are all positive. It follows that the integral over $a_0$ runs from $0$ to infinity. Once these replacements have been made, the integral over $a_i$ (with $i\neq0$) are Gaussian integrals and may just be performed. 
\footnote{ The width of $a_i$ that contribute to the integral are given by
\begin{equation} \label{widthai} 
    \delta a_i \approx \sqrt{2\frac{\sum_{\mu} \zeta_0^\mu a_0}{\lambda_i}}
\end{equation} } 
yielding 
\beq\label{bsg3}
\begin{split}
G =   A (i \pi)^{(D+3)/2 }\int_0^\infty da_0 \   &\frac{a_0^{4(\frac{D-2}{2}-1)}(\prod_\mu \zeta_0^\mu)^{\frac{D-2}{2}-1}}{{(a_0\sum_\mu \zeta^\mu_0)}^{D/2}}\left(\prod_{i=1}^{3}\sqrt{2\frac{\sum_{\mu} \zeta_0^\mu a_0}{\lambda_i}}\right) \ \exp{\left[i\left(\frac{a_0 \lambda_0}{2\sum_\mu \zeta^\mu_0}\right)\right]} \\
&a_0^{2(m+n)}\left( T(\tilde s, \tilde t)+ T(\tilde t, \tilde u)+ T(\tilde u, \tilde s) \right)
\end{split}
\eeq
where $2(m+n)$ is the total number of derivatives in the interaction and 
$$\tilde s= 2 \zeta_0^1\zeta_0^2 N_{12}, ~~~~\tilde t= 2 \zeta_0^1\zeta_0^3 N_{13}~~~~\tilde u= 2 \zeta_0^1\zeta_0^4 N_{14}$$

It remains to perform the integral over $a_0$. Clearly this integral is dominated by $a_0$ of order 
\beq
a_0 \approx \frac{2\sum_{\mu} \zeta_0^\mu }{\lambda_0}
\eeq
The important point here is that $a_0$ scales like the inverse of smallest eigenvalue $\lambda_0$ while rest of $a_i$'s scaled like inverse of $\sqrt{\lambda_i}$. 

We now perform the $a_0$ integral to obtain
\beq\label{bsgf}
G =   \mathcal{C}\frac{(\prod_\mu\zeta_0^\mu)^{\frac{D-2}{2}-1}}{(\lambda_0)^{ \frac{3D-11}{2}+ 2(m+n)}}  \prod_{i=1}^{3}\frac{1}{\sqrt{\lambda_i}} \ \left(\sum_\mu \zeta_0^\mu\right)^{D-4 + 2(m+n)}\left( T(\tilde s, \tilde t)+ T(\tilde t, \tilde u)+ T(\tilde u, \tilde s) \right)
\eeq
where 
$$\mathcal{C} = (-1)^{m+n}\frac{2^{-12 + \frac{3D}{2} + 2(m+n)}}{\pi^{\frac{3}{2}(D-1)}} \Gamma\left(\frac{3D-11}{2}+ 2(m+n)\right)$$
From above expression, it is clear that when $x_i$'s are chosen on the singular manifold i.e. when $\lambda_0 = 0$, the correlator develops a singularity. The power of the singularity depends on the energy scaling of the S-matrix.

It is now a simple matter to self consistently justify 
our estimates for the width and center of the integral 
over $y$ (see under \eqref{expbsgo}). Our estimate from the width follows from the fact that $a_0$, hence $\sum_i \omega_i$ is dominantly of order $\frac{1}{\lambda_0}$. 
Our estimate for the center for the $y$ integral can be 
obtained as follows. When $\omega^\mu$, the $\mu$th entry in the vector $\omega$, is proportional to   $\zeta_0^\mu$, the shift is proportional to 

\begin{equation}
    \sum_{i} \omega_{i}  (x_i-{\tilde y}) =\sum_{i} \omega_{i}  (\tilde{x}_i-{\tilde y}+\delta x_i) 
    =a_0\sum_{\mu} \zeta_0^\mu  \, \delta x^\mu
\end{equation}
where, in the last step, we have used momentum conservation in the form \eqref{momcons} and that $\omega^\mu=a_0 \zeta_0^{\mu}$. We can relate the shift in insertions $\delta x^\mu$ to the change in the zero eigenvalue as follows.  Momentum conservation at the bulk point $\tilde{y}$ gives
\begin{equation}
\sum_{j=1}^{n} \tilde{N}_{ij}=-2(\tilde{x_i}-\tilde{y}).\sum_{j=1}^{n} (\tilde{x_j}-\tilde{y})=0.
\end{equation}
This means that $\tilde{\zeta}_0=\frac{1}{\sqrt{n}} (1,1,\ldots 1)^T$ is the unit normalized eigenvector of $\tilde{N}$ with eigenvalue zero. Thus we have, to first order in perturbations,
\beq
 \lambda_0= \tilde{\zeta}_0^T \delta N \tilde{\zeta}_0 =\frac{1}{n} \sum_{i,j} \delta N_{ij}.
\eeq

As an example, let us now specialize to the case of the contact four point interaction $g \phi^4$ in four dimensions. We substitute $D=4$ and $m = n = 0$ in \eqref{bsgf} to obtain
\beq\label{bsgf4d}
G =   \frac{i g}{2^6\pi^4} \frac{1}{(\text{det} N)^{1/2}}  
\eeq
where $\text{det} N = \prod_{i = 0}^{3} \lambda_i$. Using \eqref{4detn}, we can express the above expression in terms of 4d conformal cross ratios to obtain
\beq\label{bsgfinal}
G =    \frac{i g}{2^6\pi^4}\frac{1}{N_{13} N_{24} (Z-\zbar)} 
\eeq
Hence, as long as no two insertions coincide, the singularity condition ${\rm Det}N=0$ is met precisely when $Z=\zbar$.

\subsubsection{Multi-bulk point singularity ($c>1$)}
In this subsection, we will explore the cases in which the co-dimension of the singular manifold is greater than one i.e. $c>1$. This happens when the number of particles $m$ is greater than $D+1$, where $D$ is spacetime dimension (see \eqref{Pform}). Hence in this subsection we will consider $m> D+1$. We will first give the general structure of the bulk Greens function and then show the explicit relationship between S-matrix and the singularity of Greens function in a particular example.

As discussed in \S \ref{Nmatrix}, the co-dimension of singularity is related to the number of zero eigenvalues of the distance matrix $N_{ij}$. Hence for $m> D+1$, the distance matrix will develop $m-D$ zero eigenvalues. The Greens function for $m$ point scattering in $D$ dimensions is given by
\beq\label{bsg1m}
G =   A \int ~ d^{D}y~ \left(\prod_{i = 1}^{m} d\omega_i~ \omega_i^{\frac{D-2}{2}-1} \right) e^{i\sum\limits_{i=1}^m\omega_i((x_i-y)^2)}
 T(\omega_i, x_i, y) 
\eeq
where
\beq
A= \frac{e^{-\frac{i \pi (D-2) m }{4} }}{4^m\pi^{\frac{m D}{2}}}\\
\eeq
where we have used Schwinger parametrization to express the propagators as exponentials. This parameterization is valid only for $D\geq 4$. After doing the $y$ integral as explained in previous subsection, we obtain the analog of \eqref{bsg2}, which is given by
 \beq\label{bsg2m}
G =   A (i\pi)^{D/2} \int ~ \left(\prod_{i = 1}^{m} d\omega_i~ \frac{\omega_i^{\frac{D-2}{2}-1}}{{\sum_i \omega_i}^{D/2}}\right) \exp \left[\frac{i}{\sum\limits_{i=1}^{m} \omega_i} \sum\limits_{\substack{i,j=1 \\ i \le j}}^{m} \omega_i \omega_j N_{ij}\right]
T(\omega_i, \tilde{x}_i, \tilde{y}) 
\eeq

 Recall that, at generic insertion points, the matrix $N_{ij}$ has $m-D-2$ zero eigenvalues but the corresponding eigenvectors are non positive. On the other hand, when the boundary insertion points lie on singular manifold, the matrix $\tilde{N}_{ij}$ has $m-D$ zero eigenvalues with positive eigenvectors. 
 
 For the configurations close to the singular manifold $m-D$ eigenvalues of $N_{ij}$ are close to zero. We we label those eigenvalues by $\lambda_i$ and the corresponding eigenvector by $\zeta_i$ where $i = 1, \cdots, m-D$. In addition to these small eigenvalues, $N_{ij}$ has $D$ large eigenvalues which we label by $\chi_i$ and the corresponding eigenvector by $\eta_i$ where $i = 1, \cdots, D$. We will again expand  $\vec{\omega}$ (the vector with entries $\{\omega_i\}$) in the basis of eigenvectors of $N_{ij}$ as 
$$\vec{\omega}=\sum\limits_{i=1}^{m-D} a_i \vec{\zeta_i} + \sum\limits_{i=1}^{D} c_i \vec{\eta_i} $$
for some constants $a_i$ and $c_i$. The integral over $\omega_i$ can now be expressed as integral over $a_i$ and $c_i$. When the insertion points are near to the singular manifold, the constant $a_i \gg c_i$. Hence for the purpose of determining leading singularity, the exponent in \eqref{bsg2m} can be replaced by 
$$\frac{i}{2\sum\limits_{i=1}^{m-D} \left(a_i \sum\limits_\mu\zeta^\mu_i\right)} \left(\sum\limits_{i=1}^{m-D} a^2_i \lambda_i + \sum\limits_{i=1}^{D} c^2_i \chi_i\right)$$
As discussed in the previous subsection, we can replace $\omega^\mu \sim \sum\limits_{i=1}^{m-D} a_i \zeta^\mu_i$ in the T-matrix as well as in the prefactor of exponent in \eqref{bsg2}. 

The replacement $\omega^\mu \sim \sum\limits_{i=1}^{m-D} a_i \zeta^\mu_i$ can only legitimately be made when all entries of $\zeta_i^\mu$ are positive (this follows as all the Schwinger parameters $\omega^\mu$ are positive. 
For this reason this replacement is only valid in the neighborhood of the scattering manifold (which, recall, 
has a $m-D$ zero parameter family of zero eigenvalues with eigenvectors of positive coefficients). We emphasize 
that the replacement made above cannot be affected at 
a generic point in insertion space, despite the fact that the matrix $N$ has $m-D-2$ zero eigenvalues at such a point.

After making these replacements, the integrals over $c_i$'s are Gaussian integrals which can be performed to obtain

\beq\label{bsg3m}
G =   A (i \pi)^{D/2}\left(\prod_{i=1}^{D}\sqrt{\frac{2 i \pi}{\chi_i}}\right) \int \prod_{i = 1}^{m-D} da_i \  \left(\prod_{\mu = 1}^{m}\sum\limits_{i=1}^{m-D} a_i \zeta^\mu_i\right)^{\frac{D-2}{2}-1} \ \exp{\left[i\left(\frac{\sum\limits_{i=1}^{m-D} a^2_i \lambda_i}{2\sum\limits_{i=1}^{m-D} \left(a_i \sum\limits_\mu\zeta_i^\mu\right)}\right)\right]}T(\omega^\mu, \tilde{x}_i, \tilde{y})
\eeq
The next step is to perform the $a_i$ integrals.
We introduce $ 1= \int_0^{\infty} ds \, \delta(s-\sum\limits_{i=1}^{m-D} \gamma_i a_i ) $, for arbitrary $\gamma_i >0$, and then take $a_i \rightarrow s \ b_i$ to find
\beq
\begin{split}\label{spower}
    G =   A (i \pi)^{D/2}\left(\prod_{i=1}^{D}\sqrt{\frac{2 i \pi}{\chi_i}}\right) &\left[\prod_{i = 1}^{m-D} \int  db_i \right]  \left(\prod_{\mu=1}^{m}\sum\limits_{j=1}^{m-D} b_j \zeta^{\mu}_j\right)^{\frac{D-2}{2}-1}\delta(1-\sum_{i=1}^{m-D} \gamma_i b_i)\\
    &\times \int ds \, s^{m-D+m\left(\frac{D-2}{2}-1\right)+\ell-1}\exp{\left[i s\left(\frac{\sum\limits_{i=1}^{m-D} b^2_i \lambda_i}{2\sum\limits_{j=1}^{m-D} \left(b_j \sum\limits_{\mu=1}^m \zeta^{\mu}_j\right)} \right)\right]} T(b_i, \tilde{x}_i, \tilde{y}),
\end{split}
\eeq
where $\ell$ equals the total number of derivatives in the contact interaction.
Here, $s$ sets the overall energy scale of the scattering process, and the $b_i$ correspond to the directions of approach to the scattering singularity. We perform the integral over $s$ to find
\beq
\begin{split}\label{bsg6m}
    G =   A' \left(\prod_{i=1}^{D}\sqrt{\frac{1}{\chi_i}}\right) &\left[\prod_{i = 1}^{m-D} \int  db_i  \right]\left(\prod_{\mu=1}^{m}\sum\limits_{j=1}^{m-D} b_j \zeta^{\mu}_j\right)^{\frac{D-2}{2}-1}  \delta(1-\sum_{i=1}^{m-D} \gamma_i b_i)\\
    &\times \left(\frac{\sum\limits_{j=1}^{m-D} \left(b_j \sum\limits_{\mu=1}^m \zeta^{\mu}_j\right)}{\sum\limits_{i=1}^{m-D} b^2_i \lambda_i}\right)^{m-D+m\left(\frac{D-2}{2}-1\right)+\ell} T(b_i, \tilde{x}_i, \tilde{y})
\end{split}
\eeq
where $$A' = (-1)^{\frac{m}{2}(D-2)}i^{\frac{m}{2}(D-2)+ D}\frac{(2)^{l + d/2(m-1) - 3 m}}{\pi^{\frac{D}{2}(m-2)}}.$$
As expected for $m> D+1$, the above expression develops a singularity when all $\lambda_i \rightarrow 0$. To perform the rest of the integrals explicitly, we need to know the details of the insertion points and the interaction vertex. 

 We will turn to the $b_i$ integrals. For the purpose of evaluation we find it convenient to  make the choice $\gamma_i =\delta_{i,1}$ in \eqref{bsg6m} so that the $b_1$ integral can be evaluated using the $\delta$ function. After performing the integral over $b_1$  we find,
\beq\label{bsg5m}\begin{split}
G =   A' \left(\prod_{i=1}^{D}\sqrt{\frac{1}{\chi_i}}\right) & \int \prod_{i =2}^{m-D} db_i \  \left(\frac{\sum_\mu\zeta_1^\mu + \sum\limits_{i=2}^{m-D} \left(b_i \sum\limits_\mu\zeta^\mu_i\right)}{\lambda_1 + \sum\limits_{i=2}^{m-D} b^2_i \lambda_i} \right)^{m-D+ m(\frac{D-2}{2}-1)+ \ell} \\
&\left(\prod_{\mu = 1}^{m} \left(\zeta_1^\mu + \sum\limits_{i=2}^{m-D}b_i \zeta^\mu_i\right)\right)^{\frac{D-2}{2}-1} T(b_i, \tilde{x}_i, \tilde{y})
\end{split}
\eeq

In the rest of this subsection, we study 
\eqref{bsg5m} for the special case of 6 particle scattering in $D=4$. In this case, the singular manifold has co-dimension 2, and so the index $i$ in \eqref{bsg5m} runs from $2$ to $2$, i.e. only one integral remains to be performed in \eqref{bsg5m}. 

In this case (i.e. 6 particle scattering in $D=4$) the total number of parameters in the space of possible boundary points is 18. As we have mentioned above, the scattering manifold is a codimension 2 (hence 16 dimensional) submanifold of this 18 dimensional space. 
As dimensionality of these spaces is large, for purposes of illustration we work
on a 5 dimensional submanifold of the 18 dimensional space of boundary points, which (as one would generically expect) intersects the scattering submanifold on a 
3 dimensional submanifold. \footnote{The 3 dimensional rotational and translational symmetry group is 9 dimensional. It follows that the 18 parameter collection of boundary points can be organized in a $18-9=9$ parameter set of symmetry inequivalent boundary configurations. 
In the example below we probe a particular 5 dimensional submanifold of this 9 dimensional space.} 
Our 5 dimensional manifold consists of the boundary insertion points 
\beq\label{insbdry}
\begin{split}
x_1 &= T(-1,-1,0,0)\\
x_2 &= T(-1,1,0,0)\\
x_3 &= T(-1, \cos \theta, \sin \theta +\tau ,0)\\
x_4 &= T(1,-\cos \theta,-\sin \theta ,0)\\
x_5 &= T(1, - \cos \psi + \rho  ,- \sin \psi \sin \phi ,- \sin \psi \cos \phi)\\
x_6 &= T(1,  \cos \psi, \sin \psi \sin \phi, \sin \psi \cos\phi)
\end{split}
\eeq
As we see, our 5 dimensional submanifold 
is labelled by the coordinates 
$\theta, \phi, \psi, \tau$ and $\rho$.
It is straightforward to compute the $6\times 6$ matrix $N_{ij}$ corresponding to the collection of insertion points 
\eqref{insbdry}, and to verify that this (generically maximal rank) matrix develops two zero eigenvalues (and hence is of rank 4) when $\tau$ and $\rho$ equal zero. 
It follows that three dimensional singular  submanifold is given by the conditions 
$\tau=\rho=0$.

On the scattering submanifold $\tau=\rho=0$ we use the notation of previous subsection to denote $N_{ij}$ as $\tilde{N}_{ij}$. 
It is possible to check that zero eigenspace of the matrix $\tilde{N}_{ij}$ can be spanned by the two eigenvectors 
\beq
\begin{split}\label{eve}
\tilde{\zeta}_1 = (1,1,0,0,1,1)^T , \qquad \qquad \tilde{\zeta}_2 = (0,0,1,1,0,0)^T. 
\end{split} 
\eeq
(the basis use in the presentation of this result is that of \eqref{insbdry}:  note that the zero eigenspace is 
independent of $\theta$, $\phi$ and $\psi$). We have used the particular basis 
in \eqref{eve} so that the linear term in the Taylor expansion of $N_{ij}$ (in $\rho$ and $\tau$) about the singular submanifold is `diagonal' in this two dimensional subspace: i.e. 
\begin{equation} \label{twoddiag}
\begin{split} 
&{\tilde \zeta}_a^T N {\tilde \zeta}_b
= \delta_{ab} \lambda_a \\
& \lambda_1 = (16 T \cos \psi) \rho , \qquad \qquad \lambda_2 = (8 T \sin \theta) \tau
\end{split}
\end{equation} 
In other words the two zero eigenvalues 
are lifted at linear order in $\tau$ and 
$\rho$; ${\tilde \zeta}_a$ are the corresponding eigenvectors, with eigenvalues $\lambda_a$ presented above. 

In the rest of this subsection we will 
evaluate the remaining integral in \eqref{bsg5m} (i.e. the integral over $b_2)$
for three simple examples of the $T$ matrix; the 
 a constant contact interaction and two distinct four derivative interactions.
 
 \textbf{Example 1}: Consider the  interaction term $g\phi^6$. Since our interaction has no derivative, we have $\ell=0$. 
Plugging $m=6$, $D=4$ and $\ell=0$ into  \eqref{bsg5m} and always working to first order in $\rho$ and $\tau$, 
we obtain
\beq\label{bsg8m}
G =   A' \left(\prod_{i=1}^{4}\sqrt{\frac{1}{\chi_i}}\right)  \int_0^\infty db_2 \  \left(\frac{4 + 2 b_2}{(16 T \cos \psi) \rho + b^2_2 \tau(8 T \sin \theta)  } + \right)^{2}  (i g)
\eeq
(we have inserted $T(b_i, \tilde{x}_i, \tilde{y}) = i g$ into \eqref{bsg5m}). Recall that  $\chi_i$ $i=1 \ldots 4$ are the four nonzero eigenvalues of the matrix ${\tilde N}_{ij}$ \footnote{These eigenvalues are easy to work out explicitly- for instance on Mathematica - but the corresponding expressions are cumbersome, and appear to be uninstructive, so we do not list them. The eigenvalues go along for the ride (as multiplicative factors) in the analysis of singularities below.}. The integral over $b_2$ in \eqref{bsg8m} is elementary; performing this integral we obtain 
\beq \label {mbpeo}
G =   -\frac{1}{(2 \pi)^8} \left(\prod_{i=1}^{4}\sqrt{\frac{1}{\chi_i}}\right)\frac{i g}{16 T^2 \sin \theta  \cos \psi}\left(\frac{\pi  }{2 \sqrt{2} \rho ^{3/2} \tau^{1/2} }\sqrt{\frac{\sin \theta }{\cos \psi }}+\frac{\pi  }{4 \sqrt{2} \rho^{1/2} \tau ^{3/2}}\sqrt{\frac{\cos \psi }{\sin \theta }}+\frac{1}{\rho  \tau }\right)
\eeq
The singularities in $\rho$ and $\tau$ are 
the multi bulk point singularities. Note 
that all terms in \eqref{mbpeo} are of homogeneity $1/\epsilon^2$ ($\tau$ and 
$\rho$ are both taken to be of order $\epsilon$). Note also that, in this case 
the maximum singular power of $\rho$ ($-3/2$) is the same as the maximal singular power of $\tau$ (again $-3/2)$. 
Infact these singular terms are symmetric 
under interchange of the variables 
$\cos \psi \rho$ and $\sin \theta \tau$.

As we have explained in the main text, the 
precise structure of this singularity encodes the $T$ matrix (simply $i g$ in this simple case). We now turn to the study of the bulk point singularity corresponding to a slightly more elaborate $T$ matrix. 

\textbf{Example 2}: In this case we evaluate the $G$ that is generated by following interaction term 
\beq\label{dint}
\frac{g_{4}}{4}\int d^4 x \ \left(\partial_{\mu} \partial_{\nu}\phi_1\right) \left(\partial^{\mu}\phi_2\right) \left(\partial^{\nu}\phi_3 \right)\ \phi_4 \phi_5 \phi_6.
\eeq
This interaction term contains $4$ derivatives. One of the derivatives on $\phi_1$  is contracted with derivative on $\phi_2$ and other is contracted with the one on $\phi_3$. The $T$ matrix generated  by the interaction \eqref{dint} is 
\beq 
T(s_{12}, s_{13}) = i g_{4} \ s_{12} \ s_{13}
\eeq
where
$$s_{ij} = 2 \omega_i \omega_j (x_i -y). (x_j -y).$$
As our interaction term has four derivatives $\ell = 4$ , so \eqref{bsg5m} reduces to  
\beq\label{bsg8deri}
G =   -\frac{1}{(2 \pi)^8} \left(\prod_{i=1}^{4}\sqrt{\frac{1}{\chi_i}}\right)  \int_0^\infty db_2 \  \left(\frac{4 + 2 b_2}{(16 T \cos \psi) \rho + b^2_2 \tau(8 T \sin \theta)  } \right)^{6}  \left(i (2 T) ^{4} g_{4}\ b_2 \cos^{2} \left(\frac{\theta}{2}\right)\right)
\eeq
where we have used $T(b_i, \tilde{x}_i, \tilde{y}) = i g_{2k} b_1^{3} b_2 \tilde{N}_{12} \tilde{N}_{13}$. Performing the (elementary) integral over $b_2$ in \eqref{bsg8deri} we obtain 
\beq
\begin{split}\label{eg1}
G = &  -\frac{1}{(2 \pi)^8} \left(\prod_{i=1}^{4}\sqrt{\frac{1}{\chi_i}}\right)\left(\cot \frac{\theta}{2} \cos^2 \psi \right) \left(\frac{i g_{4}}{(2)^{13} T^2 \rho^2 \tau}\right) \bigg(\frac{16 \sec^3 \psi}{5 \rho^3}+\frac{21 \pi \sec^{5/2} \psi \csc^{1/2} \theta}{8 \sqrt{2} \rho^{5/2} \tau^{1/2}}\\
& + \frac{6 \csc \theta \sec ^2 \psi}{\rho ^2 \tau } +\frac{15 \pi  \csc ^{3/2}\theta \sec^{3/2} \psi}{8 \sqrt{2} \rho^{3/2}  \tau ^{3/2}} +\frac{2 \csc ^2\theta \sec \psi }{\rho  \tau ^2}+\frac{9 \pi  \csc ^{5/2}\theta \sec^{1/2} \psi }{32 \sqrt{2} \rho^{1/2} \tau ^{5/2}} +\frac{\csc ^3 \theta}{10 \tau ^3}\bigg) 
\end{split}
\eeq
Once again the singularities in $\rho$ and $\tau$ are 
the multi bulk point singularities and all terms in \eqref{eg1} are of homogeneity $1/\epsilon^6$ ($\tau$ and 
$\rho$ are both taken to be of order $\epsilon$). As the interaction term in this example had four more derivatives than
the interaction in the previous example, 
the singularity in $\epsilon$ is greater (by the factor $1/\epsilon^4$ as compared to the previous example. 

 Note also that, in this case 
the maximum singular power of $\rho$ ($5$) is the not the same as the maximal singular power of $\tau$ ($4$). This is because the derivatives in the interaction given in \eqref{dint} are not distributed symmetrically among various $\phi_i$. 

\textbf{Example 3}: In this example, we will again consider four derivative interaction but with different derivative contractions among $\phi_i$. We will consider the following interaction term  \beq\label{dint1}
\frac{g'_{4}}{4}\int d^4 x \ \left(\partial_{\mu} \partial_{\nu}\phi_1 \right)\left(\partial^{\mu}\partial^{\nu}\phi_2\right) \phi_3  \phi_4 \phi_5 \phi_6.
\eeq
In this case, we obtain the following bulk Greens function
\beq
\begin{split}\label{eg2}
G = & -\frac{1}{(2 \pi)^8} \left(\prod_{i=1}^{4}\sqrt{\frac{1}{\chi_i}}\right)\left(\cos^4 \frac{\theta}{2} \cos^2 \psi \csc^2 \theta \right) \left(\frac{i g'_{4}}{(2)^{13} T^2 \rho^2 \tau^{3/2}}\right) \bigg(-\frac{7 \pi  \csc ^{3/2}\theta \sec ^{5/2}\psi(\sin (3 \theta ) \csc \theta -3)}{32 \sqrt{2} \rho ^{5/2}}\\
&+\frac{24 \sec ^2\psi}{5 \rho ^2 \tau^{1/2} } +\frac{45 \pi  \csc^{1/2} \theta  \sec^{3/2} \psi} {64 \sqrt{2} \rho^{3/2}  \tau}+\frac{16 \csc \theta \sec \psi }{3 \rho  \tau ^{3/2}} +\frac{45 \pi  \csc ^{3/2}\theta \sec^{1/2} \psi}{32 \sqrt{2} \rho^{1/2}\tau^{2}} +\frac{6 \csc ^2\theta }{5 \tau ^{5/2}}\bigg)
\end{split}
\eeq
Here all terms in \eqref{eg2} are again of homogeneity $1/\epsilon^6$ ($\tau$ and $\rho$ are both taken to be of order $\epsilon$). This is because the total number of derivatives in \eqref{dint1} is same as those in \eqref{dint2}. The detailed structure of above answer is different  from \eqref{eg1} because the distribution of derivatives in the interaction terms is different.

\subsection{Exact computation of correlators}
\label{exactcor}
In subsubsection \ref{ceq1}  we have presented an exact characterization of the leading singular behavior of the four point 
Greens function generated by an arbitrary interaction in any dimension $D \geq 3$. 
In this subsection, we specialize to a special case in this class of Greens functions - namely to the case of contact interactions. Later in 
this subsubsection we also specialize to 
$D=4$. Working in Euclidean space in the context of this special example we are able to find a completely exact closed form expression for the Greens function everywhere (i.e. not just in the neighborhood of singularities). After 
performing an analytic continuation back to Lorentzian space we verify that our exact 
result does, indeed, have a singularity of exactly the form predicted by subsubsection \ref{ceq1}. 

Consider the Euclidean $n$-pointfunction generated by a $g \phi^n$ contact interaction  
\begin{equation} \label{ncontact}
	    \begin{split}
	        \gf(x_1, \ldots , x_n)&={\mathcal{A} \over \pi^{D/2}} \int d^D y \prod_{i=1}^{n} \frac{\Gamma(\Delta_i)}{((x_i-y)^2)^{\Delta_i}}, 
	    \end{split}
	\end{equation}
where $\mathcal{A} = \frac{i g}{4^4 \pi^{3D/2}}$. Integrals of this kind in a conformally invariant theory have been treated in the literature beginning with the work of Symanzik \cite{Symanzik:1972wj}. Some of the methods employed there will be useful to us too. For rest of this section, we will drop the numerical factor $\mathcal{A}$ and will restore it in the final expression. In the situation under study $\Delta_i=\dm2$; however, we leave $\Delta_i$ arbitrary for now (specializing to this value only later), as the answer for   general values of $\Delta_i$ will turn out to be useful in computing the n-point functions for massive fields in Appendix 
\ref{dflat}. 
Just as in subsection \ref{ceq1}, we introduce Schwinger parameters $\omega_i$ for each of the denominators in \eqref{ncontact}, and perform the bulk integral over $y$ to obtain
\begin{equation} \label{eq:schpara}
	    \begin{split}
	        \gf(x_1, \ldots , x_n)&= \prod_{i=1}^n \int_{0}^{\infty} d\omega_i \omega_i^{\Delta_i-1}{1 \over \Omega^{D/2}} \exp(-{1\over \Omega} \sum\limits_{\substack{i,j=1 \\ i \le j}}^{n} \omega_i \omega_j N_{ij}),
	    \end{split}
\end{equation}
 where we have defined $\Omega\equiv\sum\limits_{i=1}^n {\omega_i}$. Henceforth, we will not explicitly display the range of summation of $i,j$ and the $i\leq j$ condition on the sum $\sum \omega_i \omega_j N_{ij}$. It will turn out to be convenient to introduce unity in the form of $$ 1= \int_0^{\infty} ds \delta(s-\sum\limits_{i=1}^n \lambda_i \omega_i) $$ for arbitrary $\lambda_i \geq 0$ and obtain the correlator in the form

\begin{equation} 
	    \begin{split}
	        \gf(x_1, \ldots , x_n)&= \prod_{i=1}^n \int_{0}^{\infty} d\omega_i \omega_i^{\Delta_i-1}{1 \over \Omega^{D/2}} \exp\left(-{1\over \Omega} \sum \omega_i \omega_j N_{ij}\right) \int_0^{\infty} ds \delta(s-\sum\limits_{i=1}^n \lambda_i \omega_i)
	    \end{split}
\end{equation}
Now we scale $\omega_i \rightarrow s \omega_i$ and define $\Delta \equiv \sum\limits_{i=1}^n \Delta_i$ to arrive at
\begin{equation} \label{eq:lamexp}
	    \begin{split}
	        \gf(x_1, \ldots , x_n)&= \prod_{i=1}^n \int_{0}^{\infty} d\omega_i \omega_i^{\Delta_i-1}{1 \over (\sum\limits_{i=1}^n \omega_i)^{D/2}} \delta(1-\sum\limits_{i=1}^n \lambda_i \omega_i)\\
	        &\times \int_0^{\infty} ds  s^{\Delta-1-D/2} \exp\left(-{s\over \sum\limits_{i=1}^n \omega_i} \sum \omega_i \omega_j N_{ij}\right)
	    \end{split}
\end{equation}
We perform the integral over $s$  to obtain
\begin{equation} \label{lampow}
	    \begin{split}
	        \gf(x_1, \ldots , x_n)&= \Gamma\left(\Delta-{D\over2}\right)\prod_{i=1}^n \int_{0}^{\infty} d\omega_i\omega_i^{\Delta_i-1} \delta(1-\sum\limits_{i=1}^n \lambda_i \omega_i)  {(\sum\limits_{i=1}^n \omega_i)^{\Delta-D} \over \left(\sum \omega_i \omega_j N_{ij}\right)^{\Delta-D/2}}
	    \end{split}
\end{equation}
While the above expression might look complicated, we can actually use it to perform the integrals over the Schwinger parameters and obtain closed form answers for any $n$-point correlator. In Appendix \ref{dflat}, we demonstrate this procedure for $3$ point and $4$ point functions in any dimension. Here, we work out the special case of the $4$ point correlator in four dimensions. Later, we will use this exact answer and provide an explicit check of our general result that the $S$ matrix can be obtained from an appropriate smearing of the boundary correlator. 

Before turning to an exact evaluation of the integrals in \eqref{lampow} (for the special case mentioned above) we pause to note a useful structural feature of \eqref{lampow}: \eqref{lampow} can be used to determine when (i.e for what values of the matrix of distances of external insertions,  $N_{ij}$) the Greens function develops a singularity.

 Let us define $F=\left(\sum_{i,j} \omega_i \omega_j N_{ij}\right)$. We see that \eqref{lampow} is potentially singular only if the integrand in \eqref{lampow} goes singular somewhere on the contour of integration, i.e. only if there exist positive numbers ${\omega^*_i}$ such that $F|_{\omega=\omega^*}=\sum_{i,j} \omega^*_i \omega^*_j N_{ij}=0$. When this condition 
 is met, the integral over $\omega_i$ is only singular if it is impossible to  deform the contour in the complex ${\omega_i}$ space in a direction normal to the singularity surface and avoid it. This impossibility arises when the singular manifold `pinches' the integration contour. This happens when the singular manifold approaches the contour from two different ends. When this happens the normal to the singular manifold is necessarily ill-defined at the pinch point. 
 Following this line of reasoning, it has been rigorously demonstrated \cite{Eden:1966dnq} that it is impossible to deform the integration contour away from the singular manifold, if and only if where $F$ vanishes, it is also true that \footnote{In the current context,  $F$ is a degree two homogenous polynomial in $\omega$. Consequently  $2 F=\sum_i{\omega_i }\frac{\partial}{\partial \omega_i} F$, so that the pinching condition \eqref{eigvec} automatically implies $F=0$.}
\begin{equation} \label{eigvec}
\frac{\partial}{\partial \omega_i} F|_{\omega=\omega^*}=\sum_{j=1}^n \omega^*_j N_{ij}=0 \qquad\text{for each} \, i
\end{equation}
 Thus, we find that the correlator suffers a singularity if and only if the matrix $N$ has a zero eigenvalue with the corresponding eigenvector having all positive (or all negative) entries\footnote{The last condition of positivity (or negativity) comes from the fact that the Schwinger parameters run from $0$ to  $\infty$ (or $-\infty$ to $0$).}. 

While we have presented our analysis for the special case of a contact interaction, our conclusion remains valid for the S matrix generated by any local interaction. The reason for this is that the part of \eqref{lampow} that is responsible for the singularity - namely the denominator in 
that expression - is the same for every contact (or any other short range) interaction
(See \S \ref{ceq1} for an explicit demonstration in the case of four points). 

In summary, we have demonstrated - as promised in \S \ref{Nmatrix} - that our Greens function is singular if and only if our $N$ matrix has atleast one zero eigenvalue whose corresponding zero eigenvector has all positive (or all negative) entries. 

In four dimensions, $\Delta_i=\dm2=1$ and $\Delta = \sum\limits_{i=1}^n \Delta_i=4$ so that the $4$- point coorelator takes the form  
\begin{equation} 
	    \begin{split}
	        \gf(x_1, \ldots , x_4)&= \prod_{i=1}^4 \int_{0}^{\infty} d\omega_i \delta(1-\sum\limits_{i=1}^4 \lambda_i \omega_i)  {1 \over \left(\sum \omega_i \omega_j N_{ij}\right)^{2}}
	    \end{split}
\end{equation}
In the above expression, the parameters $\lambda_i$ are arbitrary. We choose $\lambda_i= \delta_{i,4}$.  This sets $\omega_4=1$ and we obtain
\begin{equation} 
\begin{split}
\gf(x_1, \ldots , x_4)&= \prod_{i=1}^{3} \int_{0}^{\infty} d\omega_i   {1 \over \left(\sum\limits_{i=1}^{3}\omega_i N_{4i}+\sum\limits_{\substack{ i,j \neq 4}} \omega_i \omega_j N_{ij}\right)^{2}}
\end{split}
\end{equation}


Using the same trick as before, we introduce unity again in the form
$$ 1= \int_0^{\infty} ds \delta(s-\sum\limits_{i=1}^{3} \lambda_i \omega_i) $$ for arbitrary $\lambda_i \geq 0$, and then scale $\omega_i \rightarrow s \omega_i$ for $i=1$ to $3$ to find 

\begin{equation} \label{eq:nminusone}
\begin{split}
\gf(x_1, \ldots , x_4)&= \prod_{i=1}^{3} \int_{0}^{\infty} d\omega_i \delta(1-\sum\limits_{i=1}^{3} \lambda_i \omega_i) \int_0^{\infty} ds {1 \over \left(\sum\limits_{i=1}^{3}\omega_i N_{4i}+s \sum\limits_{\substack{ i,j \neq 4}} \omega_i \omega_j N_{ij}\right)^{2} }
\end{split}
\end{equation}

This time, we make the choice $\lambda_i= \delta_{i,3}$ which sets $\omega_3=1$. 
Performing the elementary integral over $s$, we obtain
\begin{equation} \label{nminustwo}
\begin{split}
\gf(x_1, \ldots , x_4)&= \prod_{i=1}^{2} \int_{0}^{\infty} d\omega_i \frac{1}{\left(N_{34}+\sum\limits_{i=1}^{2}\omega_i N_{4i}\right) } \frac{1}{\left(\omega_1 \omega_2 N_{12} + \sum\limits_{i=1}^{2}\omega_i N_{3i} \right)} 
\end{split}
\end{equation}
We will now make use of the following remarkable identity 
\begin{equation} \label{dengamma}
    \frac{1}{(A+B)^n} ={1 \over \Gamma(n)} \int_{c-i \infty}^{c+i \infty} {ds \over {2\pi i}} \Gamma(n+s) \Gamma(-s) {A^s \over B^{s+n}},
\end{equation}
where $c$ is a small positive number.
It is easy to prove this identity as follows. The contour in \eqref{dengamma} hugs the imaginary axis from the positive real side. We close the contour with a large semicircular arc on the right. Then, the integral picks up the residues of all the poles of the gamma functions at positive integer arguments, and reproduces the binomial expansion of the left hand side.

Using \eqref{dengamma} with $n=1$ twice for each of the numerators in \eqref{nminustwo}, we find
\begin{equation} 
\begin{split}
\gf(x_1, \ldots , x_4)= &\int_{c-i \infty}^{c+i \infty} {ds \over {2\pi i}} \int_{c-i \infty}^{c+i \infty} {dr \over {2\pi i}} \Gamma(-s)\Gamma(-t) \Gamma(1+s) \Gamma(1+t) N_{24}^q N_{23}^t  \\
&\int_{0}^{\infty} d\omega_1 {\omega_1^{-t-1} \over \left(N_{34}+\omega_1 N_{14}\right)^{1+q} } \int_{0}^{\infty} d\omega_2 {\omega_2^{q+t} \over \left(N_{13}+\omega_2 N_{12}\right)^{1+t} }
\end{split}
\end{equation}
The $\omega_1$ and $\omega_2$ integrals are now elementary and can be performed to arrive at the following representation of the correlator,
\begin{equation} \label{4dgreen}
\begin{split}
\gf(x_1, \ldots , x_4)= & \frac{i g}{4^4 \pi^{6}}{1 \over N_{12} N_{34}} {\cal D}_{1111}(u,v)
\end{split}
\end{equation}
where 
\begin{equation} \label{4dfunc}
\begin{split}
{\cal D}_{1111}(u,v)= &  \int {ds \over 2\pi i} \int {dt \over 2\pi i}  u^s v^t \, \left(\Gamma(-s)\right)^2 \,\left(\Gamma(-t)\right)^2 \,\left(\Gamma(1+s+t)\right)^2
\end{split}
\end{equation}
where the integrals over both $s$ and $t$ 
are taken over the contours described above, namely contours parallel to the imaginary axis, at a fixed real value between zero and unity. 

The quantities $u$ and $v$ in \eqref{4dfunc} are the usual conformal cross ratios 
 \beq\label{cratio}
 u=\frac{N_{13}N_{24}}{N_{12}N_{34}}, \qquad v=\frac{N_{14}N_{23}}{N_{12}N_{34}}.
 \eeq
 In \eqref{4dgreen} we have restored all the numerical factors.
 
 In summary, our correlator is expressed in terms of the CFT 4 point ${\cal D}$ function in 4 dimensions ($\cal{D}$ functions were introduced in \cite{DHoker:1999kzh}, Appendix A) for four identical operators of scaling dimension one each.  In \cite{Dolan:2001}, it was shown that this particular $\cal{D}$ function has the following closed form expression (Eq. C.15 in \cite{Dolan:2001})
\begin{equation}\label{dfunc4d}
    {\cal D}_{1111}(Z,\zbar)=\frac{1}{Z-\zbar} \left(2 \text{Li}_2(Z)-2\text{Li}_2(\zbar)+\ln(Z\zbar) \ln \left(\frac{1-Z}{1-\zbar}\right) \right)
\end{equation}
with the usual redefinition of the cross ratios to $Z,\zbar$ given by $u=Z\zbar$ and $v=(1-Z)(1-\zbar)$.

The reason we find a conformally invariant answer for the correlator is simply the fact that the $\phi^4$ operator is marginal in four dimensions. This means that the contact contribution to the four point $\phi$ correlator obtained by adding this term to the Lagrangian gives the same (conformal) contribution as we have computed.

The general $n$-point correlator in $D$ dimensions can be expressed in terms of a generalization of the CFT ${\cal D}$ functions which we dub as $\dflat$ functions. We compute the $3$-point and $4$-point correlators in $D$ dimensions and the corresponding $\dflat$ functions in Appendix \ref{dflat}.

In the next subsubsection, we analytically continue the exact answer for the Euclidean 4-point function in 4 dimensions \eqref{4dgreen} to Lorentzian space and find its behavior around the singularity.

\subsubsection{Analytic Continuation for  the four point function in four dimensions}

In the previous subsubsection, we have worked out the 
expression for the Euclidean four point Green's function (coming from a non derivative contact interaction) in 4 dimensions. In order to apply this result to the study of scattering, we need to analytically continue this result to Lorentzian space - more particularly to one of the Lorentzian scattering configurations (two points on the future boundary the slab - located at $t=T$-  and the other two points on the past boundary of the slab - located 
at $t=-T$ two future points each lying in the causal future of the two past points). In this brief 
subsubsection, we explicitly perform the necessary 
analytic continuation on the Green's \eqref{4dgreen}. Our task 
here is made very easy by the fact that the Green's \eqref{4dgreen} is precisely proportional to the 
correlator of four dimension 1 operators computed 
from a bulk contact interaction of the dual $m^2=-3$ fields in a putative $AdS_5$ bulk dual \cite{Dolan:2001} (see also \cite{Gary:2009ae}), and so 
has been extensively studied and we are able simply to borrow relevant results in the literature.  The analytic continuation we need is studied in \cite{Gary:2009ae, Chandorkar:2021viw} in the context of bulk point singularity in AdS/CFT. More particularly, the analytic continuation that occurs in our context (see below for details) turns out to be precisely the continuation described in great detail in  \S 2.4 and \S 2.5 \cite{Chandorkar:2021viw}: see particularly Fig. 5 in \S 2.5 there for the motion in cross ratio space.

 For non-coincident insertions, potential singularities of the Green's function in \eqref{4dgreen} are given by the singularities of the ${\cal D}$ function given in \eqref{4dfunc}. 
 On the principal or Euclidean sheet, the  ${\cal D}$ function is regular everywhere including at $Z=\zbar$ \footnote{That is, in Euclidean space, at any real value of $Z$.}  (see Appendix B of \cite{Gary:2009ae}). \footnote{The fact that the function is regular at $Z=\zbar$ is not immediately visually apparent from \eqref{4dfunc}, but follows from the fact that the divergence in the prefactor $\frac{1}{Z-\zbar}$ is canceled by the vanishing - at the same point - of  the big bracket on the RHS of \eqref{4dfunc}.} 
 
 Our strategy is the following. We first consider insertions that are all spacelike related to each other (and so lie on the Euclidean sheet) and then continuously deform the location of our insertions 
 so that we finally reach the scattering configuration 
 of interest. The trajectory of insertion points we study is given by 
\beq \label{anains}
\begin{split}
x_1 &= \{-T, 0, 0 , T\} \\ 
x_3 &= \{-T, 0, 0 , -T\} \\  
x_2 &= \{-T + \tau, -T \sin \theta, 0 , -T\cos \theta\} \\  
x_4 &= \{-T+ \tau, T \sin \theta, 0 , T\cos \theta\}. 
\end{split}
\eeq
for a fixed value of $\theta <\frac{\pi}{2}$ (in the plots below we choose the particular example $\theta=\frac{\pi}{3}$).  In \eqref{anains}, the first 
coordinate is time, while the remaining three are spatial coordinates (let us call the $x, y,$ and $z$). All through the trajectory, the $y$ coordinate of every insertion point vanishes. Points $x_1$ and $x_3$ lie on the $z$ axis (they have vanishing $x$ coordinates). Points $x_2$ and $x_4$ lie on the $xz$ plane and make an angle $\theta$ with the $z$ axis. 

Our starting configuration is at $\tau =0$ (at this point all our insertions are located at $t=-T$ and are spacelike related as required). Our final configuration is at $\tau = 2T$; at this point insertions $2$ and $4$ 
have reached the `final' surface $t= T$. 

\begin{figure}
    \centering  \includegraphics[width=100mm]{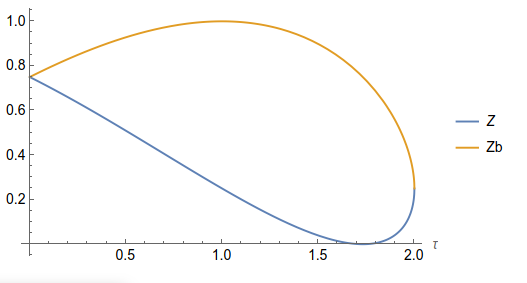}
    \caption{Motion of $Z$ and $\zbar$ as insertions move from Euclidean ($\tau=0$) to scattering ($\tau=2$) configuration}
    \label{zzbre}
\end{figure}

As we move from the initial to the final configuration 
the conformal cross ratios evolve as follows. On the initial configuration (at $\tau=0$)
\beq
Z =  \zbar = \cos^2\left(\frac{\theta}{2}\right)
\eeq
As $\tau$ increases, ${Z}$ decreases while 
${\bar Z}$ initially increases. At $\tau = 2 T \sin\left(\frac{\theta}{2}\right)$,  ${\bar Z}$ reaches 
its maximum value on the trajectory, namely ${\bar Z}=1$. \footnote{More precisely, at this point \beq
Z =\cos^2\theta, \qquad \zbar = 1.
\eeq}
At this special point $x_4$ intersects the lightcone of $x_1$, and $x_2$ simultaneously intersects the lightcone of $x_3$. As ${\bar Z=1}$ is a branch point of the function \eqref{4dfunc}, it is important to resolve the approach of ${\bar Z}$ to unity very carefully. In order to do this we use the fact that the insertion locations for our time ordered correlators have a small imaginary part given by the formula $t_i \rightarrow t_i - i \epsilon t_i$ (this prescription guarantees time ordering). Inserting these imaginary pieces into the formula for the cross ratio, and proceeding exactly 
as in subsection 2.5.1 of \cite{Chandorkar:2021viw}, 
we conclude that the path in cross ratio space circles around the branch point located at ${\bar Z}=1$ in the counterclockwise direction.

On further increasing $\tau$, we now find that 
$Z$ and ${\bar Z}$ both decrease. At the special value 
$\tau = 2 T \cos\left(\frac{\theta}{2}\right)$,  $Z=0$.  At this points the insertions at  $x_2$ and $x_4$ respectively cross the lightcones of the insertions at $x_1$ and $x_3$. \footnote{At this value of $\tau$, the cross ratios become
\beq
Z = 0, \qquad \zbar = \sin^2 \theta
\eeq
}
Once again the point $Z=0$ is a branch point of \eqref{4dfunc}, and so the approach to this branch point
needs to be carefully resolved. Including the imaginary parts of the insertion locations as above, and proceeding exactly as in subsubsection 2.5.2 of \cite{Chandorkar:2021viw}, we conclude that our trajectory circles around the branch point at $Z=0$ in a counterclockwise manner. 

As $\tau$ is further increased, $Z$ now increases while 
${\bar Z}$ continues to decrease. At the final value 
$\tau=2T$
\beq
Z =  \zbar = \sin^2\left(\frac{\theta}{2}\right)
\eeq

In summary, one reaches the scattering configuration - starting from the Eucldiean configuration - by traversing counterclockwise around the branch point ${\bar Z}=1$ 
and also traversing counterclockwise around the branch point at $Z=0$. 

It is not difficult to apply these branch moves on the particular function at hand, namely \eqref{4dfunc}: after performing both monodromy moves (see Appendix B of \cite{Gary:2009ae}) we find 
\beq\label{ldfun}
  \mathcal{D}_{1111}=\frac{1}{Z-\zbar} \left(2 \text{Li}_2(Z)-2\text{Li}_2(\zbar) + 4 \pi i \ln \zbar+\left(\ln(Z\zbar)+ 2 \pi i\right) \left(\ln \left(\frac{1-Z}{1-\zbar}\right)- 2\pi i \right) \right)
\eeq
The monodromy around ${\bar Z}=1$ produces both the shift of $-2 \pi i$ in the last bracket on the RHS of \eqref{ldfun}, and the shift $4 \pi i \ln \zbar$, again in the RHS of \eqref{ldfun} (from the monodromy of the function $-2\text{Li}_2(\zbar)$. The monodromy around $Z=0$ produces the shift of $2\pi i$ in $\left(\ln(Z\zbar)+ 2 \pi i\right)$. 

Taking the limit $Z \rightarrow {\bar Z}$ in \eqref{ldfun}, we find that $D_{1111}$ on the scattering sheet now has a singularity at $Z={\bar Z}$. The precise structure of this singularity is 
\beq\label{ldfunfin}
  \mathcal{D}_{1111} \approx \frac{4 \pi^2}{Z-\zbar} 
\eeq

Using \eqref{4dgreen}, around the singularity full four point correlator in four dimensions is given by
\beq\label{exactfinal}
\gf(x_1, \ldots , x_4) \approx \frac{i g}{2^6 \pi^{4}}{1 \over N_{12} N_{34}}\frac{1}{Z-\zbar} 
\eeq
This exactly matches the expression \eqref{bsgfinal} computed in \S\ref{ceq1}.

\subsection{Relation between $G$ and $G_{\rm bdry}$ in the neighbourhood of the singular manifold} \label{relgbdy}

While dealing with massive particles we established in the previous section that 
\begin{equation}\label{gfmo}
G_{\rm bdry} = \left(\prod_i 2 i {\hat n_i}. \nabla_i \right) G.
\end{equation} 
In the case of the Dirichlet path integral, we can rework the analysis of Appendix \ref{apgrn} for the case of massless particles, it is easy to verify that \eqref{gfmo} does not, in general, apply to the case of massless particles (even for the specially simple choice of the cut off spacetime being the slab). In this section, we will demonstrate, however, that as we approach the bulk point singularity \eqref{gfmo} starts to hold.

Let us first focus on the case of the 'slab' boundary. In this case, the exact Greens function can be computed using the method of images (as computed in Appendix \ref{geuc}). The Greens function is given by an infinite sum given in \eqref{gfpb}. For the case of massless particles, all terms in this series are equally relevant (there is a power law suppression between various terms as compared to exponential suppression in the case of massive particles). As we approach the bulk point limit, the spacetime distance between the bulk point and the boundary point (the point just close to the boundary) and its first image outside the boundary, approaches zero. Hence these two terms in the infinite series diverges whereas all the other terms still remain finite. Hence in this limit, the divergent part of the boundary Greens function is given exactly by \eqref{gfmo} (of course \eqref{gfmo} 
does not correctly capture the terms that remain analytic
in the neighborhood of the scattering configuration). \footnote{One may ask what happens when the spacetime distance between other image charges (not the first one) and the bulk point diverge. It is possible that such a situation arises, giving rise to `spurious divergences' 
in the boundary to bulk Greens functions. Spurious divergences are those that occur at locations at which 
the  Greens function on $R^{D-1,1}$, $G$,  itself is finite. Such divergences appear to be unphysical; 
our prescription is simply to ignore them (i.e. not to 
assign them any physical meaning). }

The same result holds for the case of the in-out path integral follows 
even more simply from the observation that the discussion of Appendix \ref{digch} 
(see around \eqref{ndelz}) applied uniformly both to massive as well as massless scattering (ignoring possible issues related to IR divergences). 

Using the expression of the bulk two point function given in \eqref{bulkgr}, we obtain the following expression for bulk to boundary propagator between a boundary point $x$ and a bulk point $y$
\beq \label{ngpro}
G_{\partial B}(x , y) = \frac{\Gamma \left(\frac{D-2}{2}\right)}{4 \pi^{\frac{D}{2}}} \frac{2 (D-2) n.(x-y)}{((x-y)^2 )^{\frac{D}{2}}}.
\eeq
where $n$ is the unit vector normal to the boundary. Note that the bulk to boundary propagator and the bulk two point function differ by an overall constant and change in power of the denominator. We can repeat the analysis of the previous subsection by using $G_{\partial B}$ instead of $D(x,y)$. The first step is to rewrite the RHS of 
\eqref{ngpro} using Schwinger parameters: 
\begin{equation}\label{ngprosch}
\begin{split}
G_{\partial B}(x , y) &=  2 (D-2) n.(x-y)\ \frac{\Gamma \left(\frac{D-2}{2}\right)}{4 \pi^{\frac{D}{2}}} \ \frac{e^{-\frac{ i \pi (D-2)}{4} }}{\Gamma\left(\frac{D}{2}\right)} \int_0^\infty ~ d \omega ~ \omega^{\frac{D}{2}-1}e^{i\omega((x-y)^2 )} \\
& = n.(x-y)\frac{e^{-\frac{ i \pi (D-2)}{4} }}{\pi^{\frac{D}{2}}} \int_0^\infty ~ d \omega ~ \omega^{\frac{D}{2}-1}e^{i\omega((x-y)^2 )}
\end{split}
\end{equation} 
Stitching the propagators together via the interaction vertex, and proceeding as in subsubsection \ref{ceq1}, 
we find that in the case of the codimension one singularity ($c=1$), the leading singularity of the boundary to bulk Greens function 
is given by 

\beq\label{bsgf12}
\begin{split}
G_{\rm bdry} =   &\mathcal{C'}\frac{(\prod_\mu\zeta_{0,b}^\mu)^{\frac{D}{2}-1} \prod\limits_{i}n_i.(x_{i,b} - \tilde y)}{(\lambda_{0,b})^{ \frac{3D-11}{2}+ 2(m+n)+4}}  \prod_{i=1}^{3}\frac{1}{\sqrt{\lambda_{i,b}}} \ \left(\sum_\mu \zeta_{0_,b}^\mu\right)^{D+ 2(m+n)}\left( T(\tilde s, \tilde t)+ T(\tilde t, \tilde u)+ T(\tilde u, \tilde s) \right)
\end{split}
\eeq
where 
$$\mathcal{C'} = (-1)^{m+n}\frac{2^{-8 + \frac{3D}{2} + 2(m+n)}}{\pi^{\frac{3}{2}(D-1)}} \Gamma\left(\frac{3D-11}{2}+ 2(m+n)+4\right)$$
and the subscript $b$ reminds us that our insertion points are all located on the boundary manifold 
\footnote{This means that the coordinates that appear in  $N_{ij}$ are all on the boundary, appropriately constraining the eigenvalues and eigenvectors of $N$,  appear in  
\eqref{bsgf12}.}  Note that the principal effect of each of the  boundary derivatives,  $2 n_i.\nabla$,  is to increase the power of the singularity by unity.

As a check, we can reobtain \eqref{bsgf12} by directly evaluating $\prod_i (2 n_i. \nabla_i)$ of our final result for the leading singularity of Greens function $G$,  \eqref{bsgf}. The leading singularity in 
$\prod_i (2 n_i. \nabla_i) G$ is obtained when all the boundary derivatives act on the singular term, 
$$\frac{1}{\lambda_{0}^{\frac{3D-11}{2}+ 2(m+n)}}$$
in \eqref{bsgf}. We find 
\beq\label{bsgf22}
\begin{split}
G_{\rm bdry} =   &\mathcal{C''}\frac{(\prod_\mu\zeta_{0,b}^\mu)^{\frac{D-2}{2}-1}}{(\lambda_{0,b})^{ \frac{3D-11}{2}+ 2(m+n)+4}}  \prod_{i=1}^{3}\frac{1}{\sqrt{\lambda_{i,b}}} \ \left(\sum_\mu \zeta_{0,b}^\mu\right)^{D-4 + 2(m+n)}\left(\prod\limits_i n_i.\frac{\partial \lambda_0}{\partial x_i}\bigg|_{x_i = x_{i,b}}\right)\\
& \left( T(\tilde s, \tilde t)+ T(\tilde t, \tilde u)+ T(\tilde u, \tilde s) \right)
\end{split}
\eeq
where 
$$\mathcal{C''} = (-1)^{m+n}\frac{2^{-8 + \frac{3D}{2} + 2(m+n)}}{\pi^{\frac{3}{2}(D-1)}} \Gamma\left(\frac{3D-11}{2}+ 2(m+n)+4\right).$$

In order to evaluate the derivative of $\lambda_0$ with 
respect to $x_i$, we need to track how a particular eigenvalue of the matrix $N$ changes when we make an infinitesimal change in its matrix elements. This can be evaluated using the familiar quantum mechanical formulae of first order perturbation theory. We find
\begin{equation}
\begin{split}   \label{evchange}
\partial_{x_i^\mu} \lambda_0 &=
\sum_{m,n} \zeta_0^m   \partial_{x_i^\mu} N_{mn} \zeta_0^n\\
&= 4 \sum_{n} \zeta_0^{i}(x_i-x_n)^\mu \zeta_0^n = 4 \sum_{n} \zeta_0^{i}\left( (x_i-{\tilde y})^\mu-(x_n-{\tilde y})^\mu\right) \zeta_0^n \\
&= 4 \zeta_0^{i}\left( (x_i-\tilde y)^\mu\right) \left( \sum_{n} \zeta_0^n \right) 
\end{split}
\end{equation}
In going from the second to the third line of 
\eqref{evchange}, we have used the fact that when 
the external insertions are located on the singular 
manifold (and so to leading order in the neighborhood of the singular manifold)
\begin{equation}\label{atsp}
\sum\limits_{j}\zeta_0^j (x_j - \tilde y)^\mu = 0.
\end{equation} 
Dotting both sides of \eqref{evchange} with $2 n_i^\mu$ 
we obtain 
\beq
 n_i.\frac{\partial \lambda_0}{\partial x_i} = \left(4\zeta_0^i \ n_i.(x_i - \tilde y)\right)\left( \sum_{j} \zeta_0^j\right)
\eeq
It follows that 
\beq \label{lpro}
\prod_i n_i.\frac{\partial \lambda_0}{\partial x_i} = \prod_i \left(4\zeta_0^i \ n_i.(x_i - \tilde y)\right)\left( \sum_{j} \zeta_0^j\right)^4 
\eeq
Substituting \eqref{lpro} into \eqref{bsgf22} we obtain
\eqref{bsgf12}, completing our consistency check. 

\subsection{Comparison with \eqref{smatrixl}}

In \S \ref{sec:massless}, we derived a relationship between the boundary correlator and the bulk S-matrix involving the scattering of massless particles. We saw that the coefficient of leading singularity in the boundary correlator captures the bulk S-matrix. In this section, we discuss the consistency of \eqref{bsgf12} and \eqref{smatrixl}. 

\eqref{bsgf12} determines the singularity of boundary correlators in terms of bulk S matrices. On the other hand
\eqref{smatrixl}, can roughly be thought of as a sort of Fourier transform of the boundary correlator, and so  depends 
on the boundary correlators everywhere - and not just at its singularities. These two facts may, at first, appear to be in tension with the expectation that the Fourier transform \eqref{smatrixl} simply pick out the coefficient of the singularity in the boundary correlator. 

The resolution of this puzzle goes along the following lines.  The S matrix that we expect includes a full $D$ dimensional energy momentum conserving $\delta$ function. Focussing on the simple special case of a slab type boundary, we will now explain that while the the contribution to the coefficient of this $D$ dimensional delta function comes entirely (in the integral in \eqref{smatrixl} ) from the neighborhood of the singularities of $G_{\rm bdry}$. The regular part of $G_{\rm bdry}$ gives a less singular - hence subdominant - contribution to the Fourier transform in \eqref{smatrixl}.  

To see how this works, let us (as a mathematical exercise) perform the  Fourier transform in \eqref{smatrixl}, with $G_{\rm bdry}$ taken to be a smooth function. It follows from translational invariance that the result of the Fourier transform includes a $D-1$ dimensional momentum conserving delta function. However, the additional `energy conserving' $\delta$ function will, in general, be absent. This additional piece can only appear if $G_{\rm bdry}$ itself has singularities. It follows, in other words, that the coefficient of the full energy- momentum conserving $\delta$ function in $G_{\rm bdry}$ receives contributions only from the neighborhood of the singularities of $G_{\rm bdry}$. As the residue singularities are determined by the S matrix (see \eqref{bsgf12}),  the consistency between \eqref{bsgf12} and \eqref{smatrixl} may now seem less surprising. 

In order to see how the full $D$ dimensional delta function is produced from the singularity in $G_{\rm bdry}$
let us focus on the special case of four-particle scattering with $g \phi^4$ interaction in $D=4$.  In this case, the leading singularity of the boundary correlator is given by
\beq\label{consch}
\begin{split}
G_{\rm bdry} =   &\frac{1}{4\pi^{\frac{9}{2}}} \Gamma\left(\frac{9}{2}\right) \ \frac{(\prod_\mu\zeta_{0,b}^\mu) \prod\limits_{i}n_i.(x_{i,b} - \tilde y)}{(\lambda_{0,b})^{ \frac{9}{2}}}  \prod_{i=1}^{3}\frac{1}{\sqrt{\lambda_{i,b}}} \ \left(\sum_\mu \zeta_{0_,b}^\mu\right)^{4}\left( i g\right)
\end{split}
\eeq
`Fourier transforming' as instructed by \eqref{smatrixl}  we obtain
\beq\label{conschft}
\begin{split}
G_{\rm bdry} = \frac{1}{4\pi^{\frac{9}{2}}} \Gamma\left(\frac{9}{2}\right) & \int \prod_{\rm in} \left( dx_{i,b} e^{i p_i^{\rm in} . x_{i,b}} \right)\ \int \prod_{\rm out} \left( dx_{i,b} e^{-i p_i^{\rm out} . x_{i,b}} \right)\ \frac{(\prod_\mu\zeta_{0,b}^\mu) \prod\limits_{i}n_i.(x_{i,b} - \tilde y)}{(\lambda_{0,b})^{ \frac{9}{2}}}  \\
&\prod_{i=1}^{3}\frac{1}{\sqrt{\lambda_{i,b}}} \ \left(\sum_\mu \zeta_{0_,b}^\mu\right)^{4}\left( i g\right)
\end{split}
\eeq
The integrand becomes singular at $\lambda_{0,b} =0$, hence the integral receives a major contribution from the neighborhood of that region. Let us perform the $x_{i,b}$ integral using saddle point approximation. The saddle point for $x_{i,b}$ is given by the following equation
\beq
i \vec{p}^{\rm in}_i - \frac{9}{2 \lambda_{0,b}} \frac{\partial \lambda_{0,b}}{\partial x_{i,b}} - \frac{1}{2 \lambda_{i}} \frac{\partial \lambda_{i}}{\partial x_{i,b}} + \prod_i \frac{\partial (n_i.(x_{i,b}-\tilde y)}{\partial x_{i,b}} =0
\eeq
As the eigenvalue $\lambda_{0,b}$ is close to zero, this equation can be approximated as
\beq\label{sap}
i \vec{p}_i^{\rm in} \sim \frac{9}{2 \lambda_{0,b}} \frac{\partial \lambda_{0,b}}{\partial x_{i,b}} 
\eeq
Recall that the singularity (in the neighborhood of where $\lambda_{0,b} \to 0$) happens because our correlator receives large contributions from a scattering process that takes place at a unique bulk point - let us call it ${\tilde y}^\mu$. Using \eqref{evchange} we obtain a formula for the derivative of $\lambda_{0,b}$ w.r.t. $x^\mu$, in terms of the bulk point $y^\mu$, i.e. 
\beq\label{sadd}
i \vec{p}_i^{\rm in} \sim \frac{18}{\lambda_{0,b}} \zeta_0^i \left(\vec{ x}_{i,b} - \vec{\widetilde y}\right) \left(\sum_j \zeta_0^j\right)
\eeq
This equation tells us that our Fourier transform at momentum ${\vec p}_i^{\rm in}$ receives dominant contributions only from those points ${\vec x}_i$
that are on the singular manifold and are so located that 
$(x_i-y) \propto {\vec p}_i^{\rm in}$. A similar equation is valid for the outgoing momenta $\vec{p}_i^
{\rm out}$. Recall that the residue of the singularity 
on the singular manifold was given by the S matrices of scattering particles with momenta proportional to $(x_i-y)$. It is satisfying, therefore, the only singularities that contribute to \eqref{smatrixl}
at Fourier momentum ${\vec p}_i$ are those whose residues are the 
S matrices with momenta proportional to ${\vec p}_i$. These saddle points occur at boundary locations given, in terms of the bulk scattering point $y^\mu$, by 
\beq \label{xi}
\vec{x}_{i,b} = \frac{i}{\zeta_0^i} \frac{\lambda_{0,b}}{18}\frac{\vec{p}_i^{\rm in}}{\sum_j \zeta_0^j} + \vec{\tilde y}
\eeq

How many such boundary locations satisfy the condition described above? The answer to this question is very clear. Consider any bulk point 
$y$. A scattering process with momenta proportional to  ${\vec p_i}$
leads us to a unique collection of boundary points. It follows that every choice of the bulk point $y$ (and associated boundary points) gives a solution to \eqref{sadd}. In order to find the full saddle point 
contribution to \eqref{smatrixl} at momenta ${\vec p}_i$, we simply need to integrate over all bulk points $y^\mu$. 

The saddle point contribution corresponding to any given 
$y$ can easily be verified to be  
$$e^{i \sum_i \omega_i T-  \sum_i\frac{\vec{p_i}^2}{\zeta_0^i}\frac{\lambda_{0,b}}{18}\frac{1}{\sum_j \zeta_0^j}+ i \left(\sum_i \vec{p}_i^{\rm in}-\sum_i \vec{p}_i^{\rm out}\right). \vec{\tilde y}} $$
Let us denote the time part of $y^\mu$ by $t$ and the spatial part of the vector $y^\mu$ by ${\vec y}$.  It follows that the integral over the manifold of saddle points gives 
\beq\label{no}
\int dt \ d \vec{\tilde y} \ e^{i \sum_i \omega_i T-  \sum_i\frac{\vec{p_i}^2}{\zeta_0^i}\frac{\lambda_{0,b}}{18}\frac{1}{\sum_j \zeta_0^j}+ i \left(\sum_i \vec{p}_i^{\rm in}-\sum_i \vec{p}_i^{\rm out}\right). \vec{\tilde y}}
\eeq
 The $\vec{\tilde y}$ integral gives the spatial momentum conservation i.e.
\beq\label{no1}
\int dt  \ e^{i \sum_i \omega_i T-  \sum_i\frac{\vec{p_i}^2}{\zeta_0^i}\frac{\lambda_{0,b}}{18}\frac{1}{\sum_j \zeta_0^j}}\ \delta^D\left(\sum_i \vec{p}_i^{\rm in}-\sum_i \vec{p}_i^{\rm out}\right)
\eeq
 Using the fact that $(x_{i,b} - \tilde y)^2 = 0$, $p_i^2=0$,  and \eqref{xi}, we see that 
\beq \label{pisq}
\begin{split}
(\vec{p}_i^{\rm in})^2=-(T + t)^2\left(\frac{18}{\lambda_{0,b}} \zeta_0^i \right)^2\left(\sum_j \zeta_0^j\right)^2, &~~~~\omega_i^{\rm in}= i(T + t)\left(\frac{18}{\lambda_{0,b}} \zeta_0^i \right)\left(\sum_j \zeta_0^j\right)\\
(\vec{p}_i^{\rm out})^2=-(T - t)^2\left(\frac{18}{\lambda_{0,b}} \zeta_0^i \right)^2\left(\sum_j \zeta_0^j\right)^2, &~~~~\omega_i^{\rm out}= i(T - t)\left(\frac{18}{\lambda_{0,b}} \zeta_0^i \right)\left(\sum_j \zeta_0^j\right)
\end{split}
\eeq 
Inserting \eqref{pisq} into \eqref{no} we obtain 
\beq
\int dt  \ \exp{\left( \sum_i \left(-(T+t)T + (T+t)^2\right)\left(\frac{18}{\lambda_{0,b}} \zeta_0^i \right)\left(\sum_j \zeta_0^j\right) +  (p^{\rm out}~ {\rm terms})\right) }\ \delta^D\left(\sum_i \vec{p}_i^{\rm in}-\sum_i \vec{p}_i^{\rm out}\right)
\eeq
The quadratic terms in $T$ cancel and using \eqref{pisq}, the exponent can be expressed in terms of $\omega_i$. After performing $t$ integral, we obtain
\beq
 \delta\left(\sum_i \omega_i^{\rm in} - \sum_i \omega_i^{\rm out}\right)\delta^D\left(\sum_i \vec{p}_i^{\rm in}-\sum_i \vec{p}_i^{\rm out}\right)
\eeq

We see that the Fourier transform over the singularities in $G_{\rm bdry}$, at momentum ${\vec p}_i$, give us a spacetime energy momentum conserving delta function, whose coefficient is also proportional to the 
S matrix with scattering momenta ${\vec p}_i$. We have not kept track of factors of order unity, and so have not verified that the proportionality constant for this formula is unity, as we expect. We leave a careful verification of this expectation to the interested reader.

\section{Relation with Celestial CFT} \label{ccft}

Consider Minkowski space cut off on the `slab', i.e. 
at $t=T$ and $t=-T$. Consider a correlator involving two 
insertions on the future boundary and two insertions 
on the past boundary. As we have explained earlier 
in this paper, boundary correlators are singular 
on a surface of codimension unity (the equation for this 
surface is determined by the condition $det (N)=0$). 
The leading singularity of the correlator in the neighborhood of any point on this `singular manifold' 
is completely determined by a scattering process that 
occurs at a particular bulk point ${\tilde y}^\mu$. 

Given this situation, it is natural to ask the following 
question. If we change the cutoff surface (i.e. change the value of $T$), how must we change our boundary insertion points so as to ensure that the correlator at the new insertion points is also singular, and that the singularity is captured by the same bulk scattering 
process.

The answer to the question described above is easily given. Let us choose our spatial coordinates, so that 
the scattering process we wish to track occurs at 
the spatial origin, but at time $u$. The lightrays 
that emerge out of our scattering process then always 
live at a fixed angle on the boundary $S^{D-2}$. The part of the future lightcone emerging from this scattering point, that is at radius $r$,  lives at 
time $t$ where
\begin{equation}\label{flcr}
t-u= r
\end{equation} 
On the other hand, the part of the past lightcone at radius $r$ lives at 
\begin{equation}\label{plc} 
u-t= r
\end{equation}
It follows that if we wish to track this particular 
scattering process as we change $T$, we must scale the 
radius of insertions on the future/past boundary of the slab 
so that 
\begin{equation}\label{fbsscale}
r= T-u~~~~{\rm future},\qquad  r=u+T, ~~~~{\rm past}
\end{equation}
The fact that we are working at fixed $u/v$ as $T$, 
hence $r \to \infty$, tells us that our boundary effectively goes to ${\mathcal I}^\pm$ as $T \to \infty$
\footnote{This can be seen as follows. Because we want to work at fixed $u/v$, it is convenient to change the coordinates 
from $(t, r)$ to $(u, r)$ ($u=t-r$) in the neighbourhood 
of the future boundary, and to the coordinates $(v, r)$
($v=t+r$) in the neighborhood of the past boundary. 
Once we have made this change of coordinates, we work 
at a fixed value of $u$ on the future boundary and at a
fixed value of $v$ on the past boundary. 
The metric in these two choices of coordinates is given by 
\begin{equation}\label{metuv} 
\begin{split} 
&ds^2=-du^2+ 2 du dr + r^2 d \omega_{D-2}^2\\
&ds^2=-dv^2 -2 dv dr + r^2 d \omega_{D-2}^2\\
\end{split}
\end{equation} 
As $r \to \infty$ this yields the standard metric in the neighbourhood of ${\mathcal I}^\pm$.} 

Working in this way, it is not difficult to determine 
the precise location of the singular manifold. Let 
the two future and two past boundary points be located 
at (we use Cartesian coordinates $(t, {\vec x})$ in our listing)
\begin{equation}\label{fourbund}
\begin{split} 
x_1 =&(T, (T-u_1){\hat n}_1)\\
x_2 =& (T, (T-u_2){\hat n}_2)\\
x_3 =& (-T, (v_3+T){\hat n}_3)\\
x_4 =& (-T, (v_4+T){\hat n}_4)\\
\end{split}
\end{equation} 
In the limit $T\to \infty$ with $u_i$ and $v_i$ held 
fixed we find that the conformal cross ratios 
$u$ and $v$ (see \eqref{cratio} for definitions) are given by 
\begin{equation}\label{uvval}
u = \frac{(1- \hat n_1 .\hat n_3)(1- \hat n_2 .\hat n_4)}{(1- \hat n_1 .\hat n_2)(1- \hat n_3 .\hat n_4)}~~, \qquad \qquad v = \frac{(1- \hat n_2 .\hat n_3)(1- \hat n_1 .\hat n_4)}{(1- \hat n_1 .\hat n_2)(1- \hat n_3 .\hat n_4)}.
\end{equation} 

The striking aspect of \eqref{uvval} is that $u$ and 
$v$ - hence the cross ratios $Z$ and ${\bar Z}$ - 
all depend only on the angular locations of our insertion points and are completely independent of $u_i$ and $v_i$. It follows that the singular manifold is $R^4$ 
$\times$ codimension one surface in $\left(S^{D-2} \right)^4$. \footnote{The $R^4$ is parameterized by 
$u_1$, $u_2$, $v_3$ and $v_4$.}. 

Let us now turn to the precise nature of the singular 
manifold on $\left(S^{D-2} \right)^4$. The equations 
\eqref{uvval} are precisely the formulae for the 
conformal cross ratios for four insertions on $S^{D-2}$ in a $D-2$ dimensional CFT. It follows that the singular
manifold is given precisely by the locations of four 
insertions on $S^{D-2}$ that obey the condition 
$Z={\bar Z}$. 

This condition can be made completely explicit in the 
case of $D=4$. In this case, we can parameterize our three dimensional unit vectors as 
$${\hat n_i}= \left(\frac{\bar{z}_i+z_i}{1+z_i \bar{z}_i},-\frac{i \left(z_i-\bar{z}_i\right)}{1+z_i \bar{z}_i},\frac{1-z_i \bar{z}_i}{1+z_i \bar{z}_i}\right)$$
With this parameterization we find the standard result for $Z$ and ${\bar Z}$ in two dimensional CFTs,  
\begin{equation}\label{scr}
    \begin{split}
        Z&= \frac{\left(z_1-z_2\right)\left(z_3-z_4\right)}{\left(z_1-z_3\right) \left(z_2-z_4\right)}+\mathcal{O}\left(\frac{1}{T}\right) \qquad\qquad \bar Z= \frac{\left(\bar z_1-\bar z_2\right)\left(\bar z_3-\bar z_4\right)}{\left(\bar z_1-\bar z_3\right) \left(\bar z_2-\bar z_4\right)}+\mathcal{O}\left(\frac{1}{T}\right)
    \end{split}
\end{equation}
and the equation for the singular manifold is completely 
explicit. 

Clearly, the structures described above have many similarities with those encountered in the study of the 
so called `Celestial Holography' programme \cite{Pasterski:2021rjz,Raclariu:2021zjz}. There is, 
however, one apparently striking point of difference. 
The bulk (and boundary) correlators encountered in our work are analytic functions on the future and past boundaries (in the $T \to \infty$ limit these boundaries approach ${\mathcal I}^\pm$). These analytic functions develop pole type singularities on the singular manifold $Z={\bar Z}$ described above. In this sense, the analytic structure of our correlators is very similar to 
those of conformal correlators (encountered in the study 
of the usual AdS/CFT correspondence) in the neighborhood of bulk point singularities \cite{maldacena2017looking}. 

In contrast, it is our understanding that the correlators encountered in the study of the Celestial holography programme are claimed to be strictly vanishing 
away from $Z={\bar Z}$. As in our study, the values 
of the celestial correlators at $Z={\bar Z}$ capture 
bulk S matrices. Atleast at first sight it would appear 
that the celestial correlators are more closely analogous to the imaginary part (hence singularities) of the correlators of this paper rather than our correlators themselves. It would be very interesting to better understand this connection. We leave this to future work. 
The discussion of this subsection can be illustrated 
in the exactly solvable example of a contact interaction in $d=4$.

As a concrete example of the discussion above, consider the  bulk Greens function for $g \phi^4$ interaction in 4d is given exactly by (see \eqref{exactfinal})
\beq\label{ccft1}
\gf(x_1, \ldots , x_4) \approx \frac{i g}{2^6 \pi^{4}}{1 \over N_{12} N_{34}}\frac{1}{Z-\zbar} 
\eeq
 Using \eqref{scr} in the above equation, we obtain
 \begin{equation} \label{4dgreen3}
\gf(x_1, \ldots , x_4) \approx \frac{i g}{2^6 \pi^{4}}{1 \over N_{12} N_{34}}\frac{1}{z-\bar z} 
\end{equation}
where $z$ and $\bar z $ are the cross ratios on celestial sphere i.e.
\beq
z= \frac{\left(z_1-z_2\right)\left(z_3-z_4\right)}{\left(z_1-z_3\right) \left(z_2-z_4\right)}, \qquad \qquad \bar z = \frac{\left(\bar z_1-\bar z_2\right)\left(\bar z_3-\bar z_4\right)}{\left(\bar z_1-\bar z_3\right) \left(\bar z_2-\bar z_4\right)}
\eeq
We note that our correlator is indeed an analytic function (as claimed above): moreover its singularities are of the pole type, and occur at $z={\bar z}$, as anticipated above.

\section{Discussion}

In this paper, we have investigated the AFS prescription for obtaining 
S matrices from an in-out path integral, and have also we have presented an alternative framework for studying the S-matrix in terms of a Dirichlet path integral. While we believe that the analysis presented in this paper applies in some generality,  most of the concrete computations we have performed have been at tree level, and especially for contact interactions (see however Appendix \ref{dflat1}). It is possible that new complications that we have not foreseen will occur at the loop level (this is almost certain to be the case in massless theories that have infrared divergences). It would thus be useful to perform some explicit computations of loop contributions to the path integral as a functional of boundary values defined in this paper. 

For massless particles, at tree level,  the dependence on the size of the cut-off surface,  of the path integral as a functional of boundary values, is 
(diagram by diagram) a simple power law determined by dimensional analysis. At the loop level, however, this dimensional analysis is typically corrected by UV renormalization as well as nontrivial IR effects. It would be interesting to investigate the impression that nontrivial $\beta$ functions and infrared divergences leave on our action as a functional of boundary values.

IR issues are maximally severe in conformal field theories. Paradoxically, scale invariance ensures that the power law dependence of the action as a functional of boundary values is exact in such theories. This fact suggests that the observable studied in this paper develops a particularly simple structure in the case of conformal field theories. In this case, it will not be possible to relate the path integral as a functional of boundary values to a traditional S matrix, as traditional S matrices are believed not to exist in 
conformal field theories. Perhaps there is a useful sense in which the path integral as a functional of boundary values is the well defined CFT observable that replaces the S matrix in conformal theories (this would be particularly interesting if the unitarity of this observable could be expressed in a useful form, perhaps an interesting deformation of our equation \eqref{unitsm})\footnote{We note, in this context, that even though S matrices are supposed not to exist in CFTs, they are routinely computed in CFTs like ${\cal N}=4$ Yang-Mills theory.}. We leave further investigation of the provocative suggestion of this paragraph to the future. 

 In the massive case, the correlation functions are dominated by a bulk saddle point corresponding to bulk scattering in a small elevator region. The location of this elevator region is dynamically determined by the boundary points. The boundary correlation function diverges in the asymptotic limit, and suitable ``holographic renormalization'' is necessary. This happens in AdS as well, but there is an important difference: in AdS, the renormalization factors are local from the boundary point of view, while in the case of flat spacetime, we find that the renormalization factors are dynamically determined, and as such, depend on the location of all the boundary points. In the language of the extrapolate dictionary, holographic renormalization in flat spacetime works ``mode-by-mode'' in momentum space:\footnote{In contrast, in AdS the renormalization is local $\phi_{\vec{k}} (z)\sim z^{\Delta}\,O_{\vec{k}}$, i.e., does not depend on $\vec{k}$.}
\beq 
\phi_{\vec{k}}(T) \sim e^{-i\omega_{\vec{k}}T}\,O_{\vec{k}},\;\;\;\cdots \;\;\;(\omega_{\vec{k}} = \sqrt{\vec{k}^2 + m^2}).
\eeq 
Nevertheless, once we strip off these dynamically determined renormalization factors, the boundary correlation function reduces to the S-matrix, much like in AdS \cite{Komatsu:2020sag}. The situation is slightly different for massless fields: in this case, the boundary correlation functions have singularities at special loci corresponding to bulk light-like scattering, and the coefficients of these singularities are given by the bulk S-matrix. Thus, we land on a beautiful analytic structure of correlation functions on the boundary, albeit one that has hints of non-locality in it. Perhaps, this is appropriate for flat space holography: in AdS, the gravitational potential of AdS keeps localized wavepackets from spreading out too much in the asymptotic region, while in flat spacetime such a spread is inevitable.  It may be that the structure we arrive at on the boundary is an inherently non-local one (see also \cite{Marolf:2006bk, Li:2010dr} for other reasons to believe this). This may sound worrying, because much of what we know about quantum field theory lies within the realm of local QFT. However, at least in low enough dimensions, non-local QFTs can be approached in a controlled way by flowing along irrelevant directions, such as is the case in $T\overline{T}$-deformed conformal field theories. Indeed, one could optimistically hope that the structure of boundary correlation functions we have landed on lies on some such controlled, irrelevant trajectory starting from a CFT. 
The well-studied program of celestial holography is one approach which attempts to circumvent the above trouble by defining the boundary theory in one-lower dimension (i.e., $\text{Mink}_{d+1}/\text{QFT}_{d-1}$) than would be natural in the AdS context. At least in the massless case, the idea is to take bulk operators to future null infinity. This, per se, does not help with the problem of ``mode-by-mode'' holographic renormalization. The essential idea is that since the renormalization factor depends on $\omega$, we should regard it not as momentum/energy in the dual theory, but as an independent quantum number. This then reduces the number of spacetime dimensions in the dual theory to $(d-1)$ rather than $d$. Having done this, one gets a more standard, ``local'' holographic renormalization, but the price to pay is that now in the structure one has defined, local operators are labeled (in addition to their positions on the celestial sphere) by a real-valued, internal quantum number.  We could take this approach in our case, and define boundary operators as 
$$ O^{(\omega)}(\hat{n})\sim \lim_{T\to \infty}\,e^{i\omega T}\,\phi_{\vec{k} = \omega\,\hat{n}},$$
where $\hat{n}$ is a point on the celestial sphere, and now $\omega = |\vec{k}|$ is regarded as an internal quantum number. From this point of view, the path-integral as a function of fixed boundary values turns into a generating functional for boundary correlation functions of the operators $O^{(\omega)}(\hat{n})$. It would be a good check to reproduce the correlation functions known in celestial holography from the path-integral with fixed boundary values. However, it is not clear to us whether the resulting celestial structure fits with the realm of local QFT.

In this work, we have focused our attention on scalar fields in the bulk. It would be particularly interesting to generalize our construction to include dynamical gravity in the bulk. In this case, one would want to fix the non-normalizable mode of the graviton on the asymptotic boundaries, or equivalently, fix the induced metric $h_{ij}$ on the boundary. In principle, this allows one to define a stress tensor in the dual theory and compute its boundary correlation functions. The partition function $Z[h_{ij}]$ as a function of the induced metric must satisfy the Wheeler-de Witt constraint equation, which we expect to translate to a Ward identity on the stress tensor in the dual theory. It would be interesting to explore whether this Ward identity admits solutions within QFT. 

While our discussion was entirely focused on quantum field theory on a fixed, non-dynamical Minkowski spacetime, we view this work as an attempt towards understanding the holographic principle in asymptotically flat spacetimes. Whether lessons from the AdS/CFT correspondence can be imported to asymptotically flat and de Sitter spacetimes is an interesting open question. In the absence of a top-down construction coming from string theory, it seems natural to attempt to imitate the rules of the AdS/CFT dictionary in the context of asymptotically flat/de Sitter spacetimes, and hope that this gives rise to an elegant structure on the asymptotic boundaries. In the case of asymptotically AdS spacetimes, of course, the dictionary famously gives rise to the structure of a local, conformally invariant quantum field theory on the asymptotic boundary of AdS. For correlation functions, the AdS/CFT dictionary comes in two forms: the Gubser-Klebanov-Polyakov-Witten (GKPW) dictionary \cite{Gubser:1998bc,Witten:1998qj},  and the extrapolate dictionary \cite{Harlow:2011ke}. The GKPW dictionary instructs us to fix the boundary values of the non-normalizable bulk modes, and then differentiate with respect to these boundary values to obtain boundary correlation functions. In the AdS/CFT context, these boundary correlation functions turn out to be the correlation functions of local operators in a dual CFT ``living'' on the boundary. On the other hand, the extrapolate dictionary instructs us to take the boundary limit of bulk correlation functions, and after stripping off a suitable renormalization factor, this once again reproduces boundary CFT correlation functions. In exporting these dictionary elements to flat/de Sitter spacetimes, one could thus attempt to mimic the above instructions to produce a set of boundary correlation functions on the corresponding asymptotic boundaries. In this work, we have tried to explore the analog of the GKPW dictionary in flat spacetime. We have studied two different path integrals - the in-out and the Dirichlet path integrals, both of which have some points of 
similarity with the AdS path integral as a function of boundary values. We leave the extension of the above framework to de Sitter spacetime to future works.

Another motivation for this work, from a purely QFT point of view, is that the Euclidean path-integral with fixed boundary values is a natural object to consider if one is interested in the spatial entanglement structure of states in the QFT. For instance, the wave function of the vacuum state of the QFT is an object of this type, and the entanglement properties of interest can then be obtained from the replica trick, which involves gluing copies of this path integral in various ways. An explicit map between the S-matrix and the Euclidean path integral with fixed boundary values could thus conceivably lead to a formula for the spatial entanglement/R\'enyi entropies of the vacuum state in terms of the S-matrix of the QFT. 

It is natural to wonder whether several outstanding questions relating 
to the S matrix can be addressed more efficiently in the framework of the current paper. Could, for instance, the construction of this paper (with timelike boundaries) lead to a more intuitive understanding of 
crossing invariance \cite{Bros:1965kbd}? Could either of the path integrals studied in this paper lead to a clearer understanding of the CRG conjecture \cite{Camanho:2014apa,  Chowdhury:2019kaq, Chandorkar:2021viw}? Could the framework described in this paper be used to more clearly understand the strange crossing symmetry properties of S matrices in theories with anyons \cite{Mehta:2022lgq}?  We leave all these exciting sounding questions to future work. 

Finally, we have emphasized in this paper that, especially when dealing with massless particles,  the path integral as a functional of boundary values appears to contain more information than simply the S matrix (recall that the Taylor coefficients of the path integral are analytic functions with poles, and the residues of these poles are the S matrices). If the path integral constructed in this paper is really 
a natural object, it should be possible to construct it (in the case of gravitational scattering) directly from the string world sheet. 
In order to do this we would have to learn to do string theory in 
target spacetimes with a boundary, a problem of interest from several 
other points of view as well (see e.g. \cite{Kraus:2002cb}). Any progress in this direction would be of tremendous interest. 

\section*{Acknowledgments}

We would like to thank S. Bhatkar, A. Gadde, A. Laddha, G. Mandal, P. Mitra, C. Patel, V. Talasila, and S. Trivedi for very useful discussions. We would also like to thank P. Mitra,  B. Van Rees, S. Raju, and especially P. Kraus and R. Myers, for comments on the manuscript. The work of all authors was supported by the Infosys Endowment for the study of the Quantum Structure of Spacetime. The work of D.J., S.M., and S.G.P. is partially supported by the J C Bose Fellowship JCB/2019/000052. S.G.P. acknowledges support by SERB under the grant PDF/2021/004777. We would all also like to acknowledge our debt to the people of India for their steady support to the study of the basic sciences.

\appendix
\section*{Appendix}

\section{Euclidean action as a function of boundary values in perturbation theory} \label{eapt}  

In this Appendix, we explicitly compute the `path integral as a function of 
boundary values' to a few orders in perturbation theory for the theory of a massive 
scalar field interacting with a $\phi^4$ interaction. In the context of this example, we also explore the connection between this object (path integral as a function of boundary values), Euclidean Greens function in flat space, and the S matrix. 

Consider the theory of a real scalar field governed by the Euclidean action 
\begin{equation}\label{actform} 
S= \int d^{D} x 
\left( \frac{(\partial \phi)^2}{2} +  m^2 \frac{\phi^2}{2} + g^2 \frac{\phi^4}{4}  \right) 
\end{equation}
on a Euclidean space with a boundary, such that the length scale $T$ associated with the boundary satisfies $m T \gg 1$. Let $x$ denote coordinates in the bulk, and $y$
denote coordinates on the boundary. The boundary surface is given by 
$x= x(y)$. 
\footnote{For instance, one concrete choice would be to choose our region to be a spatial solid ball $B^{D-1}$ of radius $R$ times an interval of length $2T$. The boundary of our spacetime is then $S^{D-2} \times I$ where the sphere has radius $R$ and the interval $I$ has length $2T$, together with two `flat caps' which are each $B^{D-1}$ (solid balls of radius $R$). On the cylinder, $y$ can be taken to be the Cartesian time coordinate $t$ on $R$ together with a set of $D-2$ angles $\theta^j$ on $S^{D-2}$, in terms of which 
\begin{equation}\label{bulboundcoord} 
(t, x^i)= \left(t, r {\hat n}^i(\theta^j)  \right) 
\end{equation} 
where ${\hat n}$ is the unit vector in $R^{D-1}$, parameterized by angles on $S^{D-2}$. On the flat caps, the $y$ can be chosen to be radius and angles.
One special case of this space is obtained by taking $R$ to infinity at finite $T$.} 
We wish to evaluate Euclidean action as a functional of the given boundary value of $\phi$, $\beta(y)$, in a perturbative expansion in the parameter $g^2$. 

\subsection{The bulk to bulk  Greens Function} 

In this section, we will remind the reader of some completely standard manipulations involving two point functions. Most of these manipulations and their derivations can be found in introductory texts on electromagnetism, for instance in the text of Jackson. 

Consider the `bulk to bulk'  Greens function $G(x_1,x_2)$ defined by the equations 
\begin{equation}\label{ugf} 
\left(\nabla_2^2 -m^2 \right) G(x_1,x_2)= \delta^D (x-y), ~~~~~~G(x_1, x_2(y))=0\\
\end{equation} 
where $D$ is the bulk spacetime dimension.
The fact that \eqref{ugf} defines a unique Greens function may be seen as follows. 
Let the difference between any two Greens functions that obey \eqref{ugf} be $A(x_1, x_2)$. Clearly \begin{equation}\label{ugfa} 
\left(\nabla_2^2 -m^2 \right) A(x_1,x_2)= 0, ~~~~~~A(x_1, x_2(y))=0\\
\end{equation} 
Now 
\begin{equation}\label{gfu} 
\int d^D {x_2} \left( \nabla_2 A(x_1, x_2) \right)^2 
= \int d^D {x_2} \left(\nabla \left( A(x_1, x_2) \nabla A(x_1, x_2) \right)  -A(x_1, x_2) \nabla_2^2 A(x_1, x_2) \right)
\end{equation} 
The RHS of \eqref{gfu} vanishes. \footnote{The second term vanishes by the equation of motion - the first of \eqref{ugfa}, while the first term is a total derivative, hence a boundary term, which in turn vanishes because of the second of \eqref{ugfa}.} 
As the integrad of the LHS of \eqref{gfu} is pointwise positive, it follows that  $A(x_1, x_2)$ must be constant, and hence (given the second of \eqref{ugfa}) vanishes. 

The Greens function defined in \eqref{ugf} is also symmetric in its arguments. To see this we note that 
\begin{equation}\label{gfsym} 
\nabla_3 \left( \nabla_3 G(x_1, x_3) G(x_2, x_3) -   G(x_1, x_3) \nabla_3 G(x_2, x_3) \right) = \delta(x_1-x_3) G(x_2, x_3) -  G(x_1, x_3) \delta(x_2-x_3) 
\end{equation} 
(where we have used the first of \eqref{ugf}). We now integrate both sides of the equation over $x_3$. The LHS is a surface term that vanishes using the second of \eqref{ugf}. The RHS evaluates to 
$$G(x_2,x_1)-G(x_1, x_2)$$
It follows that $G$ is symmetric in its two arguments. 

\subsection{The solution at leading order in coupling} 

We now turn to the problem of determining the solution of the 
equations of motion that follow from \eqref{actform}, subject to the boundary conditions 
\begin{equation}\label{boundcond} 
\phi(x(y))= \beta(y)
\end{equation} 
We expand 
\begin{equation}\label{expphi}
\phi= \phi_0+ g^2 \phi_1 + g^4 \phi_2 \ldots 
\end{equation} 
where 
\begin{equation}\label{boundcondcomp} 
\phi_0(x(y))= \beta(y), ~~~~\phi_1(x(y))=0, ~~~~\phi_2(x(y))=0, \ldots 
\end{equation} 
The equations of motion obeyed by these fields are 
\begin{equation}\label{eom} 
\left( -\nabla^2+m^2 \right)  \phi_0(x)= 0, ~~~~\left( -\nabla^2+m^2 \right) \phi_1(x)=-\phi_0^3, ~~~~
\left( -\nabla^2+m^2 \right) \phi_2(x)= -3 \phi_1 \phi_0^2, \ldots 
\end{equation} 
The first of these equations is easily solved using the identity 
\begin{equation}\label{gfsym} 
\nabla' .\left( \nabla' G(x, x') \phi_0(x') -   G(x, x') \nabla' \phi_0(x) \right) = \delta(x-x') \phi_0(x') 
\end{equation} 
(where we have used \eqref{eom} and the first of \eqref{ugf}).
Integrating both sides over $x'$, using the second of \eqref{ugf} and the boundary conditions \eqref{boundcond}, we obtain 
\begin{equation}\label{solnofeq0} 
\phi_0(x)= \int_{\mathcal {B}} dA ~{\hat n}'. \nabla' G(x, x'(y))  \beta(y)
\end{equation} 
where $dA$ is the area element on the boundary of our spacetime, 
${\hat n}$ is the outward pointing normal to the boundary.

Let us define the  bulk to boundary propagator by the equation  (`extrapolate limit' 
in the language of AdS/CFT)
\begin{equation}\label{bbprop} 
G_{\partial B}(x,x')= {\hat n}'.\nabla'G(x, x'). 
\end{equation} 
The first argument in \eqref{bbprop} lies on the boundary of our spacetime, while the second is an arbitrary bulk point.  
\eqref{solnofeq0} can be rewritten using the bulk to boundary propagator as  
\begin{equation}\label{solnofeq} 
\phi_0(x)= \int_{\mathcal{B}} ~G_{\partial B} (x, x'(y)) ~ \beta(y)
\end{equation} 
Specializing to the case the bulk point $x$ in \eqref{solnofeq} is taken to be a boundary, we see that $G_{\partial B}(x, x')$ reduces to the boundary delta function when $x$ approaches the boundary 

The action corresponding to this solution is given by 
\begin{equation} \label{actioncorsol} 
S_0= \int \frac{ (\nabla \phi_0)^2 + m^2 \phi_0^2}{2} 
= \int \nabla \left( \frac{\nabla \phi_0 \phi_0}{2} \right) =
\int dA_1 dA_2  ~ \frac{ \beta(y_1)~ ~\left( {\hat n}. \nabla_1 {\hat n}.{\nabla}_2 G(x(y_1), x(y_2)) \right) ~~\beta(y_2) }{2} 
\end{equation} 
In going from the second to the third expression in \eqref{actioncorsol}, we have used the fact that $\phi_0$ obeys the free equation of motion. Note that the symmetry between points $y_1$ and $y_2$, which was obscured in the middle expression in \eqref{actioncorsol}, is restored in the final answer.

In terms of the  boundary to boundary propagator defined by  
\begin{equation}\label{boundbound} 
G_{\partial \partial}(y_1, y_2)=({\hat n}_1. \nabla_1 )({\hat n}_1 . \nabla_2) G(y_1, y_2)
\end{equation} 
(where $y_1$ and $y_2$ are points on the boundary, $S_0$ can be 
rewritten as 
\begin{equation} \label{actioncorsolBB} 
S_0= 
\int~ \frac{ \beta(y_1)~ G_{\partial \partial}(y_1, y_2) ~\beta(y_2)}{2} 
\end{equation} 
 
Our knowledge of the leading order solution $\phi_0$ allows us to evaluate the action to first order in $g^2$ (even without evaluating the first correction to the solution $\phi_1$).  At order $g^2$, the action appears to receive two contributions; first from the quartic term evaluated on the zero order solution and second from the quadratic term evaluated on $\phi_0+g^2 \phi_1$ keeping only terms linear in $\phi_1$. However, the change in the quadratic part of the action vanishes at this order
(because we are expanding the action, to first order, about 
a solution to the equations of motion of the quadratic action, and because the modified solution obeys the same boundary conditions as the zero order solution). It follows that 
\begin{equation}\label{actone}
\begin{split}
S_1&= \int d^D x_5  \frac{ \phi^4_0(x_5)}{4}\\
&= \frac{1}{4}\int d^D x_5  \int dA_1 dA_2 dA_3 dA_4~G_{\partial B} (x_5, y_1)G_{\partial B} (x_5, y_2)G_{\partial B} (x_5, y_3)G_{\partial B} (x_5, y_4) ~ \beta(y_1)\beta(y_2)\beta(y_3)\beta(y_4) 
\end{split}
\end{equation} 
 
Note that 
\begin{equation}\label{fdoso} 
\begin{split} 
\frac{\delta^4 S_1}{\delta \beta(y_1)\delta \beta(y_2)\delta \beta(y_3) \delta \beta(y_4)} 
&= \frac{4!}{4} \int d^D x_5G_{\partial B} (x_5, y_1)G_{\partial B} (x_5, y_2)G_{\partial B} (x_5, y_3)G_{\partial B} (x_5, y_4)\\
&\frac{4!}{4} \int d^D x_5  n_1. \nabla_1 n_2. \nabla_2 n_3. \nabla_3 n_4. \nabla_4 G (x_5, y_1) G (x_5, y_2)G (x_5, y_3)G (x_5, y_4)\\
&=  n_1. \nabla_1 n_2. \nabla_2 n_3. \nabla_3 n_4. \nabla_4 
G^{(1)}_{\rm con}(x_1,x_2, x_3, x_4) \big|_{x_i=y_i} 
\end{split} 
\end{equation} 
where the bulk Greens function is defined in general by   
\begin{equation}\label{gdef} 
{G}_{\rm con} (x_1, \ldots x_n )= \langle \phi(x_1) \ldots  \phi(x_n) \rangle_{\rm Connected} ,
\end{equation}
where $x_i$ are bulk points and the expectation value in \eqref{gdef}  is computed subject to the condition that $\phi$ vanishes on the boundary of our spacetime. In \eqref{fdoso}, $G^{(1)}_{\rm con}(x_1,x_2, x_3, x_4)$ is the coefficient of $g^2$ in the expansion of the Greens function ${G}_{\rm con}(x_1, x_2, x_3, x_4)$ in a power series expansion in  $g^2$. 

\subsection{The solution at first subleading order} 

The solution of the second of \eqref{eom} (subject to the boundary conditions listed in the second of \eqref{boundcondcomp}) is given by 
\begin{equation}\label{solfo} 
\phi_1(x)= \int d^D x_4 G(x_4, x) \phi^3_0(x_4)
\end{equation} 
Once again, in order to avoid writing lengthy equations we have not substituted for $\phi_0$ and displayed our answer explicitly in terms of $\beta$;  the answer is given by three bulk to boundary propagators, sewn together at the bulk point $x_4$ which itself then connects to a bulk to bulk propagator ending at $x$. The bulk point $x_4$ is integrated over, as are the boundary points (weighted with boundary values of the fields).  

As in the previous subsection, the knowledge of $\phi_1$ (even without knowing $\phi_2$) is sufficient to evaluate the action to second order. At this order, $\phi_2$ could only have appeared in the quadratic part of the action in the schematic combination $\phi_0 \phi_2$. Once again the coefficient of this contribution vanishes because $\phi_0$ is a solution to the equation of motion 
that follows from the quadratic part of the action. The action at this order is simply given by 
\begin{equation}\label{acttwo}
\begin{split}
 S_2&= \int d^D x_5  ~\phi_1(x_5) \phi^3_0(x_5)\\
&= \int d^D x_5 \int d^D x_4 \int dA_1 dA_2 dA_3 ~ G(x_5, x_4)G_{\partial B} (x_4, y_1)G_{\partial B} (x_4, y_2)G_{\partial B} (x_4, y_3)~\beta(y_1)\beta(y_2)\beta(y_3)\\
& \int dA_4 dA_5 dA_6 ~G_{\partial B} (x_5, y_4)G_{\partial B} (x_5, y_5)G_{\partial B} (x_5, y_6) ~ \beta(y_4)\beta(y_5)\beta(y_6) 
\end{split}
\end{equation} 
Note that 
\begin{equation}\label{fdoso} 
\begin{split} 
\frac{\delta^6 S_2}{\delta \beta(y_1)\cdots \delta \beta(y_6)} 
&= 6! \int d^D x_5 \int d^D x_4 \int dA_1 \cdots dA_6~ G(x_5, x_4)G_{\partial B} (x_4, y_1)G_{\partial B} (x_4, y_2)G_{\partial B} (x_4, y_3)\\
  &~G_{\partial B} (x_5, y_4)G_{\partial B} (x_5, y_5)G_{\partial B} (x_5, y_6)\\
&=  n_1. \nabla_1 n_2. \nabla_2 n_3. \nabla_3 n_4. \nabla_4  n_5. \nabla_5  n_6. \nabla_6 
G^{(1)}_{\rm con}(x_1,x_2, x_3, x_4, x_5, x_6) \big|_{x_i=y_i} 
\end{split} 
\end{equation} 
We obtain a sum of 10 such exchange graphs (three boundary points leading to a bulk point connected to another bulk point and going back to three other boundary points)  with both bulk points integrated over all possible values,  and all six boundary points smeared against $\beta$ and then integrated over all values.  

\subsection{Generalization to arbitrary order and to the quantum theory} 

The generalization of the low order computations presented above, both to higher orders and also to the quantum theory is clear.
 Let us define 
\begin{equation}\label{pi} 
Z=\int {\cal D} \phi ~e^{-S} = e^{ \sum\limits_{n} \frac{1}{n!} \int \left ( \prod_{i=1}^n dy_i \beta(y_i) \right) G_{\rm bdry, con }(y_1 \cdots y_n)} = \sum_{n}  \int \prod_{i=1}^n dy_i \beta(y_i)   \frac{ G_{\rm bdry}(y_1 \cdots y_n)}{n!}  
\end{equation}
Note the boundary Greens function $G_{\rm bdry}$ appeared in \eqref{gb}, which is the full Greens function including the disconnected diagrams. Then 
\begin{equation}\label{bulbitd} 
G_{\rm bdry, con}(y_1 \cdots y_n)= \prod_{i=1}^n {\hat n}_i. \nabla_i { G}_{\rm con}(x_1, x_2 \ldots, x_n) 
\end{equation} 
where the function ${G}_{\rm con} (x_1, x_2 \ldots, x_n) $ was defined in 
\eqref{gdef}. The explicit computations presented above may be viewed as an explicit verification of \eqref{rhsder}. 

\subsection{Propagators in the limit $mT \gg 1$} 

The analysis presented so far in this Appendix has been exact; in particular, it 
applies to every choice of cut-off surface. Given a general choice of cut-off surface, however, the two point Greens functions (in terms of which we have presented our analysis) 
would, in general, be very difficult to compute. 

Of course, this problem is ameliorated in especially symmetrical situations. For instance, 
if our spacetime is the `slab' defined by $-T \leq t \leq T$ (here $t$ is Euclidean time) 
then it is possible to use the method of images to find reasonably explicit formulae 
for all Greens functions (see Appendix \ref{geuc} below). However even in this situation, the explicit results are given in terms of infinite sums, and so are relatively complicated. 

There is one situation, in which, however, all Greens functions can be well approximated 
in a relatively simple manner even for arbitrary smooth cut-off surfaces.  This situation arises when spacetime bounded by the cut-off is convex, the cut-off surface of our spacetime only varies on the length scale of the IR cut-off $T$,  and finally when $mT \gg 1$.
Let us consider two bulk points which are separated from all boundary points by a distance that is held fixed in units of $T$ as $T$ is taken to infinity. As we explain in more detail 
in subsection \ref{apgrn} (see the paragraph above \eqref{goodapprox}) the bulk to bulk propagator in such a situation is well approximated by the Euclidean Greens function on $R^D$. Moreover, 
the bulk to boundary propagator in the same situation (again assuming that the bulk point 
in question is separated from all boundary points by a distance that is held fixed in 
units of $T$) is well approximated by evaluating $n. \nabla G$ after accounting for 
the contribution of a single `image' charge: the reflection of the `source' that approaches 
the boundary. In order to evaluate $n.\nabla G$ we only need to reflect a source that is 
infinitesimally separated from the boundary. In such a situation the effective 
image charge is obtained by reflecting in the boundary pretending it is flat. 
The error generated by this process - stemming from the fact that the boundary is curved rather than flat is of order  $e^{-\alpha mT}$ where $\alpha$ is a positive number.  Implementing this procedure, it follows that in this situation the bulk to boundary Greens function is well approximated by 
\begin{equation}\label{formbb} 
G_{\rm bdry}(x, x') = -\frac{d}{da} \left( {\widetilde G}(x, x'-a{\hat n'}) - {\widetilde G} (x, x'+a{\hat n'}) \right) = 2 {\hat n}'.\nabla' {\widetilde G} (x, x')
\end{equation} 
where ${\widetilde G}$ the Euclidean propagator. 

In summary, both the bulk to bulk and the bulk to boundary propagators have a simple explicit form in the limit $m T \gg 1$. 

\subsection{Propagators in Momentum space} \label{eaaafobv}
In this section, we give momentum space expressions for the bulk-to-boundary and boundary-to-boundary propagators discussed previously in the case of a Euclidean `slab' $-T < \tau < T$. It is convenient to parametrize time by shifting it by $\frac{T}{2}$, so we will take $0 \leq \tau \leq 2T$. The equation of motion is given by
\beq 
\delta^{\mu\nu}\pa_\mu\pa_\nu \phi - m^2 \phi =0,
\eeq
which we can write in momentum space along the spatial directions as
\beq 
\phi(\tau,\vec{x})= \int \frac{d^d\vec{k}}{(2\pi)^d}e^{i\vec{k}\cdot \vec{x}}\phi_{\vec{k}}(\tau),
\eeq 
\beq 
\pa_{\tau}^2 \phi_{\vec{k}} - \omega_{\vec{k}}^2 \phi_{\vec{k}}=0,\;\;\;\cdots \;\;\;(\omega^2_{\vec{k}}=  \vec{k}^2 + m^2). 
\eeq
The solutions are linear combinations of the form:
\beq 
\phi_{\vec{k}}(\tau) = a_{\vec{k}} \sinh(\omega_{\vec{k}}\tau) + b_{\vec{k}} \cosh(\omega_{\vec{k}}\tau).
\eeq
\subsubsection*{Bulk to boundary propagator}
In order to get the bulk to boundary propagator, we need to impose boundary conditions. Let us first consider the bulk to boundary propagator $K(X|\vec{y})$, where $X=(\tau,\vec{x})$ is the bulk point and $\vec{y}$ is a boundary point at either the future or past boundary where the propagator approaches a delta function. More explicitly, if the source point is on the past boundary, the bulk to boundary propagator satisfies the boundary condition:
\beq 
K_E^{(-)}(\tau=0,\vec{x}|\vec{y}) = \delta^d(\vec{x}-\vec{y}),\;\;K_E^{(-)}(\tau=2T,\vec{x}|\vec{y}) = 0,
\eeq 
while if the source point is on the future boundary, the propagator satisfies
\beq 
K_E^{(+)}(\tau=0,\vec{x}|\vec{y}) = 0,\;\;K_E^{(+)}(\tau=2T,\vec{x}|\vec{y}) = \delta^d(\vec{x}-\vec{y}).
\eeq 
Thus, we get
\beq \label{bulkb1}
K_E^{(+)}(\tau,\vec{x}|\vec{y}) = \int \frac{d^d\vec{k}}{(2\pi)^d}e^{i\vec{k}\cdot (\vec{x}-\vec{y})}\frac{\sinh(\omega_{\vec{k}}\tau)}{\sinh(2\omega_{\vec{k}}T)},
\eeq 
\beq \label{bulkbi}
K_E^{(-)}(\tau,\vec{x}|\vec{y}) = \int \frac{d^d\vec{k}}{(2\pi)^d}e^{i\vec{k}\cdot (\vec{x}-\vec{y})}\frac{\sinh(\omega_{\vec{k}}(2T-\tau))}{\sinh(2\omega_{\vec{k}}T)}.
\eeq 
\subsection*{Bulk to bulk propagator}
In order to get the bulk to bulk propagator, we first solve the eigenvalue equation:
\beq 
(\pa_{\tau}^2 - \omega_{\vec{k}}^2) \psi_{E,\vec{k}}(\tau) = -E^2 \psi_{E,\vec{k}}(\tau), 
\eeq 
with Dirichlet boundary conditions at $\tau =0$ and $\tau = 2T$. For $E^2 > \omega_{\vec{k}}^2$, we get
\beq 
\psi_{E,\vec{k}}(\tau) = \alpha_{E,\vec{k}} \sin(\sqrt{E^2 - \omega^2_{\vec{k}}} \tau) + \beta_{E,\vec{k}} \cos(\sqrt{E^2 - \omega^2_{\vec{k}}} \tau).
\eeq 
This can only satisfy Dirichlet boundary conditions for 
\beq 
\sqrt{E^2_n -\omega_{\vec{k}}^2} = \frac{\pi n}{2T},\;\;\; \cdots \;\;\;(n \in \mathbb{Z}).
\eeq 
Thus, we get a complete set of (normalized) eigenfunctions for $n >0$:
\beq 
\psi_{n,\vec{k}}(\tau) = \frac{1}{\sqrt{T}}\sin\left(\frac{\pi n \tau}{2T}\right),
\eeq 
There are no non-trivial solutions for $E^2 \leq \omega_{\vec{k}}^2$. Thus, the bulk to bulk propagator is  given by
\beq 
G_E(\tau_1, \vec{x}_1|\tau_1, \vec{x}_2) = -\frac{1}{T}\sum_{n=1}^{\infty}\int \frac{d^d\vec{k}}{(2\pi)^d} e^{i\vec{k}\cdot (\vec{x}_1-\vec{x}_2)} \frac{\sin(\frac{\pi n \tau_1}{2T})\sin(\frac{\pi n \tau_2}{2T})}{\omega_{\vec{k}}^2+ \frac{\pi^2 n^2}{4 T^2}}.
\eeq 
Note that in the large $T$ limit, the sum becomes an integral over the intermediate energy. 
\subsubsection*{Example: Tree level contact diagram}
As an example of diagrams one can compute using these propagators, consider the boundary correlation function where we take two derivatives with respect to the source on the future boundary (labeled by the points $\vec{y}_{1,2}$ below) and two derivatives on the past boundary (labeled by the points $\vec{y}_{3,4}$ below). In $\phi^4$ theory, this boundary correlation function has a tree-level contact diagram contribution, which we can compute as
\beq 
G_{\text{bdry}}\Big|_{\text{contact}} = \lambda_4 \int_0^{2T} d\tau \int d^d\vec{x} K^{(+)}(\tau,\vec{x}|\vec{y}_1)K^{(+)}(\tau,\vec{x}|\vec{y}_2)K^{(-)}(\tau,\vec{x}|\vec{y}_3)K^{(-)}(\tau,\vec{x}|\vec{y}_4).
\eeq 
The integral over spatial coordinates gives a momentum-conserving delta function. The integral over $\tau$ is of the form:
\beq 
\int_{-T}^{T}d\tau\,\frac{\sinh(\omega_{1}(T+\tau))}{\sinh(2\omega_{1}T)}\frac{\sinh(\omega_{2}(T+\tau))}{\sinh(2\omega_{2}T)}\frac{\sinh(\omega_{3}(T-\tau))}{\sinh(2\omega_{3}T)}\frac{\sinh(\omega_{4}(T-\tau))}{\sinh(2\omega_{4}T)},
\eeq
where $\omega_1 = \omega_{\vec{k}_1}$ and so on, and we have shifted the integration variable. In the large $T$ limit, the denominators give a suppression factor of $e^{-2T\sum_i\omega_i}$. The only term that survives in the integral in this limit is the exponentially growing part of the $\sinh$, which gives
\beq
e^{-T\sum_i\omega_i}\int_{-T}^{T}d\tau\,e^{\tau(\omega_1 + \omega_2 - \omega_3 - \omega_4)}+\cdots =e^{-T\sum_i\omega_i}\frac{2\sinh\left((\omega_1 + \omega_2 - \omega_3 - \omega_4)T\right)}{(\omega_1 + \omega_2 - \omega_3 - \omega_4)} + \cdots,
\eeq 
where $\cdots$ denotes terms that are exponentially subleading in $T$. Continuing analytically in $T$ to the imaginary axis, we get\footnote{The terms we dropped were exponentially smaller in $T$, and even after analytic continuation, they remain so. This is because under Wick rotation we send $T \to T(i+\epsilon)$, and we take $T\to \infty$ along this contour before sending $\epsilon \to 0$.}
\beq 
e^{-iT\sum_i \omega_i}2\pi \frac{\sin((\omega_1 + \omega_2 - \omega_3 - \omega_4)T)}{\pi(\omega_1 + \omega_2 - \omega_3 - \omega_4)},
\eeq
which in the $T\to \infty$ limit gives
\beq 
e^{-iT\sum_i \omega_i}2\pi \delta(\omega_1 + \omega_2 - \omega_3 - \omega_4).
\eeq
The overall oscillating factor is precisely canceled by the wavefunctions in equation \eqref{smatrixl}, and thus, this boundary correlation function when inserted in \eqref{smatrixl} precisely reproduces the expected result for the S-matrix. 

\section{Euclidean Two Point Functions in position space: massive particles} \label{geuc}

In this Appendix, we first review the result for the exact Euclidean 
massive two point function on $R^D$, and then present 
a series expansion for the two point function on the `slab' of 
$R^D$ defined by $-T < t < T$ where $t$ is one of the Cartesian coordinates in $R^D$ and the two point function is computed after 
imposing Dirichlet boundary conditions.

\subsection{Euclidean Greens Function on $R^d$ equals the bulk to bulk propagator}

The Euclidean Greens function on $R^D$,  ${\widetilde G}$, 
is the solution of the equation 
$$\left( \nabla^2 -m^2 \right)  \widetilde G( \vec{r}_1, \vec{r}_2)= 
\delta^{D}(\vec{r}_1 -\vec{r}_2)$$
that vanishes at infinity. Translational and rotational invariance tell us that 
$\widetilde G= \widetilde G(|\vec{r}_1-\vec{r}_2|)$. 
Let $y=|\vec{r}_1-\vec{r}_2|$. It follows that at $y \neq 0$
\begin{equation}\label{diffeq1}
 \frac{1}{y^{D-1}} \partial_y y^{D-1} \partial_y {\widetilde G}(y) - m^2 {\widetilde G}=0
\end{equation} 
The general solution to \eqref{diffeq1} is,
\begin{equation}\label{diffeqsol}
    {\widetilde G}(y)=a (my)^{\frac{2-D}{2}} I_{\frac{D-2}{2}}(m y)+ b (my)^{\frac{2-D}{2}} K_{\frac{D-2}{2}}(m y)
\end{equation}
At small and large values of the arguments 
\begin{equation} \begin{split} 
&I_{\frac{D-2}{2}}(z) \approx \frac{1}{\Gamma(\frac{D-2}{2}+1)} \left( \frac{z}{2} \right) ^{\frac{D-2}{2}}, ~~~{\rm small} ~~~~~~~~~~~~~~~~ \approx \frac{e^z}{\sqrt{ 2 \pi z}} ~~~~~{\rm large}\\
& K_{\frac{D-2}{2}}(z) \approx \frac{\Gamma(\frac{D-2}{2})}{2}  \left( \frac{2}{z} \right) ^{\frac{D-2}{2}}, ~~~{\rm small}~~~~~~~~~~~~~~~~~~~ \approx   \pi \frac{e^{-z}}{\sqrt{2 \pi z}}  ~~~~~{\rm large} \\
\end{split}
\end{equation}
Since we are interested in the decaying solution,  conclude that $a=0$ in \eqref{diffeq1}. The value of $b$ is fixed 
by the requirement that the action of $\nabla^2-m^2$ produces a delta function at zero on the right hand side. Using the fact that the volume of the unit $D-1$ sphere is $\frac{\pi^{\frac{D}{2}}}{\Gamma(\frac{D}{2}+1)}$ 
we conclude that 
\begin{equation}\label{whaty}
{\widetilde G} (y) = b_D (my)^{\frac{2-D}{2}} K_{\frac{D-2}{2}}(m y)
\end{equation}
where $y(t,t', \vec{r}, \vec{r'}) = \sqrt{(t - t')^2 + |\vec{r} - \vec{r'}|^2}$ and $b_D = \frac{m^D}{(2 \pi)^{\frac{D}{2}}}$.\\

In the large $y$ limit 
\begin{equation}\label{gfly}
{\widetilde G}(y)\approx \left( \frac{\pi}{(2 \pi)^{\frac{D+1}{2}} } m^{\frac{D+1}{2}} \right) ~e^{-my}  y^{\frac{1-D}{2}} 
\end{equation} 

\subsection{Greens function for the `slab'}

We now work in a space in which the Euclidean time direction has a cutoff at $t = \pm T$ and the Greens function satisfies $G(y)|_{t = \pm T} = 0$. The Greens function in this space is easily obtained using 
the method of images. The total number of images is infinite (as 
the image in one boundary must also be reflected in the other boundary, and so on). The final answer for the Greens function is given by:
\beq \label{gfpb}
\begin{split} 
G(t, t') &= \widetilde G(|t-t'|)  +\\
&-\sum_{n = 0}^{\infty} \left[\widetilde G \left( |t+t'+ (4n+2) T| \right ) + \widetilde G \left( |t+t' -(4n+2)T | \right )\right] \\
& +\sum_{n = 1}^{\infty} \left[\widetilde G \left( |t-t'- 4n T| \right ) + \widetilde G \left( |t-t' + 4nT | \right )\right] \\
\end{split} 
\eeq
where we have suppressed the dependence of the Greens functions on ${\vec r}$ and ${\vec r}'$ (these dependence just go along for the ride).

\subsection{Approximate Greens Functions when $mT \gg 1$}\label{apgrn}

In the limit $m T \gg 1$, and for all values of $t'$ and $t$ lying within the slab, we obtain a good approximation to \eqref{gfpb} by keeping 
the terms in \eqref{gfpb} with $n \leq 2$. 

In fact when $t$ and $t'$ are separated from both boundaries by a finite fraction of $T$, we get an excellent approximation
\footnote{For the purposes of this Appendix, we define an approximation to the Greens 
function to be good if the fractional error in the Greens function is of order $e^{- \alpha m T}$ where $\alpha$ is any positive number that is held fixed as $T$ is taken to infinity.} 
to the Greens function by retaining only those terms in \eqref{gfpb} with $n=0$. That this is the case is easily seen from the exact formula \eqref{gfpb}, but may also be understood physically in the following terms. The Greens function $G_{\text{full}}(t, t')$ may be 
thought of as the `potential' created at $t$ by a `charge' placed at $t'$. This potential has two contributions. The first comes directly from the charge. The second comes from the 
`rearrangement of surface charges' that are needed in order to enforce the Dirichlet boundary conditions of our problem. As the theory we are dealing with is massive, the direct effect is of order $e^{- m d}$ where $d$ is the distance between the source and the measuring point. 
On the other hand, the second effect is proportional to $e^{-m(d_1+d_2)}$ where $d_1$ is the distance between the source and any point on the boundary, while $d_2$ is the distance between the same boundary point and the measuring point. If the source and measuring point 
are both separated by the boundary by a finite fraction of $T$, it is obvious that $d_1+d_2 -d$ is positive and is also a finite fraction of $d$. It follows that the `direct' contribution (first line of the RHS of \eqref{gfpb}) is a good approximation to the full result.  In other words, in this case
\begin{equation}\label{goodapprox} 
G \approx  \widetilde G(t, t')
\end{equation} 

When $t'$ approaches one of the boundaries but $t$ is in the bulk (and is separated 
from both boundaries by a distance that is a finite fraction of $T$), the argument 
presented above would suggest that, in addition to the `direct' term on the first line of the RHS of \eqref{gfpb}, we need to include one of the two $n=0$ terms on the second line of the RHS of \eqref{gfpb} \footnote{This term can be thought of as capturing the reflection of the 
source at $t'$ in the boundary that it is tending towards.};  all other terms in this expansion may be neglected. It is easy to verify directly from \eqref{gfpb} that this approximation is indeed accurate. In this situation, in other words, if either $t$ or $t'$
or bout approaches the boundary at time $T$
\begin{equation}\label{goodapproxtwo} 
G \approx \widetilde G(t, t')  - \widetilde G \left(t, 2T - t' \right )
\end{equation} 
while if $t$ or $t'$ or both approach the boundary at time $-T$
\begin{equation}\label{goodapproxtwop} 
G \approx \widetilde G(t, t')  - \widetilde G \left(t,  -2T - t'\right)
\end{equation} 

Finally, if $t$ approaches the boundary at time $T$ while if $t'$ approaches the boundary 
at time $-T$ then 
\begin{equation}\label{goodapproxthree} 
G \approx \widetilde G(t, t')  - \widetilde G \left(t, 2T - t' \right ) - \widetilde G \left(t,  -2T - t'\right) + \widetilde G \left( |t-t'- 4 T| \right )
\end{equation} 
On the other hand of $t$ approaches the boundary at time $-T$ while $t'$ approaches the 
boundary at time $T$, 
\begin{equation}\label{goodapproxthreeprime} 
G \approx \widetilde G(t, t')  - \widetilde G \left(t, 2T - t' \right ) - \widetilde G \left(t,  -2T - t'\right) + \widetilde G \left( |t-t'+4 T| \right )
\end{equation}

Assuming that all bulk points are separated from both boundaries by a finite fraction of 
$T$, therefore, it follows that the bulk to bulk propagator is well approximated by 
\eqref{goodapprox}. On the other hand, the bulk to boundary propagator is well approximated by \begin{equation}\label{bbin}
G_{\rm bdry}= \partial_{t'} G(t, t')|_{t'=t}
=2\partial_{t'} {\widetilde G}(t, t')|_{t'=t} 
=2 {\hat n}. \nabla' \widetilde G(x, x')
\end{equation}

In Lorentzian signature, this relationship is given by
\begin{equation}\label{bbinl}
G_{\rm bdry} =2 i {\hat n}. \nabla' \widetilde G(x, x')
\end{equation}
This is because in the usual definition of Lorentzian Greens function, $- i \delta (x-x')$ appears on the RHS whereas in the Euclidean case discussed in the previous Appendix, we defined Greens function with just the delta function on RHS. Tracking the $i$'s in our Euclidean derivation, it is easy to verify \eqref{bbinl}.

Using the explicit expression for ${\widetilde G}$ presented in the previous Appendix, it is easy to explicitly evaluate the RHS of \eqref{bbin}. Depending on whether the boundary is at time $T$ or time $-T$ we find 
\beq
\begin{split}
G_{\partial B} (r,t, r', T) &= \partial_{t'}G(y)|_{t' = T} = 2 b_D m^2 (t-T) \frac{K_{\frac{D}{2}}\left(m y_{T-t} \right)}{\left(m y_{T-t}\right)^{\frac{D}{2}} }\\
G_{\partial B}(r,t, r', -T) &= -\partial_{t'}G(y)|_{t' = -T} = - 2 b_Dm^2 (t + T) \frac{K_{\frac{D}{2}}(my_{T+t})}{(m y_{T+t})^{D/2}}
\end{split}
\eeq
where, $y_t=\sqrt{|\vec{r}-\vec{r'}|^2+t^2}$. In the limit that $m y_T$ is large we obtain:
 \beq \label{bblargeT}
 \lim_{T\to\infty} G_{\partial B}(r,t, r', T)= \lim_{T\to\infty} G_{\partial B}(r,t, r', -T) = b_D \sqrt{2\pi m^3 T}{e^{-my_T} \over \left(my_T\right)^{D\over 2}}
 \eeq
 
 Finally, the boundary to boundary propagators, \eqref{boundbound}, are also easily evaluated. 
 We have two types of boundary to boundary propagators, one in which both points are at the same boundary (we denote them by $G_{\partial \partial}^{++}$ and $G_{\partial \partial}^{--}$) and the other one in which one point is at boundary $T$ and other point is at $-T$ (we denote them by $G_{\partial \partial}^{+-}$).
 \beq \label{bdybdy}
 \begin{split}
G_{\partial \partial}^{++}(r,T, r', T) &=   2 b_D m^2\frac{K_{\frac{D}{2}}(m y_0)}{( m y_0)^{\frac{D}{2}}}  \\
G_{\partial \partial}^{+-}(r,T, r',-T) &= {4b_D m^2 \over \left(my_{2T}\right)^{D+2\over 2} }  \left(4 (m T)^2 K_{\frac{D+2}{2}}\left(m y_{2T}\right)-my_{2T} K_{\frac{D}{2}}\left(my_{2T}\right)\right)
 \end{split}
 \eeq
 In the large $T$ limit these expressions simplify to
 \beq
 \begin{split}
 G_{\partial \partial}^{++}(r,T, r', T) &=   2 b_D m^2\frac{K_{\frac{D}{2}}(m y_0)}{( m y_0)^{\frac{D}{2}}}  \\
 G_{\partial \partial}^{+-}(r, T, r', -T) &= 8b_D m^2(mT)^2 \sqrt{2\pi } { e^{-m y_{2T}} \over \left(my_{2T}\right)^{\frac{D+3}{2}}}
 \end{split}
 \eeq

\section{Toy Model: Harmonic oscillator}\label{Harmonic}
 
In this Appendix, we study the coherent state path integral for the `anharmonic' harmonic oscillator defined by the Hamiltonian 
\begin{equation}\label{harmham}
 \mathcal{H} = \frac{P^2}{2} + \frac{\omega^2 X^2}{2} + {\tilde V}(X)
\end{equation} 
where $X$ and $P$ are the usual position and momentum operators. 
We also recall the usual position space path integral for the same system, and comment on the relationship between these two path integrals.

 \subsection{Coherent State Quantization for a Harmonic Oscillator} \label{Harmonic12}

Recall that coherent states for the Harmonic oscillator are defined so that  
\beq\label{codef}
a | z\rangle = z | z \rangle \qquad \text{where} \qquad | z \rangle =  e^ {z a^{\dagger} }|0\rangle =\sum_{n=0}^{\infty}\frac{z^n}{\sqrt{n!}}|n\rangle
\eeq
where $|n\rangle$ is the n-particle state. The inner product is given by:
\beq
\langle z' | z \rangle = e^{z' z}
\eeq
and obey the (over) completeness relation 
\begin{equation}\label{ovecomp} 
\int \frac{ d^2z}{\pi} e^{- z {\bar z}}  |z \rangle \langle z|=I
\end{equation} 

Recall that the operator $X(t)$ can be expanded in terms of $a$ and $a^\dagger$
\begin{equation}\label{xexp} 
X= \frac{1}{\sqrt{2 \omega}}\left(a e^{-i\omega t} + a^\dagger e^{i \omega t}\right) 
\end{equation} 
It follows that $z$, the eigenvalue of the operator $a$, is `positive energy data', while ${\bar z}$ is `negative energy data'. 

Let us define the matrix elements (between coherent states) of the time evolution operator $e^{- i Ht}$ ($H$ is the Hamiltonian operator) as 
\beq\label{pathint}
U(\bar{z}_f, t_f; z_i, t_i) = \langle z_f| e^{-i H (t_f-t_i)}|z_i\rangle
\eeq
Using the usual Feynman time slicing method (and inserting \eqref{ovecomp} into each time step), it is not difficult to demonstrate that $U(\bar{z}_f, t_f; z_i, t_i)$ is given by the Lorentzian path integral \cite{Itzykson:1980rh} 
\beq\label{cin}
U(\bar{z}_f, t_f; z_i, t_i) = \int \mathcal{D}(z,\bar z) \, e^{i \mathcal{S}}
\eeq
where
\beq\label{actionvar}
\mathcal{S} = -i\frac{\bar{z}_f z_f+\bar{z}_iz_i}{2} - i \int_{t_i}^{t_f} dt \left[ \frac{\Dot{\bar{z}}z-\Dot{z}\bar{z}}{2} - i \mathcal{H}(z,\bar z) \right] 
\eeq
\footnote{The action \eqref{actionvar} can be rewritten as 
\beq\label{actionvarn}
\begin{split}
\mathcal{S} &= -i\frac{\bar{z}_f z_f+\bar{z}_iz_i}{2} - i \int_{t_i}^{t_f} dt \left[ \frac{({\Dot z} + {\Dot {\bar z}}) (z -{\bar z}) -z {\Dot z} +{\bar z} {\Dot {\bar z}}}{2}
 - i \mathcal{H}(z,\bar z) \right]  \\
    & = i  \frac{ z_f^2-z_i^2 -{\bar z}_f^2 + {\bar z}_i^2 -2\bar{z}_f z_f -2\bar{z}_iz_i}{4}  - i \int_{t_i}^{t_f} dt \left[ \frac{({\Dot z} + {\Dot {\bar z}}) (z -{\bar z}) }{2}
 - i \mathcal{H}(z,\bar z) \right] \\
\end{split}
\eeq 
The point of this rewriting is that the integral that appears in the last of \eqref{actionvarn} is simply 
 \begin{equation}\label{formofintpart}
 \int_{t_i}^{t_f} dt \left[ p {\dot x}
 - \mathcal{H}(x, p) \right]
\end{equation}
(where we have used $z=x+i p$ and ${\bar z}=x-ip$). \eqref{formofintpart} is (upto normal ordering issues, see below) the `usual' form of the action - i.e. the action 
appropriate for the computation of the action (or the path integral) as a function of $x$ at initial and final times. }
The function ${\cal H}$ given by 
The Hamiltonian is given by 
\beq \label{hamv1}
 \mathcal{H} =\omega z {\bar z} + V\left(\frac{z+ \bar z}{\sqrt{2 \omega}}\right)
\eeq
\footnote{Roughly speaking, $z$ and $\bar z$ are related to the variables 
$x$ and $p$ in the phase space path integral via
\begin{equation} \label{zzbardef}
\omega x+ip = \sqrt{2\omega} z, \qquad \qquad \omega x-ip = \sqrt{2\omega} {\bar z}
 \end{equation} } 
The function $V(x)$ is related to the potential ${\tilde V}(x)$ that appears in \eqref{harmham} as
\begin{equation}\label{vharm}
{\tilde V} (x)= e^{-\frac{\partial_x^2}{4 \omega}} V(x), ~~~{V} (x)= e^{+\frac{\partial_x^2}{4 \omega}} {\tilde V}(x)
\end{equation}
\footnote{The nontrivial relationship between $V(x)$ and ${\tilde V}(x)$ is a manifestation of operator ordering issues. Consider a Harmonic oscillator with potential ${\tilde V}(X)$ (here $X$ is the position operator). This potential may be rewritten in terms of the operators $a$ and $a^\dagger$, but the expression one obtains is not normal ordered. Wicks theorem tells us that 
\begin{equation}\label{wtherm}
{\tilde V}(X)= ~ :e^{+\frac{\partial_X^2}{4 \omega}} {\tilde V}(X) : 
\end{equation} 
where the symbol $: :$ denotes normal ordering (recall that $\langle X X\rangle = \frac{1}{2 \omega}$). Once we have normal ordered the potential, the repeated insertion of \eqref{ovecomp} into \eqref{pathint} converts the potential as a function of $a$ and $a^\dagger$ into a 
function of ${\bar z}$ and $z$. It is crucial that this simple replacement rule applies only once we have normal ordered.  These facts, taken together, explain \eqref{vharm}.}

The path integral in \eqref{cin} is evaluated subject to the boundary conditions that 
$z(t_i)=z_i$ and ${\bar z}(t_f)={\bar z}_f$. In the above expression, $\mathcal {H} (z , \bar z)$ is the Hamiltonian of the system. Note that the boundary conditions fix positive 
energy data at the initial time $t_i$ but negative energy data at the final time $t_f$, exactly 
as in our discussion of the S matrix (this is the reason the discussion of this subsection is 
relevant to the S matrix). 

The action in \eqref{actionvar} (that comes out of time slicing the evolution operator in the Feynman manner) is the sum of two terms; the explicit counterterm  
$-i\frac{\bar{z}_f z_f+\bar{z}_iz_i}{2}$ plus an integral of the `Lagrangian' over time. 
We emphasize that it is the sum of these two terms that gives rise to an action that leads to a good variational principle (whose variation implies the usual Harmonic Oscillator equations of motion). 
In particular, the `counterterm' is needed to cancel boundary variations, as we now explain. The bulk action in $z$, ${\bar z}$ variables is given by
\beq \label{Onshellaction0}\
\mathcal{S}_{\rm bulk} =  - i \int dt \left[ \frac{\Dot{\bar{z}}z-\Dot{z}\bar{z}}{2} - i \mathcal{H} \right]
\end{equation}
We are interested in the path integral computed by specifying $z = z_i$ at initial time $t_i$ and $\bar z = \bar z_f$ at final time $t_f$. As mentioned above, these boundary conditions are equivalent to specifying positive energy data in the past and negative energy data in the future. As we have mentioned 
above, the measure for this path integral is $e^{i S}$ where $S$ is the time integral of the 
Lagrangian plus a counter-term. The form of this counter-term (which comes out independently 
from the Feynman time slice procedure) may independently have been guessed from the following 
considerations. The variation of ${\mathcal S}_{\rm bulk}$ (see \eqref{Onshellaction0})  with respect to 
$z$ and $\bar z$ is given by 
\begin{eqnarray*}
  \delta {\mathcal S}_{\rm bulk} &=&  \int dt \left[- i\frac{\dot{\bar z} \delta z - \bar{z} \dot{\delta z}}{2} - \omega \bar z \delta z - V'\left(\frac{z+ \bar z}{\sqrt{2 \omega}}\right) \delta z - i \frac{\dot{\delta \bar z}  z - \delta\bar{z}  \dot{z}}{2} - \omega  z \delta \bar z - V'\left(\frac{z+ \bar z}{\sqrt{2 \omega}}\right)\delta \bar z \right]\\
  &=&  \int dt \left[\left(- i \dot{\bar z} - \omega \bar z - V'\left(\frac{z+ \bar z}{\sqrt{2 \omega}}\right)\right) \delta z + \left(i \dot{z} - \omega z - V'\left(\frac{z+ \bar z}{\sqrt{2 \omega}}\right)\right) \delta \bar z + \frac{d}{dt} \left(i\frac{\bar z \delta z- \delta \bar z  z}{2} \right) \right] \\
  &=& \int dt \left[\left(\text{EOM of} \, \bar z\right) \delta z + \left(\text{EOM of} \,z\right) \delta \bar z \right]+ \frac{i}{2}\left( \bar z_f \delta z_f + z_i \delta \bar z_i\right)
 \end{eqnarray*}
where we have used the fact that $z(t_i) = z_i$ hence $\delta z_i = 0$, similarly $\bar{z}(t_f) = \bar z_f$ hence $\delta \bar z_f = 0$. We have also used the fact that the potential $V$ is a function of combination $z + \bar z$, hence it's derivative w.r.t. $z$ and $\bar z$ is same, which we denote by $V'\left(\frac{z+ \bar z}{\sqrt{2 \omega}}\right)$. Now we see the need to add counter-terms in the action. The variation of action should give us just the equation of motion but here we are getting some extra boundary contributions to that variation. So we need to add appropriate counter-terms to the action that cancels these boundary terms. Hence the full action, with a well-defined variational principle is given by:
\beq \label{Onshellaction}
\begin{split}
&\mathcal{S} = \mathcal{S}_{\rm bdry} + \mathcal{S}_{\rm bulk} \\
&{\rm where}\\
&\mathcal{S}_{\rm bdry} =-i\frac{\bar{z}_f z_f+\bar{z}_iz_i}{2} , \qquad \qquad 
\mathcal{S}_{\rm bulk} = - i \int_{t_i}^{t_f} dt \left[ \frac{\Dot{\bar{z}}z-\Dot{z}\bar{z}}{2} - i \omega z \bar{z} - i V\left(\frac{z+ \bar z}{\sqrt{2 \omega}}\right) \right] 
\end{split}
\end{equation}
 which matches \eqref{actionvar}. Therefore the boundary term in \eqref{actionvar} is required to make the variational principle well-defined.

\subsection{Diagrammatic Computation of the coherent state path integral}\label{digch}
In the free theory, $z$ and $\bar z$ obey the following equations
\beq \label{coherenteom}
\Dot{{z}} + i\omega z  = 0, \qquad\qquad -\Dot{\bar{z}} +i\omega  \bar z = 0
\eeq
Let us define the coherent state `bulk to bulk' propagators  as
\begin{equation}\label{btbprop}
\begin{split}
 \langle z(t) \bar z (t')\rangle &:= G_{F}^{z \bar z}(t,t') =  e^{-i\omega(t-t')}\Theta(t-t')\\
\end{split}
\end{equation}
These propagators obey the following equations
\beq \label{coherenteom1}
\begin{split}
\frac{d}{d t}G_{F}^{z \bar z}(t,t') + i \omega  G_{F}^{z \bar z}(t,t')&=  \delta(t-t')\\
-\frac{d}{d t'}G_{F}^{z \bar z}(t,t') + i\omega G_{F}^{ z \bar  z}(t,t') & =  \delta(t-t')
\end{split}
\eeq
and satisfy the following boundary conditions
\beq
G_{F}^{z \bar z}(0,t') = 0, \qquad \qquad G_{F}^{z \bar z}(t, T) = 0
\eeq
Note that the bulk to boundary propagator \eqref{btbprop} is discontinuous at $t=t'$. The value of this propagator at precisely $t=t'$ is relevant only for `tadpole' contractions in Feynman diagrams. Since 
the our potential $V$ is normal ordered, all such tadpoles must vanish. 
In order to ensure this we set 
\begin{equation}\label{tadpoles}
G_{F}^{z \bar z}(t,t)=0
\end{equation} 

Similarly, the coherent state `bulk to boundary' propagators are given by
\begin{equation}\label{propdef}
K_F^{(+)}(t)  = e^{-i\omega t},   \qquad \qquad  K_F^{(-)}(t) =  e^{i\omega (t-T)} 
\end{equation} 
These propagators obey the first and second of \eqref{coherenteom1} respectively but with 
 the following boundary conditions
\beq
K_F^{(+)}(0) = 1 , \qquad \qquad K_F^{(-)}(T) = 1
\eeq
It is easy to check that 
\begin{equation}\label{relbetprop}
\begin{split}
&K_F^{(+)}(t)=   G_F^{{z \bar z} }(t, 0)\\
&K_F^{(-)}(t)=   G_F^{z {\bar z}}(T, t)
\end{split}
\end{equation}
as could have been anticipated on general grounds. \footnote{The equations \eqref{relbetprop} can be abstractly argued for as follows. Let $z(t)$ and ${\bar z}(t)$ be solutions of the free harmonic equations \eqref{coherenteom}, and consider the two integrals 
$$ \int_{t=0}^{T} \partial_t \left( z(t) G^{z \bar z}_F(t', t) \right),~~~
\int_{t=0}^{T} \partial_t \left( {\bar z}(t) G_F^{z \bar z}(t', t) \right)$$
We can perform each of these integrals in two different ways: first by
reducing them to boundary terms, and second by opening out the derivatives and using the \eqref{coherenteom1} and \eqref{coherenteom}. This procedure gives a formula for $z$ in terms of $z_i$ and for 
${\bar z}(t)$ in terms of ${\bar z}_f$ (i.e. for the bulk to boundary 
propagators) in terms of the bulk to bulk propagators, establishing 
\eqref{relbetprop}.}

As a short diversion, it is useful to analytically study the analytic continuation of the bulk to bulk propagator \eqref{btbprop} to Euclidean space. Performing the continuation $t \rightarrow -i \tau$, we obtain the Euclidean bulk to bulk propagators 
\begin{equation}\label{btbpropeuc}
\begin{split}
 \langle z(\tau) \bar z (\tau')\rangle &:= G_{E}^{z \bar z}(\tau ,\tau') =  e^{-\omega(\tau-\tau')}\Theta(\tau-\tau')\\
\end{split}
\end{equation}
It follows that 
\begin{equation}\label{btbpropeucx}
\begin{split}
\langle \frac{z(\tau) + \bar z (\tau)}{\sqrt{2 \omega}} ~~\frac{ z(\tau')+ \bar z (\tau')}{\sqrt{2 \omega}}\rangle &= G_{E}(\tau ,\tau') = \frac{1}{2 \omega} e^{-\omega|\tau - \tau'|}\\
\end{split}
\end{equation}
But $G_E(\tau, \tau')$ is simply the well known Euclidean propagator, 
$\langle x(\tau) x(\tau') \rangle$ of the Harmonic oscillator (i.e. the propagator that is required to vanish when $|t | \to \infty$). 

The relationship between the bulk to boundary propagators \eqref{relbetprop}
and the Euclidean propagator is 
\beq\label{ndelz}
\begin{split}
&K_F^{(+)}(t)=   \theta( t) (-2 i \partial_t) G_F(t, 0)= \theta( t) (2 i{\hat n}. \partial)  G_F(t, t') \bigg|_{t'=0} \\
&K_F^{(-)}(t)=  \theta(T- t) (2 i \partial_t) G_F(T, t)= \theta(T- t) (2 i{\hat n}. \partial)  G_F(t, t') \bigg|_{t=T}
\end{split}
\eeq
where $G_F(t,t')$ is the Feynman propagator obtained by analytic continuation of the Euclidean propagator given in \eqref{btbpropeucx} and $\hat n$ is the outward pointing normal vector to the boundary.

It follows that the in out path integral as a functional of boundary conditions,
has the following simple relation to purely Euclidean Greens functions. 
If we denote the logarithm of the path integral as a functional of $z_i$ and ${\bar z}_f$ as $Z( z_i, {\bar z}_f)$, then 
\begin{equation}\label{zbeta}
Z_E( z_i, {\bar z}_f) \approx  \sum_{m, n} \frac{ (\sqrt{2 \omega} z_i)^m (\sqrt{2 \omega} {\bar z}_f)^n}{m! n!}
G_E^{m, n}(0, T)
\end{equation} 
where the Euclidean propagator $G_E^{m,n}$ is defined by 
$$G_E^{m,n}(\tau, \tau')= \langle X^m(\tau) X^n(\tau') \rangle$$

Defining $\beta=\sqrt{2 \omega} z_i$ and ${\bar \beta}=\sqrt{2 \omega} {\bar z}_f$ \eqref{zbeta} may be rewritten as 
\begin{equation}\label{zbetane}
Z_E( \beta, {\bar \beta}) \approx  \sum_{m, n} \frac{ (2 \omega \beta)^m (2 \omega {\bar \beta})^n}{m! n!}
G_E^{m, n}(0, T) \approx \sum_{m, n} \frac{ (2 \partial_t \beta)^m (2 \partial_{t'} {\bar \beta})^n}{m! n!}
G_E^{m, n}(t, t') \Bigg|_{t=0, t'= T}
\end{equation}
The approximation signs in \eqref{zbeta} and \eqref{zbetane} have the following meaning. The Euclidean Greens functions that appear on the RHS of these equations can, in perturbation theory, be computed in the usual manner using the Feynman expansion. This expansion involves an integral over the location of all 
interaction vertices. In evaluating the Euclidean Greens functions on the RHS 
of \eqref{zbeta} and \eqref{zbetane}, we are instructed to integrate the integration location of all interaction vertices over all of spacetime. 
We get the LHS of these equations, however, only if we restrict the location of these interaction vertices to the slab region, $t \in (0, T)$. The LHS is an approximation to the RHS in the following sense. The extra region of integration space does give contributions that scale like $T$ (i.e. an energy conserving $\delta$
function) and so do not contribute to the S matrix (see around \eqref{sonriw1}).

While the discussion of the last few paragraphs has focussed on the case of quantum mechanics, the generalization to quantum field theory (regarded as a theory of an infinite number of harmonic oscillators) is straightforward.

The propagators listed above can be used for a number of purposes. 
At the classical level, one can sum tree diagrams built using these propagators to (perturbatively) find the $z(t)$ and ${\bar z}(t)$
subject to the boundary conditions described above. At leading order, for instance, one finds
\beq \label{freesolzzb}
z_0(t)=  z_i K_F^{(+)}(t) , \qquad \qquad \bar z_0(t)=   {\bar z}_f K_F^{(-)}(t)
\eeq
\footnote{A similar procedure, but without the restriction to tree diagrams, yields expressions for time ordered correlators of $a(t)$ and $a^\dagger(t)$ in the quantum case. }

One can also use the propagators listed above to evaluate the logarithm of the Harmonic Oscillator path integral as a functional of $z_i$ and ${\bar z}_f$. In order to find the coefficient of $z_i^n {\bar z}_f^m$, one simply evaluates the sum of all connected Feynman diagrams with $n$ legs on the boundary at $t=0$ and $m$ legs on the boundary at $t=T$ \footnote{The restriction to tree diagrams evaluates yields the classical action as a function of this data.}.

As a simple check of this formalism (and especially of the fact that all boundary terms are working out right) we have computed the classical action resulting from the potential $V(x)=g x^3$, at order $g^2$, in two different ways: first by summing (tree exchange) diagrams, and second by 
performing the computation using the formalism of \cite{Faddeev:1980be, Kraus:2002cb}. We briefly describe this second procedure. 

The classical solutions of the nonlinear equations of motion 
\beq \label{coherenteom2}
\begin{split}
\Dot{\bar{z}} - i \omega  \bar z - i V' (z+ \bar z) &= 0\\
-\Dot{{z}} - i \omega z - i V' (z+ \bar z) &= 0
\end{split}
\eeq
that obey the boundary conditions $z(0)=z_i$ and ${\bar z}(0)={\bar z}_f$, are formally given by 
\beq\label{solzzb1}
\begin{split}
    z(t) &= z_0(t) + \tilde z (t) = \left(z_i  - i\int_{0}^{T} V'(z+\bar z) \Theta(t-t') e^{i \omega t'} dt'\right) e^{- i \omega t} \\
    &=z_i K^{(+)}_F(t)  -i \int_{0}^{T} V'(z +\bar z) G^{z\bar z}_F (t,t') dt'\\
    \bar{z}(t) &= \bar z_0 (t) + {\tilde {\bar z}}(t) = \left(\bar{z}_f e^{-i \omega T} - i\int_{0}^{T} V'(z + \bar z) \Theta(t'-t) e^{-i \omega t'} dt'\right)e^{ i \omega t} \\
    &=\bar z_f K^{(-)}_F(t)  -i \int_{0}^{T} V'(z +\bar z) G^{z\bar z}_F (t',t) dt'
\end{split}
\eeq
Here $z_0(t)$ and $\bar z_0(t)$ satisfy the free equation of motion and are given in \eqref{freesolzzb} and ${\tilde z}(t)$, ${\tilde {\bar z}}(t)$ are the corrections to the free solutions and obey trivial boundary conditions i.e. ${\tilde z}(0)=0$ and ${\tilde {\bar z}}(T)=0$.
\eqref{solzzb1} are implicit solutions to the equations of motion, as the RHS of the equations in  \eqref{solzzb1} depends on the solutions  $z$ and ${\bar z}$. This equation can be made more explicit by iterating \eqref{solzzb1}. It is easy to check that the solution obtained in this manner agrees with the solution obtained by expanding in Feynman diagrams (following the procedure described above).  

Inserting \eqref{solzzb1}  into the action \eqref{Onshellaction}, we find 
\beq
\begin{split}
    \mathcal{S}_{\rm bdry} &= -i\frac{\bar{z}_f z_f+\bar{z}_i z_i}{2}\\
    &= - i z_i {\bar z}_f e^{- i \omega T} - \frac{1}{2}\int_0^T \left(z_i e^{- i \omega t} + {\bar z}_f e^{i \omega (t-T)}\right) V'(z, \bar z) dt\\
    &=  - i z_i {\bar z}_f e^{- i \omega T} - \frac{1}{2}\int_0^T \left(z_0(t) + {\bar z}_0(t)\right) V'(z, \bar z) dt\\
    \mathcal{S}_{\rm bulk} &= \int_0^T \left(\frac{1}{2}\left(z(t) + {\bar z}(t) \right) V'(z, \bar z) - V(z+ \bar z) \right)dt 
\end{split}
\eeq
Using the above equations, we find that the full on-shell action, 
${\cal S}=\mathcal{S}_{\rm bdry}+ \mathcal{S}_{\rm bulk}$
\beq\label{fullac}
\mathcal{S} =  - i z_i {\bar z}_f e^{- i \omega T} + \int_0^T \left(\frac{1}{2}\left(\tilde{z}(t) + \tilde {\bar z}(t) \right) V'(z, \bar z) - V(z+ \bar z) \right)dt 
\eeq
As in the discussion around \eqref{solzzb1}, the expression \eqref{fullac} is implicit, as it depends on $z(t)$ and ${\bar z}(t)$.  The expression can be made explicit by inserting the iterated solutions to 
\eqref{solzzb1} into \eqref{fullac}. It is not difficult to check that this procedure reproduces the action as evaluated by summing over all tree diagrams. 

Note that, on-shell, the boundary terms and the linear piece in the free bulk action cancel each other. 

\subsection{Diagrammatic Computation of the Dirichlet path integral}\label{dirap}
Let us now study the on-shell action with Dirichlet boundary conditions for $x$ i.e.
\beq \label{xbc}
x(0) = x_i \qquad \qquad \qquad x(T) = x_f.
\eeq
In this case the variational principle is well-defined without the addition of any boundary terms. Hence the full action (in the  variable $x$) is just given by 
\beq\label{diriaction}
\mathcal{S}_{\rm bulk} = \int_0^T dt ~ \left(\frac{{\dot x}^2}{2} - \frac{\omega^2 x^2}{2}- \tilde {V}(x)\right)
\eeq
and the equation of motion  is given by
\beq
\ddot{x} + \omega^2 x + \tilde{V}'(x) = 0
\eeq
(in the equation above ${\tilde V}'$ denotes the derivative of $\tilde V$ w.r.t the variable $x$).
The `bulk to bulk' propagator 
\begin{equation}\label{xbulk}
G_D(t,t') = \frac{i}{\omega \sin\omega T} \left[ \sin \omega t \sin \omega (T-t') \Theta(t'-t) + \sin\omega t' \sin \omega(T-t) \Theta(t-t') \right]
\end{equation}
satisfies the equation of motion 
\beq\label{xgeqn}
\frac{d^2}{d t^2}G_D(t,t') + \omega^2 G_D(t,t')  =  -i\delta(t-t')
\eeq
and obeys the boundary conditions
\beq
G_D(0,t') = 0, \qquad \qquad G_D(T,t') = 0.
\eeq

Note that the function $G_D(t, t')$ is continuous at $t=t'$ (even though its first derivative is discontinuous at this point). In particular, we have
\begin{equation}\label{dirichlettadpole}
G_D(t, t)= i\frac{\sin \omega t \sin \omega (T-t) }{\omega \sin\omega T}  
\end{equation}

Similarly, the `bulk to boundary' propagators are given by
\beq \label{diribtobo}
K_D^{(-)}(t) = \frac{\sin(\omega t)}{\sin(\omega T)}, \qquad \qquad
K_D^{(+)}(t) = \frac{\sin(\omega (T-t))}{\sin(\omega T)}.
\eeq 
They satisfy \eqref{xgeqn} with the following boundary conditions
\beq
K_D^{(-)}(T) = 1 , \qquad \qquad K_D^{(+)}(0) = 1
\eeq
Note that 
\begin{equation}\label{exo}
\begin{split} 
&K_D^{+}(t)= -i\partial_{t'} G_D(t, t')\left|_{t'=0} \right.\\
&K_D^{-}(t)= +i\partial_{t'} G_D(t, t')\left|_{t'=T}\right. \\
\end{split}
\end{equation} 
in agreement with \eqref{bbprop}\footnote{In Lorentzian signature, the bulk to boundary propagator is related to bulk to bulk propagator via $$K_D(t) = i n. \nabla' G_D(t,t')\bigg|_{\rm bdry}$$}. As in the previous subsection, the propagators listed above can be used to evaluate the logarithm of the harmonic oscillator path integral as a functional of $x_i$ and $x_f$.

An implicit formula for the Dirichlet on-shell action, similar to \eqref{fullac} is given by
\beq\label{ondir}
\mathcal{S}_{\rm bulk} = \frac{\omega}{2 \sin(\omega T)} \left(\left(x_f^2+x_i^2\right) \cos (\omega T)-2 x_f x_i\right)+  \int_0^T dt ~ \left(\frac{1}{2}{\tilde x} \tilde V'(x) - \tilde V(x) \right) 
\eeq

\subsection{Relationship between the Dirichlet and Positive/Negative energy path integrals} \label{relpi}

In the previous two subsections, we have argued that 
\begin{itemize}
\item The coefficient of $z_i^m \bar z_f^n$ in the logarithm of the `In Out' path integral is given by summing all Feynman diagrams with $m$ legs on the past boundary and $n$ legs on the future boundary.
\item The coefficient of $x_i^m x_f^n$ in the logarithm of the Dirichlet path integral is given by summing all Feynman diagrams with $m$ legs on the past boundary and $n$ legs on the future boundary.
\end{itemize}
It follows that the `in-out' partition function as a function of $z_i$ and ${\bar z}_f$, is very similar to the Dirichlet partition function as a function of $x_i$ and $x_f$. The differences between these two functions are due to 
\begin{itemize}
\item[1] The `in-out' propagators differ from the Dirichlet counterparts. 
\item [2] The `in-out' partition function is given by a path integral with the potential $V(x)$, while the Dirichlet path integral uses the 
potential ${\tilde V}(x)$
\item[3] Tadpole graphs are absent in the `in-out' path integral, 
while they are present in the Dirichlet path integral 
(see \eqref{tadpoles} and \eqref{dirichlettadpole}).
\end{itemize}

The differences above tell us that the two path integrals are, indeed, distinct functions of their arguments. Nonetheless, we will now argue that an important part of the two answers is identical. 

Recall that the contribution of any Feynman diagram (to the path integral) is given by multiplying propagators, and then performing an integral over the location of all 
interaction vertices. We will now explain that the two integrands (i.e. the one for the in/out problem and the one for the Dirichlet problem) generically disagree with each other when one or more of the interaction vertices are located near one of the two boundary surfaces ($t=0$ or $t=T$) but agree with each other (upto corrections that are exponentially suppressed in $\omega T$) when all interaction vertices are 
located far from the boundary surfaces. The implication of this observation is the following. As we have seen in \eqref{s123}, the S matrix is encoded in the part of the path integral that develops an energy 
conserving delta function (see \eqref{singscat}). This delta function is a consequence of contributions that scale like $T$ when energies are conserved. This scaling comes about because the integrand becomes independent of the time of the insertion of the vertex operator precisely 
when energy is conserved. Of course terms that scale like $T$ do so precisely because they receive their dominant contribution from 
interaction vertices that are located in an `elevator' that is far from each of the boundaries. 

It follows that the demonstration (which we will turn to immediately below)
that the integrand for the in-out and Dirichlet Feynman diagrams agree 
when all insertion times are far from boundaries, guarantees that the 
terms in the path integral that are proportional to an energy conserving 
delta function (i.e. the terms in the action that contribute to scattering) are, in fact, identical between these two path integrals.

We now explain why the two integrands agree when interaction vertices lie far from the boundaries.  To see this, let us give a small Euclidean time to time coordinates appearing in the Dirichlet propagators i.e. $t \rightarrow t(1 - i \epsilon)$. When the interaction point $t$ lies deep in the bulk i.e. $0\ll t \ll T$, we can use
\beq
\sin[ \omega t(1- i \epsilon)] \approx \frac{e^{i \omega t}}{2 i}
\eeq
where we have dropped the exponentially decaying piece. Using \eqref{xbulk} and \eqref{diribtobo}, we find that in the limit, the `bulk to bulk' and `bulk to boundary' propagators in the Dirichlet case simplify to
\beq
\begin{split}
& G_D(t,t') \approx \frac{1}{2  \omega} \left[ e^{-i\omega (t'-t)} \Theta(t'-t) + e^{-i \omega (t-t')} \Theta(t-t') \right]    \\
& K_D^{(+)} (t) \approx e^{i \omega (t-T)}, \qquad \qquad K_D^{(-)}(t) \approx e^{- i \omega t}
\end{split}
\eeq
Comparing the above equations with the `in-out' propagators given in \eqref{btbprop} and \eqref{propdef}, we see that in this limit 
\begin{equation}\label{gd}
G_D(t,t')= \langle \frac{z(t) + {\bar z}(t)}{\sqrt{2 \omega} }\frac{z(t') + {\bar z}(t')}{\sqrt{2 \omega} }\rangle
\end{equation}
and that 
\begin{equation}\label{kdp}
K_D^{(+)} (t)= K_F^{(+)} (t), ~~~K_D^{(-)} (t)= K_F^{(-)}.
\end{equation}
In other words, in the above limit, all propagators in the Dirichlet and 
`in-out' problems become identical to each other. 

We now turn to a discussion of the vertex factors. The second apparent difference between the Dirichlet answer and the `in-out' answer lies in the potentials: recall  that potentials $V(x)$ and $\tilde V(x)$ are not identical but are related to each other by normal ordering \eqref{wtherm}. However, this difference is precisely compensated for by the different 
treatment of self contractions (tadpoles) of two $x(t)s$ from the same interaction vertex. Recall from \eqref{tadpoles} that such self contraction tadpoles are absent in the in out path integral (this is a consequence of the fact that the potential, in this formalism, was already normal ordered). On the other hand, the analogous tadpoles are present in the 
Dirichlet problem, and their contributions are given by \eqref{dirichlettadpole}. In the limit that $t$ and $T-t$ are both large, 
it is easily checked that the RHS of \eqref{dirichlettadpole} reduces to 
$\frac{1}{2\omega}$. In the free harmonic oscillator $<0|x^2 |0>=\frac{1}{2\omega}$. It follows that (in the relevant time window) the net effect of the tadpoles is simply to affect the replacement 
$$ V \rightarrow V_{\rm eff} =e^{ \frac{\partial_x^2}{4 \omega}} {V},$$
and then to work with the effective potential $V_{\rm eff}$ with no tadpole contractions. From \eqref{wtherm}, that $V_{\rm eff}$
equals ${\tilde V}$. In other words the effective potential for the 
Dirichlet problem (once we have accounted for Tadpoles) is identical to that for the in-out problem. 

In summary, we have demonstrated that propagators and (effective) vertex factors for the Dirichlet and in-out problems both agree exactly with each other when all vertices are inserted at times $t$ such that $\omega t\gg 1$ and 
$\omega(T-t) \gg 1$, as we set out to show.

 \subsection{Explicit computation of the On-shell Action}\label{expc}
 In this subsection, we explicitly compute the on-shell action as a function of boundary values both for the `in-out' boundary conditions and for Dirichlet boundary conditions. We consider an anharmonic oscillator with quartic potential and compute the on-shell action to the leading order in the quartic coupling.\\
 \textbf{In-out action}\\
Let us first consider the on-shell action \eqref{Onshellaction} as a function of positive energy data $z_i$ in the past and negative energy data $\bar z_f$ in the future for a quartic coupling i.e. 
 \beq\label{quarticpot}
 V(z+ \bar z) = \frac{\lambda}{16 \omega^2}(z+ \bar z)^4.
 \eeq
Using \eqref{fullac}, we find that at leading order in $\lambda$, the on shell action is given by
 \begin{equation}\label{onac12}
    \begin{split}
        i \mathcal{S} &= {\bar z}_f z_i e^{-i\omega  T }-\frac{i\lambda e^{-2 i \omega  T } }{32 \omega ^3}  \left(\left({\bar z}_f^4+z_i^4\right) \sin 2\omega T+12 \omega T  {\bar z}_f^2 z_i^2+8 {\bar z}_f z_i \left({\bar z}_f^2 + z_i^2 \right) \sin \omega T\right)+O\left(\lambda ^2\right)
    \end{split}
 \end{equation}
 In the large $T$ limit, we obtain
  \begin{equation}\label{onac1}
    \begin{split}
        i \mathcal{S} &= {\bar z}_f z_i e^{-i\omega  T }-\frac{\lambda }{64 \omega ^3} \left({\bar z}_f^4+z_i^4\right) -i\frac{3\lambda T}{8 \omega^2}   {\bar z}_f^2 z_i^2 e^{-2i\omega T}-\frac{\lambda }{8 \omega ^3}{\bar z}_f z_i \left({\bar z}_f^2 + z_i^2 \right) e^{- i \omega T}+O\left(\lambda ^2\right)
    \end{split}
 \end{equation}
As we are working only to leading order in $\lambda$, the action receives contributions from no loop diagrams (recall that tadpoles vanish in the in-out path integral). Exchange diagrams also do not contribute at this order.\\

\textbf{Dirichlet action}\\ 
Using \eqref{vharm}, we find that the potential $\tilde V(x)$ corresponding the the potential \eqref{quarticpot} is given by
\beq\label{nopot}
\tilde V(x) = \lambda\left(\frac{x^4}{4}-\frac{3 x^2}{4 \omega }+\frac{3}{16 \omega ^2}\right)
\eeq
Finally, in the large $T$ limit\footnote{To take the large $T$ limit, we give small Euclidean time to all the terms in the action}, the on-shell action \eqref{diriaction} with the above potential contains three pieces
\beq
i\mathcal{S}_{\rm bulk}  = i\mathcal{S}_{f} + i\mathcal{S}_{\rm contact} + i\mathcal{S}_{\rm tadpole} 
\eeq
where 
\beq\label{dirlarget1}
\begin{split}
i\mathcal{S}_{\rm f} &= -\frac{i \omega}{2} \left(x_f^2+x_i^2\right) +2 i \omega e^{- i \omega T} x_f x_i\\
i\mathcal{S}_{\rm contact} &=-\frac{\lambda}{8 \omega }\left(-\frac{3}{\omega} \left(x_f^2+x_i^2\right)+\frac{1}{2} \left(x_f^4+x_i^4\right)\right)- \frac{ \lambda  e^{-i T \omega }}{4 \omega }\left(x_f^3 x_i+x_f x_i^3\right)\\
&- i\frac{3}{2}  \lambda  T e^{-2 i T \omega }~  x_f^2 x_i^2 +\frac{3 i\lambda  T }{2 \omega } e^{-i T \omega }x_f x_i-i\frac{3 \lambda  T}{16 \omega ^2}\\
i\mathcal{S}_{\rm tadpole} &= -\frac{3\lambda}{16 \omega } \left(x_f^2+x_i^2\right)-\frac{3 i\lambda  T }{2 \omega } e^{-i T \omega }x_f x_i+i\frac{3 \lambda  T}{16 \omega ^2}
\end{split}
\eeq
Therefore the full action is given by
\beq\label{dirlarget}
\begin{split}
i\mathcal{S}_{\rm bulk} &= -\frac{ \omega}{2} \left(x_f^2+x_i^2\right) +2  \omega e^{- i \omega T} x_f x_i -\frac{\lambda}{16 \omega }\left(-\frac{3}{\omega} \left(x_f^2+x_i^2\right)+\left(x_f^4+x_i^4\right)\right)\\
&- \frac{ \lambda  e^{-i T \omega }}{4 \omega }\left(x_f^3 x_i+x_f x_i^3\right)
- i\frac{3}{2}  \lambda  T e^{-2 i T \omega }~  x_f^2 x_i^2 
\end{split}
\eeq\\

\textbf{Comparing S-matrix}\\
By comparing the `in-out' action given in \eqref{onac1} and the Dirichlet action given in \eqref{dirlarget}, it is easy to check that the S-matrix piece i.e. $\bar z_f^2 z_i^2$ piece in \eqref{onac1} and $x_i^2 x_f^2$ piece in \eqref{dirlarget}, match (after taking into account the factors of $\frac{1}{\sqrt {2 \omega}}$ that appear in the relationship between the variables $z$ and $x$.). \\

\textbf{Comparing Vacuum Wavefunction }\\
Using the onshell action \eqref{onac1}, we find that the vacuum wavefunctional of the harmonic oscillator receives corrections at order $\lambda$. At the leading order it is given by
\beq \label{wfon}
\Psi(\bar z)_c = 1-\frac{\lambda}{64 \omega^3} \bar{z}^4
\eeq
For the Dirichlet problem, the vacuum wavefunctional can be computed by setting $x_i$ to zero in \eqref{dirlarget}. We find that, at leading order in $\lambda$, the vacuum wavefunctional is given by
\beq\label{dirwf}
\Psi(x)_D = \left(1 - \lambda\left(\frac{1}{16 \omega} x_f^4+\frac{3}{16\omega^2} x_f^2  \right)\right) e^{-\frac{ \omega}{2} x_f^2}
\eeq

Both the answers exactly match the corrections to the wavefunction computed using Hamiltonian perturbation theory as shown below.

Using time-independent perturbation theory, at linear order, the corrections to the vacuum state are given by
\begin{equation}\label{vw}
 |\psi\rangle = |0\rangle + \sum_{m \neq 0}^{\infty}\frac{V_{m0}}{E_0^{(0)}- E_m^{(0)}} |m\rangle
\end{equation}
where $V_{m0} = \langle m| V(x)|0 \rangle$ and $V(x)$ is the potential given in \eqref{quarticpot}. Since the potential is quartic and is normal ordered, only one term in the above sum contributes and we find
\beq
 |\psi\rangle = |0\rangle +\frac{V_{40}}{E_0^{(0)}- E_4^{(0)}} |4\rangle
\eeq
Using the above formula, we can compute the coherent state wavefunction ($\Psi(\bar z)_{c}$) as well as the Dirichlet wavefunction ($\Psi (x_f)_D$), by taking the inner product with $\langle z|$ and $\langle x|$ respectively. We find the following wavefunctions
 \beq
 \begin{split}
\Psi(\bar z)_{c} &:= \langle z|\psi\rangle = 1-\frac{\lambda}{64 \omega^3} \bar{z}^4\\
\Psi (x_f)_D &:= \langle x_f|\psi\rangle = 1 -\frac{\lambda }{256 \omega ^3} H_4 (\sqrt{\omega} x_f) e^{-\frac{\omega}{2}x_f^2}
\end{split}
\eeq
where $H_4 (x)$ is the fourth Hermite polynomial. The above expressions exactly match \eqref{wfon} and \eqref{dirwf} respectively \footnote{Hermite polynomials have a constant piece which is absent in \eqref{dirwf}, but from the perspective of path integral, the constant piece just shifts the zero point energy.}.

The map between the coherent state wave function \eqref{wfon} and the 
position space wave function \eqref{dirwf} may be understood by noting that
\beq
e^{- \frac{1}{4 \omega}\partial^2_x} x^n = H_n (x)
\eeq
Therefore both wavefunctions are related to each other via the following relation
\beq
e^{- \frac{1}{4 \omega}\partial^2_x} \Psi(x)_c = \Psi(x)_D ~ e^{-\frac{\omega}{2}x^2}
\eeq
Note that the above relation is exactly the same as the relationship between the Dirichlet potential $\tilde V (x)$ and the `in-out' potential $V(x)$ given in \eqref{vharm}.

\subsubsection{Four different oscillators}\label{4ho}
In this subsubsection, we consider four different oscillators interacting via a quartic coupling and compute the `in-out' onshell action for this case. This Lagrangian can be thought of as a toy version of the field theory Lagrangian with $\phi^4$ coupling. This is because the scalar field $\phi$ can be written as an infinite sum over harmonic oscillators labeled via $\vec{k}$ and the quartic coupling couples four such harmonic oscillators. The Hamiltonian for this system is given by
\begin{equation}
     \begin{split}
          \mathcal{H} &=\sum_{k=1}^4\omega_k z_k {\bar z}_k + \lambda \prod_{k=1}^4\frac{1}{\sqrt{2\omega_k}} (z_k+\bar{z}_k)
     \end{split}
 \end{equation}

Using the `bulk to boundary' propagators given in \eqref{propdef}, we find that, at the leading order, the on-shell action takes the following form

\beq
\mathcal{S} = \mathcal{S}_f + \mathcal{S}_\lambda
\eeq
where $\mathcal{S}_f$ is the free part i.e. the $\lambda$ independent contribution to the action whereas $\mathcal{S}_\lambda$ is the first order correction to the on-shell action. Both these pieces are given by
\begin{equation}\label{4har}
    \begin{split}
    i\mathcal{S}_f =& \sum_{k= 1}^{4} z_{i_{k}} {\bar z}_{f_{k}} e^{- i \omega_k T}\\
    i\mathcal{S}_{\lambda}    =&  -\frac{ i \lambda  {\bar z}_{f_1} {\bar z}_{f_2} {\bar z}_{f_3} {\bar z}_{f_4} \left(-1+e^{-i T \left(\omega _1+\omega _2+\omega _3+\omega _4\right)}\right)}{4 \sqrt{\omega _1 \omega _2 \omega _3 \omega _4} \left(\omega _1+\omega _2+\omega _3+\omega _4\right)} + \frac{ i \lambda  z_{i_1} z_{i_2} z_{i_3} z_{i_4} \left(1-e^{-i T \left(\omega _1+\omega _2+\omega _3+\omega _4\right)}\right)}{4 \sqrt{\omega _1 \omega _2 \omega _3 \omega _4} \left(\omega _1+\omega _2+\omega _3+\omega _4\right)}\\
        &+ \frac{ i \lambda  {\bar z}_{f_3} {\bar z}_{f_4} z_{i_1} z_{i_2}  \left(e^{-i T \left(\omega _3+\omega _4\right)}-e^{-i T \left(\omega _1+\omega _2\right)}\right)}{4 \left(\omega _1+\omega _2-\omega _3-\omega _4\right) \sqrt{\omega _1 \omega _2 \omega _3 \omega _4}} -\frac{ i \lambda  {\bar z}_{f_1} {\bar z}_{f_2} z_{i_3} z_{i_4} \left(e^{-i T \left(\omega _1+\omega _2\right)}-e^{-i T \left(\omega _3+\omega _4\right)}\right)}{4 \left(\omega _1+\omega _2-\omega _3-\omega _4\right) \sqrt{\omega _1 \omega _2 \omega _3 \omega _4}}
    \end{split}
\end{equation}

In the main text, we work with a `slab' boundary which consists of two spacelike slices at initial time $-T$ and final time $T$. In this Appendix, we work with the boundaries at time $t_{in} = 0$ and $t_{f} = T$. The results derived above can be easily translated to the case where boundaries are at $-T$ and $T$. This changes the  `bulk to boundary' propagator $K_F^{(+)}(t)$ to
$$K_F^{(+)}(t) = e^{- i \omega(t+T)}$$
so that $K_F^{(+)}(-T) = 1$. This amounts to replacing $T$ by $2T$ in the on-shell action \eqref{4har}.

The results of this computation can be easily generalized to field theory, where we have an infinite number of different harmonic oscillators labeled by the momentum $\vec{k}$. We find,

\beq \label{sonriw2}
i\mathcal{S}_\lambda \propto - \int \prod_i d^d k_i ~\frac{i \lambda \left(e^{- 2i T E_{in} }-e^{- 2i T E_{out}}\right)}{(E_{in} -  E_{out})} (2 \pi)^d \delta^d  \left( \sum_{i} \vec k_i^{\rm in} -
\sum_{j} \vec k_j^{\rm out} \right) \prod_j \bar\beta_{k_j} \prod_i \beta_{k_i} 
\eeq

where we used the fact that $z_i = (2 \pi)^d \sqrt{2\omega}\beta_{k_i}$ and similarly for $\bar{z}_i$.

\section{Exchange Saddles: Lorentzian and Euclidean}\label{esap}

In subsection \ref{es} we have studied correlators with  `scattering kinematics', i.e. correlators whose insertions were located in such a manner that the equations \eqref{spexch} admit real solutions. This is not always the case. In this section, we study a particular one parameter class of insertion configurations, and study the existence (and nature) of the saddle point solutions as a function of this parameter. 

Consider a configuration in which our four insertions are located a the vertices of a rectangle, whose sides are parallel to the $x$ and $y$ axes
in spacetime. Let us label the insertions, respectively, as $1, 2, 3, 4$, moving anticlockwise starting from `southwest'. Our insertions have coordinates $(x, y)$ (all other coordinates are zero)
\begin{equation}\label{insertloc}
1: (-a, -b), ~~~2: (a, -b), ~~~3: (a, b), ~~~4: (-a, b)
\end{equation}
We assume that the external insertions are all of a field that creates a particle of mass $m$, and search for exchange saddles in which particles 1 and 2 come together and fuse into an intermediate particle, of mass $M$,  at the location $(0, -(b-y))$. The intermediate particle then propagates to 
the point $(0, b-y)$ where it, once again, bifurcates into two particles 
each of mass $m$, which propagates upto insertions 3 and 4. 

As we have explained above, in this section we focus on exchange saddles of the form $(1,2 )\rightarrow $ intermediate  $\rightarrow (3,4)$. There, of course also exist other crossed saddles (e.g. $(1, 4) \rightarrow$ intermediate  $ \rightarrow ((2,3)$. The analysis of these `crossed' saddles is very similar to that of the `direct' saddles studied in this section and is left as an exercise to the interested reader.

When $b$ is real the $y$ direction is spacelike. When $b$ is imaginary 
the $y$ direction is timelike. We will consider both possibilities. 

In the situation presented above, it is not difficult to analyze when saddle points of the sort described above exist, and how many such solutions exist. 
In this Appendix, we summarize the results of this investigation. The results turn out to depend sensitively on whether $M >2 m$ or $M< 2 m$, so we consider the two cases separately. 

\subsection{$M< 2 m$}

In this case a single real solution exists provided that $b$ is real and 
\begin{equation}\label{condone}
\frac{b}{a}> \frac{M}{\sqrt{4m^2-M^2}}
\end{equation}
The solution exists at a value of $y$ given by 
\begin{equation} \label{fsol}
y= \frac{aM}{\sqrt{4m^2-M^2}}
\end{equation}

No solutions (real or complex) exist when $b$ is real and \eqref{condone} is violated.  No solutions exist when $b$ is imaginary 

\subsection{$M> 2m$}

In this case, we have a pair of complex solutions for every real value of 
$b$. The solutions are given by the analytic continuation of \eqref{fsol}, i.e. by 
\begin{equation} \label{fsol}
y= \pm i  \frac{aM}{\sqrt{M^2-4m^2}}
\end{equation}
On this saddle point, $y^2> a^2$ so the external and intermediate particles both propagate in a complex manner (the spacetime interval for propagation, in both cases, is complex). The solution presented above, however - atleast formally - solves the saddle point equations. 

When $b$ is imaginary, on the other hand, solutions to the saddle point equation exist only when 
\begin{equation}\label{condtwo}
\left|\frac{b}{a} \right| > \frac{M}{\sqrt{M^2-4 m^2}}
\end{equation}
When this condition is met, the solution is once again given by 
\eqref{fsol}, except that the sign in the solution is chosen so that 
$\frac{y}{b} >0$. This is the scattering exchange saddle point studied in 
\S\ref{es}.

\subsection{Analytic Continuation}

As is clear from the results of this subsection, the saddle point contribution to correlators is not an analytic function of either the masses 
or of the insertion locations. Imagine starting with imaginary $b$ and in a configuration that obeys \eqref{condtwo}. If we now lower $M$ (keeping $b$
and $a$ fixed), the saddle described around \eqref{condtwo} continues to exist down to  a critical value of $M$. At values of $M$ smaller than this critical value, the saddle point has simply disappeared - it does not exist anymore. It follows, in other words, that the scattering saddle point exists only in a certain domain of parameters. Similar statements apply to the Euclidean saddle.

\section{Geometry of intersection of the lightcones}\label{lightconeapp}

In this Appendix, we discuss the geometry of the intersection of lightcones of $m$ bulk points (with coordinates $x_i$, $i=1 \ldots m$, in $(D-1,1)$ flat spacetime dimensions.

\subsection{$m \leq D+1$}

Let us first focus on the case of $c=1$ i.e. on the case when the number of insertions $m \leq D+1$. As in the main text, we define the scattering manifold to be the subset of $R^{D-1, 1}$ spanned by vectors $x_i -x_j$ (for $i, j=1 \ldots m$), i.e. the vectors that run between two insertion points. This scattering manifold is generically $m-1$ dimensional. Generically, this $m-1$ dimensional subspace is either an $R^{m-2, 1}$ or an 
$R^{m-1, 0}$ (exceptional situations, in which the subspace is null or less than $m-1$ dimensional are also possible, but we will not consider these cases here). 

We wish to find the manifold of points, $y$, that are null separated from each $x_i$, i.e. the manifold of points 
$y$ that obey the equation 
\begin{equation}\label{allnull}
(x_i-y)^2=0, ~~~~i= 1 \ldots m
\end{equation}

\eqref{allnull} is a set of $m$ nonlinear equations. By taking linear combinations, however, we can regroup these
into a set of $m-1$ linear equations and one nonlinear 
equations. The $m-1$ linear equations are 
\beq\label{equi}
\begin{split}
&(x_i - y)^2 = (x_j - y)^2~~~~~~\forall~~ i,j = 1 \ldots m\\
& 2 (x_j - x_i).y = x_j^2 - x_i^2
\end{split}
\eeq
The first line of \eqref{equi} asserts that $y$ is equidistant from each of the insertion points. The 
(algebraically equivalent) second line of \eqref{equi} 
tells us that this condition is satisfied when the dot products of the vector $y$ with a basis of vectors in the 
scattering plane, each take a certain value. This set of linear equations completely determines $y_{\parallel}$ the projection of $y$ in the scattering plane. However, these equations leave the part of $y$ that is orthogonal to the scattering plane, $y_{\perp}$ completely unfixed. 

Let us first consider the case of maximal physical interest to us, namely the case in which the scattering manifold is $R^{m-2, 1}$ (this is the case that is relevant to scattering from our slab boundary). In this case the space orthogonal to $R^{m-2, 1}$ is the Euclidean space $R^{D-m+1}$. Depending on the details of the insertions, we have three possibilities.

\begin{itemize}
    \item The insertions are all spacelike separated from $y_{\parallel}$ i.e. $(x_i - y_{\parallel})^2 = r^2>0$ 
    It follows that 
    $$(x_i-y)^2 = (x_i - y_{\parallel})^2 +   y_{\perp}^2 \geq r^2 >0.$$
    In this case, the lightcones emanating out of $\{x-i \}$ do not intersect.
    \item  The insertions are all timelike separated from $y$ i.e. $(x_i - y)^2 = -r^2 <0$  In this case
    $$(x_i-y)^2 = -r^2 +  y_{\perp}^2 \geq r^2 $$
    and so the condition $(x_i - y)^2 =0$ is met on 
    an $S^{D-m}$ of radius $r$ in $R^{D-m+1}$. 
     We see that, in this case, the lightcones intersect on a compact spacelike surface, namely an 
     $D-m$ dimensional sphere. As explained above, the radius of this sphere is $r$. This radius shrinks to 
     zero when $r^2=(x_i - y)^2 \to 0$. This is the case we turn to next.  
    \item All the insertions are null separated from $y$ i.e. $(x_i - y_{\parallel})^2 = 0$. In this case lightcones intersect only when $y_{\perp}^2=0$, i.e. 
    when $y_{\perp}=0$. In this case the $y=y_{\parallel}$
    and the common intersection of the lightcones is a single point - one that lies on the scattering plane.
    This is the case that gives rise to a scattering singularity in the main text, as, in this case, 
    it is always possible to satisfy momentum conservation.
    (momentum cannot be conserved in the previous case because the perpendicular component of the vector 
    $\sum_{i} (\omega_i (x_i-y)$ equals $(\sum_i \omega_i) y_{\perp}$, which cannot vanish if all $\omega_i$ are non negative, and at least one of them is nonzero). 
 \end{itemize}

Although this will not be of great interest to the current paper, let us also briefly consider the case that the 
scattering plane is spacelike, i.e. that it is an $R^{m-1,0}$. In this case, the orthogonal plane is $R^{D-m,1}$. 
In this situation, it is always the case that  $(x_i-y_{\parallel})^2=r^2>0$. The simultaneous intersection of 
lightcones occurs when 
\begin{equation}\label{hyp}
y_{\perp}^2= -r^2
\end{equation}
i.e. on a hyperboliod of `radius' $r$ in $R^{D-m,1}$. Note that the lightcone intersection manifold is noncompact in this case. Of course, this case is incompatible with momentum conservation as above.

\subsection{$c>1$}

We have argued above that in the $c=1$ case, whenever the lightcones intersect, there is a unique bulk point for which momentum conservation can be satisfied.

Next, we consider the $c>1$ case i.e. $m> D+1$. In this case, the scattering manifold generically covers the full D dimensional spacetime: we consider this generic case. 

Let us again try to find a bulk point $y$ which is equidistant from all the boundary insertions. We need to find solutions to equation \eqref{equi}. For $c>1$, the number of equations ($m-1$) is greater than the number of variables in $y$ ($D$), generically there are no solutions. But we can perform $m-D-1$ tunings of the boundary points such that \eqref{equi} can be solved. 
Generically, the solution of this linear equation is unique (further tuning of $\{ x_i \}$can lead to non unique solutions: we do not consider this possibility here). 

Once \eqref{equi} is solved, one additional tuning is required to make $(x_i-y)^2$ zero. The total number of tunings is, thus $m-D$, explaining the codimension of the singular manifold in the case $c$. Notice that the 
common intersection of lightcones - when it exists -
is generically unique.

\section{Zero eigenvalues of $N$ implies existence of bulk point}\label{reverse}
In this Appendix, we show that if the boundary insertions are chosen such that \eqref{zeroeig} is satisfied, then it implies that there exists a bulk point $y$ which is null separated from all the boundary insertions.

Let us assume that the boundary distance matrix $N_{ij}$ has a zero eigenvalue and all entries of the corresponding eigenvector have positive entries  i.e.
\beq
\begin{split}
&\sum_i \omega_i N_{ij} = \sum_i \omega_i (x_i -x_j)^2 = 0\\
&\sum_i \omega_i x^2_i + \sum_i \omega_i x_j^2 - 2 x_j.\sum_i x_i \omega_i =0
\end{split}
\eeq
Completing the square, we obtain
\beq\label{sw1}
\left(\sum_i \omega_i\right)\left(x_j - \frac{\sum_i \omega_i x_i}{\sum_i \omega_i}\right)^2 + \sum_i \omega_i x^2_i - \frac{(\sum_i \omega_i x_i)^2}{\sum_i \omega_i} = 0
\eeq
 Note that, in $m>D$ case, for the generic zero eigenvectors satisfying \eqref{zev1}, the above manipulations cannot be performed because
 $\sum_i \omega_i$ vanishes for those eigenvectors. Therefore, division by that quantity is not allowed. But for the insertions on the singular manifold, all the entries of the zero eigenvector $\omega$ are positive and, hence, $\sum_i \omega_i \neq 0$.

Let us next analyse the equation obtained by multiplying the eigenvalue equation by $\omega_j$ and summing over $j$
\beq\label{sw2}
\begin{split}
&\sum_{i,j} \omega_i \omega_j (x_i- x_j)^2 = 0\\
& 2 \sum_i \omega_i \sum_j \omega_j x_j^2 - 2 \left(\sum_i \omega_i x_i\right)^2 = 0
\end{split}
\eeq
Using \eqref{sw2} in \eqref{sw2}, we obtain:
\beq
\left(\sum_i \omega_i\right)\left(x_j - \frac{\sum_i \omega_i x_i}{\sum_i \omega_i}\right)^2  = 0
\eeq
Since $\sum_i \omega_i \neq 0$, we obtain
\beq\label{sw3}
\left(x_j - \frac{\sum_i \omega_i x_i}{\sum_i \omega_i}\right)^2  = 0
\eeq
Above equation implies that all boundary points $x_j$'s are null separated from a point given by $ y = \frac{\sum_i \omega_i x_i}{\sum_i \omega_i}$.

Hence, as promised, we proved that if the distance matrix has a zero eigenvalue and the corresponding eigenvector has all positive entries, there exists a point $y$, which is null separated from all the insertions.

\section{Conformal Invariance of the submanifold 
$det N=0$} \label{cis}

 We shall now show that the vanishing of $det N$ is a conformally invariant condition on the matrix elements of $N$. It suffices to show that the matrix elements of $N$ transforms homogeneously under a special conformal transformation ${x^\mu} \rightarrow x'^{\mu}$. A special conformal transformation, being decomposable as an inversion followed by a translation by an arbitrary vector $b^\mu$ and another inversion, is given by
 \begin{equation}
     \frac{x'^{\mu}}{x'^2}=\frac{x^{\mu}}{x^2}+b^\mu
 \end{equation}
 It immediately follows that under such a transformation, each insertion point $x_i \rightarrow x'_i$ such that
 \begin{equation}
     \begin{split}
         x_i'^2= \frac{x_i^2}{1+2 b \cdot x_i +b^2 x_i^2}\equiv\frac{x_i^2}{\Omega(x_i)} 
     \end{split}
 \end{equation}
 and also that
  \begin{equation}
     \begin{split}
         x_i'\cdot x_j'&=x^{'2}_i x^{'2}_j\left(\frac{x_i^{\mu}}{x_j^2}+b^\mu\right)\left(\frac{x_j^{\mu}}{x_j^2}+b^\mu\right) \\
         &=\frac{1}{\Omega(x_i)\Omega(x_j)}\left(x_i.x_j+x_j^2 (b \cdot x_i)+x_i^2 (b \cdot x_j)+b^2 x_i^2 x_j^2\right)
     \end{split}
 \end{equation}
With this, we see that under a special conformal transformation, the individual matrix elements of $N$ transform as $(x_i-x_j)^2 \rightarrow (x'_i-x'_j)^2$ given by,
\begin{equation}
    \begin{split}
        (x'_i-x'_j)^2&=\frac{1}{\Omega(x_i) \Omega(x_j)}\Big(x_i^2\left(\Omega(x_j)-2 (b \cdot x_j)- b^2 x_j^2\right)+x_j^2(\Omega(x_i)-2 (b \cdot x_i)- b^2 x_i^2) -2 x_i \cdot x_j \Big) \\
        &=\frac{1}{\Omega(x_i) \Omega(x_j)}\left(x_i-x_j \right)^2
    \end{split}
\end{equation}
 This means under the special conformal transformation, the $i^{th}$ row and the $i^{th}$ coloumn of the matrix $N$ each    pick up a factor of $\frac{1}{\Omega(x_i)}$ so the determinant of the matrix $N$ trasnforms into
 \begin{equation} \label{detNconf}
    \rm{ Det} N'=  \frac{1}{\Omega(x_1)^2 \Omega(x_2)^2 \ldots \Omega(x_m)^2} {\rm Det} N   
 \end{equation}
This proves that the condition ${\rm Det} N=0$ is left invariant by special conformal transformations. In fact, as we have already noted, \eqref{detNconf} also represents the transformation of the determinant of $N$ under every other conformal transformation (with a constant conformal factor $\Omega(x_i)$, the constant being equal to unity for Poincare transformations). Totally then, the condition on the coordinates of the insertions obtained by setting ${\rm Det} N=0$ does not change under any conformal transformation. This proves that the vanishing of the determinant of $N$ is a condition that only involves the cross ratios of the insertions.

We now illustrate this assertion for the special case of 4 and 5 points\footnote{For the case of 2 and 3 points, it is clear that $det N$ vanishes only when some of the insertions are coincident.}. For $m$ points, the number of cross ratios is $\frac{m(m-3)}{2}$\footnote{When $m<D$ these cross ratios are not all independent of each other.}. In the case of four points, we define the two independent cross ratios as follows
$$u=\frac{N_{12}N_{34}}{N_{13}N_{24}}, \qquad \qquad v=\frac{N_{14}N_{23}}{N_{13}N_{24}}.$$ where $N_{ij}$ are the Lorentzian distances between the insertion points $i$ and $j$ (the $ij^{\rm th}$ element of the matrix $N$). We find that the determinant of $N$ can be written in the following form
\begin{equation}
    {\rm Det} N=(N_{13})^2 (N_{24})^2 (1+u^2+v^2-2u-2v-2uv).
\end{equation}
Further making the redefinitions $u=Z \zbar$ and $v=(1-Z)(1-\zbar)$, we find
\begin{equation}\label{4detn}
    {\rm Det} N=(N_{13})^2 (N_{24})^2 (Z-\zbar)^2
\end{equation}
Hence, as long as no two insertions coincide, the singularity condition ${\rm Det}N=0$ is met precisely when the cross ratios satisfy $Z=\zbar$. 

In the case of five points, there are 5 cross ratios, and it turns out that they can be formed as follows. First, we pick two sets of four points each, say $\{x_1, x_2,x_3,x_5\}$ and $\{x_1, x_2,x_4,x_5\}$, and form four-point cross ratios involving these two sets of points. There are two of these ratios corresponding to each set, for a total of four ratios. We denote the four-point cross ratios involving the set of points $\{x_1, x_2,x_3,x_5\}$ as
\beq
\begin{split}
 & u_1 = \frac{N_{13}N_{25}}{N_{12}N_{35}}, \qquad v_1 = \frac{N_{15}N_{23}}{N_{12}N_{35}},
\end{split}
\eeq
and the four-point cross ratios involving the set of points $\{x_1, x_2,x_4,x_5\}$ as
\beq
\begin{split}
 &u_2 = \frac{N_{14}N_{25}}{N_{12}N_{45}}, \qquad v_2 = \frac{N_{15}N_{24}}{N_{12}N_{45}}. 
\end{split}
\eeq
The fifth cross ratio is formed from all five points and can be written as
\beq
\begin{split}
 s_5 = \frac{N_{15}N_{25}N_{34}}{N_{12}N_{35}N_{45}}. 
\end{split}
\eeq
Now, we can express the determinant of $N$ as
\beq \label{det5}
{\rm det} N = -2 \frac{N^3_{12}N^2_{35}N^2_{45}}{N_{15}N_{25}}\left(s_5 - (Y-Z)(\bar Y - \zbar\right)\left(s_5 - (\bar Y-Z)(Y - \zbar\right)
\eeq
where we used $u_1 = Z \zbar, v_1 = (1-Z)(1-\zbar)$ and $u_2 = Y \bar Y, v_2 = (1-Y)(1-\bar Y)$. Once again, we see that the vanishing of $det N$ is a condition on the conformal cross ratios only, namely it vanishes when either $s_5 - (Y-Z)(\bar Y - \zbar)=0$ or $s_5 - (\bar Y-Z)(Y - \zbar)=0$\footnote{It might seem that there are other potential zeros of $det N$, namely when the denominator of the prefactor in \eqref{det5} becomes infinite i.e $N_{15}$ or $N_{25}$ becomes infinite. However, it is easy to see that these are not zeros of $det N$ either by explicitly evaluating the rest of the expression at these points, or by noting that the elements of the matrix $N_{ij}$ are distances, so its determinant, being just a sum of products of distances, can't possibly vanish when any of the distances becomes infinite.}.

 \section{$\dflat$ functions} \label{dflat}
In this section, we will compute exact answers for the contact contributions of Euclidean $n$-point massless correlators, given by \eqref{ncontact}.
As shown in Sec. \ref{sec:massless}, we can write this in the form\footnote{As in Sec. \ref{sec:massless}, we have dropped an overall numerical factor given by $\mathcal{A} = \frac{i g}{4^4 \pi^{3D/2}}$, and we shall restore this in the final answers.} 
\begin{equation} \label{lampow}
	    \begin{split}
	        \gf(x_1, \ldots , x_n)&= \Gamma\left(\Delta-{D\over2}\right)\prod_{i=1}^n \int_{0}^{\infty} d\omega_i\omega_i^{\Delta_i-1} \delta(1-\sum\limits_{i=1}^n \lambda_i \omega_i)  {(\sum\limits_{i=1}^n \omega_i)^{\Delta-D} \over \left(\sum \omega_i \omega_j N_{ij}\right)^{\Delta-D/2}}
	    \end{split}
\end{equation}
\textbf{3-point functions}



Four point conformal correlators are expressed in terms of the so-called $\mathcal{D}$-functions (\cite{DHoker:1999kzh}) of the conformal cross ratios, invariants of the conformal symmetry. We will encounter the analogues of these functions, which we call $\dflat$ functions, already at the 3 point stage. It is simple to see that the 3-point answer is a function of two invariants. Given three points, we use translations to put one to the origin, then rotations about the origin to put a second point on a fixed coordinate axis and then choose its distance from the origin by using overall scaling invariance. Then, we use rotations about the axis defined by the two points to put the third point in the same plane. All the information is then in specifying the third point which can be done using two numbers.

We evaluate the expression Eq.(\ref{lampow}) for the case of 3 boundary points to find
\begin{equation} 
	    \begin{split}
	         \gf(x_1, \ldots , x_n)&= \Gamma\left(\Delta-{D\over2}\right)\prod_{i=1}^3 \int_{0}^{\infty} d\omega_i \omega_i^{\Delta_i-1} \delta(1-\sum\limits_{i=1}^3 \lambda_i \omega_i)  {(\sum\limits_{i=1}^3 \omega_i)^{\Delta-D} \over \left(\sum \omega_i \omega_j N_{ij}\right)^{\Delta-D/2}}
	    \end{split}
\end{equation}
We use the Dirac delta function to choose $\omega_3=1$, and write the remaining integral in the form
\begin{equation} 
	    \begin{split}
	        \int dp p^{\Delta-D/2-1} \int d\omega_1 \omega_1^{\Delta_1-1} \int d\omega_2 \omega_2^{\Delta_2-1} (1+ \omega_1+\omega_2)^{\Delta-D}  e^{-p \left(\omega_1 s_{13} +\omega_1 \omega_2 s_{12} +\omega_2 s_{23}\right)}
	    \end{split}
\end{equation}
Now, we can make use of the identity
\begin{equation}
    e^{-x}=\int_{c-i \infty}^{c+i \infty} {ds \over {2\pi i}} \Gamma(-s) x^{s}
\end{equation}
where $c$ is a positive number smaller than unity, and the contour is closed to the right so that the integral picks up the residues of all the poles of the Gamma function at positive integer values of $s$, and reproduces the series expansion of the exponential on the left. We use this identity to replace two of the exponentials in the integrand, $e^{-p \omega_1 s_{13}}$ and $e^{-p \omega_1 \omega_2 s_{12}}$ with
\begin{equation}
    \begin{split}
        e^{-p \omega_1 \omega_2 s_{12}}&=\int_{c-i \infty}^{{c+i \infty}} {dq \over {2\pi i}} \Gamma(-q) (p\omega_1 \omega_2 s_{12})^{q}\\
        e^{-p \omega_1 s_{13}}&=\int_{c-i \infty}^{{c+i \infty}} {dr \over {2\pi i}} \Gamma(-r) (p\omega_1 s_{13})^{r}
    \end{split}
\end{equation}
Henceforth, we will keep the integration limits implicit for the above integrals. The $\omega_1$ integral is then elementary so that we can easily evaluate it and find
\begin{equation} 
	    \begin{split}
	     &  \int  {dq \over {2\pi i}} \int {dr \over {2\pi i}} s_{12}^{q} s_{13}^{r}\Gamma(-q) \Gamma(-r)\frac{\Gamma(\Delta_1+q+r) \Gamma(D-\Delta-\Delta_1-q-r)}{\Gamma(D-\Delta)}\\
	     &\times \int dp p^{\Delta+q+r-D/2-1}  \int d\omega_2 \omega_2^{\Delta_2+q-1} (1+ \omega_2)^{\Delta+\Delta_1+q+r-D}  e^{-p \, \omega_2 s_{23}}
	    \end{split}
\end{equation}
We note that the $\omega_1$ integral is not convergent when the sum of the scaling dimensions $\Delta=D$. This is precisely the conformally invariant case which can be treated separately to find the usual three-point function for scalar fields in a conformal theory. We are, of course, interested in the case $\Delta\neq D$. 
Now, we can integrate over $p$ first and then over $\omega_2$, and find that
\begin{equation} 
	    \begin{split}
	        \gf(x_1, x_2, x_3)=& \frac{1}{s_{23}^{\Delta-D/2}\Gamma(D-\Delta)} \int  {dq \over {2\pi i}} \int {dr \over {2\pi i}} \,s^q \, t^r \, \Gamma(-q) \Gamma(-r) \\
	        &\times\Gamma(\Delta_1+q+r) \Gamma(\Delta-D/2+q+r) \Gamma(D/2-\Delta_1 -\Delta_3-q) \Gamma(D/2-\Delta_1 -\Delta_2-r) \\
	   \end{split}
\end{equation}
where we have defined the two cross ratios $s=\frac{s_{12}}{s_{23}}$ and $t=\frac{s_{13}}{s_{23}}$. 
We are now in a position to introduce our first $\dflat_{n,D}$ as a function of these two cross ratios,
\begin{equation} 
	    \begin{split}
	        \dflat_{3,D} (s,t)= \int  {dq \over {2\pi i}} \int {dr \over {2\pi i}}  \,s^q \, t^r \,&\Gamma(-q) \Gamma(-r) \Gamma(\Delta_1+q+r) \Gamma(\Delta-{D \over2}+q+r)  \\
	       &\times \Gamma({D \over2}-\Delta_1 -\Delta_3-q) \Gamma({D \over2}-\Delta_1 -\Delta_2-r) 
	   \end{split}
\end{equation}
so that, 
\begin{equation} 
	    \begin{split}
	        \gf(x_1, x_2, x_3)=& \frac{i g}{4^4 \pi^{3D/2}} {1 \over  \Gamma(D-\Delta)} \frac{1}{s_{23}^{\Delta-D/2}} \, \dflat_{3,D} (s,t), \\
	   \end{split}
\end{equation}
where we have restored the overall numerical factor. 

We can find an explicit form for $\dflat_{3,D} (s,t)$ as follows. We perform the $q$ and $r$ integrals by closing the contours to the right, i.e while keeping $\text{Re}(q) >0$ and $\text{Re}(r)>0$, and the points ${D \over2}-\Delta_1 -\Delta_3$ and ${D \over2}-\Delta_1 -\Delta_2$ to the left of the contours respectively. We then pick up residues of the poles of Gamma functions lying along the positive real axis i.e. when the arguments become negative integers. Doing so lands us at a closed form answer given in terms of one of the Appell Hypergeometrics of two variables, $F_4$,
\begin{equation}
    F_4(a,b,c,d;x,y)=\sum_{m,n=0}^{\infty} \frac{(a)_{m+n}(b)_{m+n}}{c_{(m)} d_{(n)}} \frac{x^m y^n}{m! n!}
\end{equation}
in terms of which we find
\begin{equation}
    \begin{split}
        \dflat_{3,D} (s,t)=&\Gamma(\Delta_1) \Gamma(\Delta-D/2) \Gamma(D/2-\Delta_{13}) \Gamma(D/2-\Delta_{12})\\&\times F_4(\Delta_1,\Delta-D/2,1-D/2-\Delta_{13},1-D/2-\Delta_{12};s,t) \\
        +&t^{D/2-\Delta_{12}} \Gamma(\Delta_3) \Gamma(D/2-\Delta_2)  \Gamma(D/2-\Delta_{13})\Gamma(\Delta_{12}-D/2) \\
        &\times F_4(\Delta_3,D/2-\Delta_2,1-D/2-\Delta_{13},1+D/2+\Delta_{12};s,t)\\
         +& s^{D/2-\Delta_{13}} \,\Gamma(\Delta_2) \Gamma(D/2-\Delta_3) \Gamma(D/2-\Delta_{12}) \Gamma(\Delta_{13}-D/2)\\& \times F_4(\Delta_2,D/2-\Delta_3,1+D/2+\Delta_{13},1-D/2-\Delta_{12};s,t)\\
         +&s^{D/2-\Delta_{13}} t^{D/2-\Delta_{12}}\, \Gamma(D-\Delta)\Gamma(D/2-\Delta_1)\Gamma(\Delta_{12}-D/2)\Gamma(\Delta_{13}-D/2) \\ \times &F_4(D-\Delta,D/2-\Delta_1,1+D/2+\Delta_{13},1+D/2+\Delta_{12};s,t) 
    \end{split}
\end{equation}
where we have defined $\Delta_{ij}=\Delta_i +\Delta_j$.

\textbf{4-point functions}
\newline
The 4-point $\dflat$ functions are functions of five invariants. We needed two numbers to specify three points. Rotations in axes perpendicular to the plane containing these three points can be used to fix a 3-plane containing all 4 points. Specifying the fourth point then requires three more numbers taking the total to 5.\\

Using similar reasoning as for the 3-point function, we find that
\begin{equation} 
	    \begin{split}
	        \gf(x_1, \ldots , x_4)&=\frac{i g}{4^4 \pi^{3D/2}} \frac{1}{\Gamma(D-\Delta)}  {1\over {s_{14}^{\Delta-D/2}}} \dflat_{4,D} (s_1,s_2,s_3,s_4,s_5)
	    \end{split}
\end{equation}
where $\dflat_{4,D}$ is a function of the simple ratios
$$
s_1={s_{34} \over s_{14}}, \ s_2={s_{23} \over s_{14}}, \ s_3={s_{13} \over s_{14}}, \ s_4={s_{24} \over s_{14}},\ s_5={s_{12} \over s_{14}}
$$
given by
\begin{equation}
\begin{split}
    \dflat_{4,D} (\{s_i\})= \prod_{i=1}^5 \int dt_i \, s_i^{t_i}\,\Gamma(-t_i) \,& \Gamma(\Delta_3+t_{123}) \Gamma(\Delta_2+t_{245}) \\ &\times \Gamma(D/2-\Delta_{234}-t_{124})  \Gamma(D/2-\Delta_{123}-t_{235}) \Gamma(\Delta-D/2+\sum_{i=1}^5 t_i)
\end{split}
\end{equation}
where $t_{ijk}=t_i+t_j+t_k$ and $\Delta_{ijk}=\Delta_i +\Delta_j+\Delta_k$. Performing the $s_i$ integrals lands us at a closed form answer which has 48 terms each involving a hypergeometric of five variables. One of these terms, for example, is given by 
\begin{equation}
    \begin{split}
        &\Gamma(\Delta_2)\Gamma(\Delta_3)\Gamma(\Delta-D/2)\Gamma(D/2-\Delta_{234})\Gamma(D/2-\Delta_{123})\\
        &\times D_5(\Delta_3,\Delta_2,\Delta-D/2,1-D/2+\Delta_{234},1-D/2+\Delta_{123},\{s_i\})
    \end{split}
\end{equation}
where we have introduced the following hypergeometric function of five variables,
\begin{equation}
    D_5(a_1,a_2,b,c,d;\{s_i\})=\sum_{i_1,\ldots i_5=0}^{\infty} \frac{(a_1)_{i_{123}}(a_2)_{i_{245}}(b)_{I}}{(c)_{i_{124}} (d)_{i_{235}}} \prod_{j=1}^{5}\frac{s_j^{i_j}}{i_j !},
\end{equation}
with the definitions $i_{jkl}=i_j+i_k+i_l$ and $I=\sum_{j=1}^{5} i_j$.

 In one special case, the answer is much simpler, namely when the conformality condition of $\Delta=D$ is satisfied. This happens in $D=4$, and can be thought of as the consequence of the $\phi^4$ operator being marginal in four dimensions. 
We have worked out this special case in \S \ref{exactcor}, leading to the correlator in \eqref{4dfunc}.

\subsection{Massive fields}
In the case of massive fields, we are interested in computing 
\begin{equation}
	    \mathcal{I}(x_1, \ldots , x_n) ={\mathcal{A} \over \pi^{D/2}}\int d^D y \prod_{i=1}^{n} G(x_i-y)
	\end{equation}
where the bulk to bulk propagator can be expressed in terms of the modified Bessel function of the second kind as 
\begin{equation}
    G(x_i-y)=\left({m_i^2 \over 4}\right)^{\Delta_i} \frac{K_{\Delta_i}(\sqrt{s_i})}{s_i^{\Delta_i/2}} 
\end{equation}
with $\Delta_i=\dm2$, $s_i=m_i^2 (x_i-y)^2$, and the overall normalization has been chosen to reproduce the massless result exactly, in the limit of all masses going to zero. As in the massless case, we drop the overall numerical factor $\mathcal{A}$ in the following.

In order to evaluate the bulk integral, we will find it convenient to use the following representation of the Bessel function, which will supply to us, a natural analog of the Schwinger parametrization employed in the massless case,
\begin{equation}
    K_{\nu}(z)=\half \left({z \over 2}\right)^{\nu} \int_0^{\infty} \frac{e^{-t-{z^2 \over 4 t}}}{t^{\nu+1}} dt
\end{equation}
With this, we get,
\begin{equation}
	    \mathcal{I}(x_1, \ldots , x_n) ={1 \over \pi^{D/2}} \half \prod_{i=1}^n {1 \over 2^{\Delta_i}} \left({m_i^2 \over 4}\right)^{\Delta_i}  \int_{0}^{\infty} dt_i t_i^{-\Delta_i-1}  e^{-t_i} \int d^D y  \, e^{-{1\over 4} \sum_{i=1}^{n} {m_i^2 \over t_i}(x_i-y)^2},
\end{equation}
so that the bulk integral is in the same form as in the massless case. Redefining ${m_i^2 \over 4 t_i} \rightarrow t_i$, introducing $T\equiv\sum\limits_{i=1}^n {t_i} $, and performing the bulk integral lands us at
\begin{equation} 
	    \begin{split}
	       \mathcal{I}(x_1, \ldots , x_n)&= \prod_i^n \int_{0}^{\infty} dt_i t_i^{\Delta_i-1} e^{- {m_i^2 \over 4 t_i} }{1 \over T^{D/2}} \exp(-{1\over T} \sum\limits_{\substack{i,j=1 \\ i \le j}}^{n} t_i t_j s_{ij}),
	    \end{split}
\end{equation}
Comparing with the massless expression in Eq. \ref{eq:schpara}, we see that the only difference is the additional factor of $e^{- {m_i^2 \over 4 t_i} }$ for each $i$, so that the massive answer smoothly goes over to the massless answer as all the masses are taken to zero. Further, we can find correlators with both massive and massless fields by taking some of the masses to zero in the above expression.
We can also express our answer as an integral over the corresponding massless $n$-point function by using
\begin{equation}
    e^{- {m_i^2 \over 4 t_i} }=\int_{c-i \infty}^{c+i \infty} {dp_i \over {2\pi i}} \Gamma(-p_i) \left({m_i^2 \over 4 t_i}\right)^{p_i}
\end{equation}
so that
\begin{equation} 
	    \begin{split}
	       \mathcal{I}(x_1, \ldots , x_n)&= \prod_i^n \int_{0}^{\infty} {dp_i  \over 2 \pi i} \Gamma(-p_i) \left({m_i^2 \over 4}\right)^{p_i} \int_{0}^{\infty} dt_i t_i^{\Delta_i-p_i-1} {1 \over T^{D/2}} \exp(-{1\over T} \sum\limits_{\substack{i,j=1 \\ i \le j}}^{n} t_i t_j s_{ij})\\
	       &=\prod_i^n \int_{0}^{\infty} {dp_i  \over 2 \pi i} \Gamma(-p_i) \left({m_i^2 \over 4}\right)^{p_i} \tgf (x_1, \ldots , x_n)
	    \end{split}
\end{equation}
where $\tgf (x_1, \ldots , x_n)$ is the $n$-point function of massless fields with scaling dimensions $\Delta_i-p_i$,  given by Eq. \ref{lampow} with the replacement $\Delta_i \rightarrow \Delta_i-p_i$.

\subsection{Singularity structure of exchanges and loops}\label{dflat1}
As a first example, consider the exchange contribution to the 4-point correlator coming from two cubic vertices, 

 \begin{equation} \label{gflate4}
	{\gf}_{e} =\int d^D y_1 \int d^D y_2\frac{1}{((x_1-y_1)^2)^{\Delta_1}} \frac{1}{((x_2-y_1)^2)^{\Delta_2}}  \frac{1}{((y_1-y_2)^2)^{\Delta_0}}   \frac{1}{((x_3-y_2)^2)^{\Delta_3}} 
	\frac{1}{((x_4-y_2)^2)^{\Delta_4}}\\
\end{equation}
 We make a change of spacetime coordinates to work in the frame of the "center of mass" and the "difference" coordinates,  
\begin{equation}
	y_{+}=\frac{y_1+y_2}{2},y_{-}=\frac{y_1-y_2}{2}
\end{equation}
so that the exchange contribution \eqref{gflate4} becomes
 \begin{equation} \label{e4y-+}
	{\cal G}_{e} =\int d^D y_- \frac{1}{(4 y_-^2)^{\Delta_0} }  \int d^D y_+\frac{1}{((x'_1-y_+)^2)^{\Delta_1}} \frac{1}{((x'_2-y_+)^2)^{\Delta_2}}   \frac{1}{((x'_3-y_+)^2)^{\Delta_3}} 
	\frac{1}{((x'_4-y_+)^2)^{\Delta_4}}\\
\end{equation}
where 
\begin{equation}
	x'_i=
	 	\begin{cases}
		  x_i+y_- &  \ i=1,2 \\
		  x_i-y_- & \ i=3,4
		\end{cases}
\end{equation}
The integral over the center of mass $y_+$ gives rise to a 4-point contact correlator with the insertion points shifted by the difference coordinate, so that we can express the exchange contribution as a spacetime integral over the difference coordinate of a $4$-point contact correlator weighted by a power of the difference coordinate thus

 \begin{equation} \label{e4y-}
	{\cal G}_{e} = 
	\int d^d y_{-}    \frac{1}{4(y_{-})^{2(\Delta_0)}} \, \gf_{\Delta_1, \Delta_2,\Delta_3,\Delta_4,}(x'_1,\ldots,x'_4),\\
\end{equation} 
 We note that the $y_-$ integral in \eqref{e4y-} is singular at $y_-=0$ and the residue at this singularity is the contact contribution to the four point correlator with insertions at $\{x_i\}$. This is an endpoint singularity in $y_-$. At finite $y_-$, the integral is singular whenever $\gf_{\Delta_1, \Delta_4,\Delta_3,\Delta_2}(x'_1,\ldots,x'_4)$ is singular. This singularity is again of the contact type with the insertions now at $\{x'_i\}$. 

Let us now consider loop contributions to the four-point function. The one-loop contribution with two quartic vertices is given by essentially the same integral as in \eqref{e4y-}. The one-loop contribution with four cubic vertices is given by
 \begin{equation} \label{gflatl4}
	\begin{split}
		{\cal G}_{\ell} =\prod_{i=1}^{4}\int d^D y_i  &\frac{1}{((x_1-y_1)^2)^{\Delta_1}} \frac{1}{((x_2-y_2)^2)^{\Delta_2}} 
		\frac{1}{((x_3-y_3)^2)^{\Delta_3}} 
		\frac{1}{((x_4-y_4)^2)^{\Delta_4}}\\
		& \frac{1}{((y_1-y_2)^2)^{\Delta_{-1}}} 
	\frac{1}{((y_2-y_3)^2)^{\Delta_{-2}}}
	\frac{1}{((y_3-y_4)^2)^{\Delta_{-3}}}
	\frac{1}{((y_4-y_1)^2)^{\Delta_{-4}}} \\
	& \\
	\end{split}
\end{equation}
Only three of the differences in $y$ coordinates that appear are linearly independent, we choose to replace $(y_4-y_1)=-(y_1-y_2)-(y_2-y_3)-(y_3-y_4)$ and move to the "center of mass" and "difference" frame defined by
\begin{equation}
	y_{+}=\frac{\sum_{j=1}^4 y_j}{4}; y_{-i}=\frac{y_i-y_{i+1}}{4}, \ \ i=1,2,3
\end{equation}

In these coordinates, we find that the one-loop contribution is given by
\begin{equation} 
\begin{split}
		{\cal G}_{\ell} =&\prod_{i=1}^{3} \int d^D y_{-i} \frac{1}{(16 y_{-i}^2)^{\Delta_{-i}} } 	\frac{1}{((y_{-1}+y_{-2}+y_{-3})^2)^{\Delta_{-4}}}\\
		& \int d^D y_+\frac{1}{((x'_1-y_+)^2)^{\Delta_1}} \frac{1}{((x'_2-y_+)^2)^{\Delta_2}}   \frac{1}{((x'_3-y_+)^2)^{\Delta_3}} 
	\frac{1}{((x'_4-y_+)^2)^{\Delta_4}}\\
\end{split}
\end{equation}
where 
$x'_i=x_i + \sum_{j=1}^3 A_{ij} y_{-j}$ with a matrix of numbers $A_{ij}$ .

As before, we recognize that the center of mass integral is the same as for a contact contribution and so, the 1-loop answer becomes

\begin{equation} 
	\begin{split}
		{\cal G}^{\flat}_{\ell} =&\prod_{i=1}^{3} \int d^D y_{-i} \frac{1}{(16 y_{-i}^2)^{\Delta_{-i}} } 	\frac{1}{((y_{-1}+y_{-2}+y_{-3})^2)^{\Delta_{-4}}} \gf_{\Delta_1,\Delta_2,\Delta_3,\Delta_4}(x'_1,\ldots,x'_4)\\
	\end{split}
\end{equation}
The integrand is singular when any one of the difference coordinates vanishes. This happens when one or more of the four sides of the loop is of zero length. When one side of the loop is contracted, we get a different one-loop diagram with three vertices - two cubic vertices and one quartic vertex. This diagram, in turn, has singularities when or more of the internal lines are contracted. If one internal line is contracted, we end up with a single exchange diagram and when both two internal lines are contracted, we find a (six-point) contact diagram. In totality then, all the singularities of the one-loop diagram with four cubic vertices arise from singularities of "sub-diagrams", i.e ones with lesser number of internal lines and same or lesser number of vertices. 

In general then, any contribution to the 4-point function can be expressed as spacetime integrals over the difference coordinates with the integrand being the product of a 4-point $\dflat$ function with a product of polynomials of difference coordinates. The integrand becomes singular when one or more of the internal lines collapse to give a "sub diagram" of the original diagram. The integrand is most singular when all the differences are zero i.e. all internal lines are collapsed, in which case one gets the contact contribution. Thus, the most singular part of any contribution to the $4$-point function is the one coming from the contact diagram, and so it suffices to study contact diagrams in order to extract the most singular part of correlators. We display this general form for a diagram with $\ell$ loops, $I$ internal lines
\begin{equation}
	\begin{split}
				{\cal G}_{\ell} =&\prod_{i=1}^{I-\ell} \int d^D y_{-i} \frac{1}{(y_{-i}^2)^{\Delta_{-i}} } 	\prod_{j=1}^{\ell}\frac{1}{((\sum_{i=1}^{\ell_{j}} y_{i})^2)^{\Delta_{-j}}} \gf_{\Delta_1,\Delta_2, \Delta_3,\Delta_4}(x'_1,\ldots,x'_4)\\
	\end{split}
\end{equation}

\newpage

\bibliographystyle{JHEP}
\bibliography{draft.bib}
 
 \end{document}